\documentclass[final,5p,times,twocolumn,authoryear]{elsarticle}
\usepackage{newsuse}
\usepackage[USenglish]{babel}

\graphicspath{{./figures/}}

\begin{document}

\begin{frontmatter}



\title{\LARGE Negative news posts are less prevalent and generate lower user engagement than non-negative news posts across six countries
}


\author[first]{Szymon Talaga}
\author[third]{Dominik Batorski}
\author[first,second]{Magdalena Wojcieszak}
\affiliation[first]{organization={Center for Excellence in Social Science, University of Warsaw}, 
            country={Poland}}
\affiliation[second]{organization={Department of Communication, University of California, Davis}, 
            country={United States}}
\affiliation[third]{organization={Interdisciplinary Centre for Mathematical and Computational Modelling, University of Warsaw}, 
            country={Poland}}

    \begin{abstract}
Although news negativity is often studied, missing is comparative evidence on the prevalence of and
engagement with negative political and non-political news posts on social media.  We use \num{6081134}
Facebook posts published between January 1, 2020, and April 1, 2024, by 97 media organizations in six
countries (U.S., UK, Ireland, Poland, France, Spain) and develop two multilingual classifiers for
labeling posts as (non-)political and (non-)negative. We show that: 
(1)~negative news posts constitute a relatively small fraction (${\sim}$12.6\%);
(2)~political news posts are neither more nor less negative than non-political news posts;
(3)~U.S. political news posts are less negative relative to the other countries on average (${\sim}$40\% lower odds);
(4)~Negative news posts get ${\sim}$15\% fewer likes and ${\sim}$13\% fewer comments than non-negative news posts. 
Lastly, (5)~we provide estimates of the proportion of the total volume of user engagement with
negative news posts and show that only between 10.2\% to 13.1\% of engagement is linked to negative posts by the analyzed news organizations.


\end{abstract}



\end{frontmatter}

Negativity is a core characteristic of news and the saying \enquote{if it bleeds, it leads} has described journalistic decisions about what to cover for 
decades~\citep{harcupWhatNewsNews2017,semetkoFramingEuropeanPolitics2000}. As a result, news outlets produce and share negative content online and on social media platforms, and this tendency has been increasing in recent years~\citep[e.g.~][]{robertsonNegativityDrivesOnline2023,watsonNegativeOnlineNews2024}. This is in part because negativity is said to generate more attention, stimulate information 
seeking~\citep{parkApplyingNegativityBias2015,sorokaNegativityDemocraticPolitics2014} and---among some citizens---promote engagement with news~\citep{trusslerConsumerDemandCynical2014} and greater news sharing on platforms~\citep{bellovaryLeftRightLeaningNews2021,schoneNegativitySpreadsMore2021,schoneNegativeExpressionsAre2023}. For example, negative words in headlines increase click-through rates on online news~\citep{robertsonNegativityDrivesOnline2023} and social media users are more likely to share links to negative than positive news articles~\citep{watsonNegativeOnlineNews2024}, especially when they are about the political out-group~\citep{yuPartisanshipSocialMedia2024,rathjeOutgroupAnimosityDrives2021}.

Here, we add to the important work on negativity in news in four key ways. First, we offer large-scale \textit{comparative} evidence on the prevalence of \textit{and} engagement with negative posts from news media organizations on social media. Because most research comes from the U.S., it is not clear whether the findings generalize to other countries. On the one hand, there should be similar levels of negativity in and user engagement with negative news posts across countries. After all, decades of research have documented a negativity bias in human judgment~\citep{baumeisterBadStrongerGood2001}.
The bias applies to a variety of domains, ranging from emotion or neurological processes, to attention and memory, with stronger activation and response to negative stimuli, and is attributed to the greater adaptive relevance of negative events and information. From the perspective of evolutionary psychology, the organism is more likely to survive by paying more attention to negative over positive stimuli. Accordingly, negativity and conflict are widespread news values~\citep{zhangNegativeNewsHeadlines2024}, audiences across cultures are similarly activated by negative news content~\citep{sorokaCrossnationalEvidenceNegativity2019}, 
and users in several countries react to and share negative news on platforms more than positive news ~\citep{deleonSadnessBiasPolitical2021,heidenreichExploringEngagementEU2022,watsonNegativeOnlineNews2024,bellovaryLeftRightLeaningNews2021}. On the other hand, negative news may be more prevalent and appealing in the U.S. than in other democracies due to its more polarized news media system and relatively weak public service 
broadcasting~\citep{hallinComparingMediaSystems2011,bruggemannHallinManciniRevisited2014}. 
For instance, U.S. news outlets published a substantially higher proportion of negative news about COVID-19 than non-U.S. news outlets (91\%~vs.~54\%) \citep{sacerdote2020all} and, in some cases, it is the positivity of the news, not negativity, that predicts news sharing and other engagement 
forms~\citep{trillingNewsworthinessShareworthinessHow2017,bergerWhatMakesOnline2012,camajAudienceLogicElection2024,muddimanNegativityBiasBacklash2020}.

Second, we examine user engagement with both political and non-political news posts. After all, the concerns about negativity in news---and especially its democratic consequences---primarily pertain to coverage of candidates, elections, the economy, social groups, among other political issues ~\citep{kleinnijenhuisNegativeNewsSleeper2006, cappellaNewsFramesPolitical1996,shehataGameFramesIssue2014,mutzNewVideomalaiseEffects2005}. Yet, the majority of content in news outlets is \textit{not} about politics~\citep{flaxmanFilterBubblesEcho2016} and only about a quarter of visits to news websites is to political news, with the bulk being to articles about sports, weather, lifestyle, or entertainment~\citep{wojcieszakNonNewsWebsitesExpose2024}. Most past work examines user engagement with \textit{all} negative articles produced by a news media organization on platforms ~\citep{trillingNewsworthinessShareworthinessHow2017,zhangNegativeNewsHeadlines2024,watsonNegativeOnlineNews2024,robertsonNegativityDrivesOnline2023} or focuses on strictly political contexts without direct comparisons against a non-political baseline~\citep{deleonSadnessBiasPolitical2021,heidenreichExploringEngagementEU2022,bellovaryLeftRightLeaningNews2021}. The few studies that differentiate between topics solely test user \textit{engagement}, not the prevalence of negative news, and rely on topic modeling to create diverse, often non-mutually exclusive, categories such as politics, sports, or local and global news ~\citep{watsonNegativeOnlineNews2024}, or entertainment, LGBTQ, government and economy, or parenting and school~\citep{robertsonNegativityDrivesOnline2023}. We aim to offer at-scale estimates of the differential prevalence of \textit{and} engagement with negativity in political vs. non-political news across countries. 

Third, we integrate three distinct engagement behaviors studied separately in past work. We examine the Facebook reactions and sharing of negative news posts \textit{and also} commenting on such content. These three behaviors reflect distinct affordances of social media platforms and have different implications for users themselves and online discourse at large. Reactions are rather passive and \enquote{private} in nature, whereas sharing and commenting represent more active engagements with news, are more public, and have greater impact on the online public 
sphere~\citep{wojcieszakMostUsersNot2022}. Diffusion through sharing increases information reach on social media~\citep{eadyHowManyPeople2019} and can generate exponential increases in user exposure to negativity. In addition, a study examining the three behaviors separately---but only for one topic and in one country---found that they are differently influenced by 
negativity~\cite{heidenreichExploringEngagementEU2022}. We offer evidence on \textit{all} engagement forms the platform provides, including comparisons between political and non-political news posts and across multiple countries.

Our fourth core contribution lies in putting in perspective extant concerns about negativity by examining the relative volume, or proportion, of engagement with negative \textit{and} positive 
news posts on platforms. Previous research \textit{a priori} zooms on negativity online, testing its 
popularity or propagation through social 
networks~\citep{deleonSadnessBiasPolitical2021,heidenreichExploringEngagementEU2022,trillingNewsworthinessShareworthinessHow2017,zhangNegativeNewsHeadlines2024,watsonNegativeOnlineNews2024,schoneNegativitySpreadsMore2021,ferraraMeasuringEmotionalContagion2015}. At the same time, multiple studies point to a general 
positivity bias in human language. This also applies to social media content. For instance, \cite{doddsHumanLanguageReveals2015} find significant positivity bias across 24 text corpora and 10 languages, without---however---accounting for exogenous or topical negativity (i.e.~reporting on negative events, such as 
disasters, violent crimes, terror attacks). Following past conceptual work, we combine 
\enquote{the mere dissemination of negative news} (exogenous negativity coming into the news from outside, that is, from the event itself) and media-initiated 
negativity~\citep[%
    \enquote{endogenous negativity imposed on news by journalists through their use of language};%
][~p.~181]{lengauerNegativityPoliticalNews2011}

We use \num{6081134} Facebook posts published between January 1, 2020, and April 1, 2024, by 97 news media organizations in six countries. Sec.~\ref{sec:methods:data} provides more details about the dataset and the data collection process. 
In particular, we use data on posts published by 37 major U.S.-based news outlets (\num{2274210} posts), 12 outlets from the United Kingdom (\num{959141} posts), 11 from Ireland (\num{451783}), 17 from Poland (\num{842321} posts), 8 from France (\num{355697} posts), and 12 news outlets from Spain (\num{1197982} posts); for a full list of outlets, see~\ref{app:sec:data}). 
For each post published by each news outlet, we obtained the text content, posting time, and the core interaction metrics --- 
numbers of Facebook reactions, comments, and shares 
(the breakdown of the three engagement metrics by country is in Table~\ref{app:tab:engagement} in \ref{app:sec:data}). 
The outlets were selected to provide a good coverage of each country's news media landscape with regard to political leaning and both quality and tabloid outlets.

To identify political and negative news posts, we develop two multilingual binary classifiers for automated labeling, both based on a moderately large transformer model known as XLM-RoBERTa-Large ~\citep{conneauUnsupervisedCrosslingualRepresentation2020}, and use them to label all posts as political vs non-political and negative vs non-negative 
(see Sec.~\ref{sec:methods:classifiers} for details on model training, performance, and validation). 
This allows us to estimate the prevalence of negativity in political and non-political posts published by all analyzed news outlets as well as the users' engagement with such news posts. Sec.~\ref{sec:methods:analysis} provides details on our methodological approach and statistical methods.

We use those data to systematically and comparatively test four progressively specific questions: (1)~Are news posts about politics more negative overall than non-political news posts? 
(2)~What are the differences in the levels of negativity in political and non-political news posts across 
the tested countries? (3)~What are the relative levels of users' engagement---in terms of reactions, 
comments, and shares---with negative news posts in general and also negative political and negative 
non-political news posts in particular? 
Lastly, (4)~what are the relative levels of user engagement with negative news posts and with political versus non-political news posts when accounting for the prevalence of negative versus non-negative news?

Our work offers five key findings: 
(1)~We demonstrate that only a relatively small fraction of 
posts published by these 97 news organizations on Facebook over four years is negative 
($12.6\%$, CI$_{0.95}$: $[10.9\%, 14.6\%]$).
(2)~News posts about political topics, such as elections, political parties, minority groups, climate change, among others, are neither systematically more nor less negative than non-political news posts about sports, celebrities, or the weather.
(3)~We find significant differences in prevalence of negativity
between countries only for political news, with the U.S. having
less negative political news than the average over the other countries
(${\sim}40\%$ lower odds, $p < 0.001$), 
similarly Poland (${\sim}37\%$ lower odds, $p \approx 0.016$),
and Spain having higher prevalence of negative political news
(${\sim}99\%$ higher odds, $p < 0.001$).
(4)~We provide evidence that negative posts get ${\sim}15\%$ 
fewer reactions ($p < 0.001$) and ${\sim}13\%$ fewer comments 
($p < 0.001$) than non-negative news 
(overall effects averaged across countries). 
This is especially the case for negative political news posts, which receive even fewer reactions and comments, 
respectively ${\sim}17\%$ ($p < 0.001$) and ${\sim}18\%$ ($p < 0.001$), than non-negative political news. For non-political news posts, negativity is not related to user engagement (reactions, comments and shares).
(5)~We put extant concerns about online negativity
in perspective and provide estimates of the proportion of the total volume
of user engagement with negative posts from news organizations. We simulate the overall share of 
engagement with negative news under different assumptions about the prevalence of political news posts, 
and consistently find that only a minority of the total engagement volume 
(roughly between 10.2\% and 13.1\%) is linked to negative news.

Before we outline the data and the results, we note that we focus on Facebook posts from news organizations and cannot test the availability of and engagement with negative news on news websites or in offline formats of these outlets. Although news organizations post only a subset of their total output on social media and the analyzed posts include only headlines and a few sentences about the articles, in \ref{app:sec:cls-consistency} we show that there is high 
(${\sim}82\%$) correspondence between the posts and the full articles in terms of their political or non-political focus and negativity wherein (see \ref{app:sec:cls-consistency}). Similarly, we do not speak to the effects of these news posts and users' reactions on polarization, outgroup hostility, and other key outcomes often attributed to negativity~\citep{boukesNewsConsumptionIts2017}, examining instead the differential patterns of posting and user engagement with posts by new media organizations. 

\section{Results}\label{sec:results}

The analyses presented below utilize the dataset and analytical approach described in
Secs. \ref{sec:methods:data} and \ref{sec:methods:analysis}. Political and negativity labels for
individual news posts were inferred from the post content using machine learning classifiers specifically developed for this analysis and detailed in~\ref{app:sec:cls}. 
All statistical tests and confidence intervals use two-sided significance level $\alpha = 0.05$ 
(and confidence level $1 - \alpha$). If not specified otherwise, confidence intervals are derived
using asymptotic normality of Maximum Likelihood estimators. Multiple related tests and 
corresponding confidence intervals are based on simultaneous
inference~\citep{hothornSimultaneousInferenceGeneral2008} 
and use family-wise Holm-Bonferroni correction~\citep{holmSimpleSequentiallyRejective1979}.
In all cases, \enquote{overall} values correspond to (uniform) averages over estimates for
subgroups, e.g.~countries and/or political and non-political posts.

\begin{figure}[tb!]
\centering

\begin{subfigure}[t]{\columnwidth}
    \centering
    \caption{}
    \vspace{-.1em}
    \includegraphics[width=\textwidth]{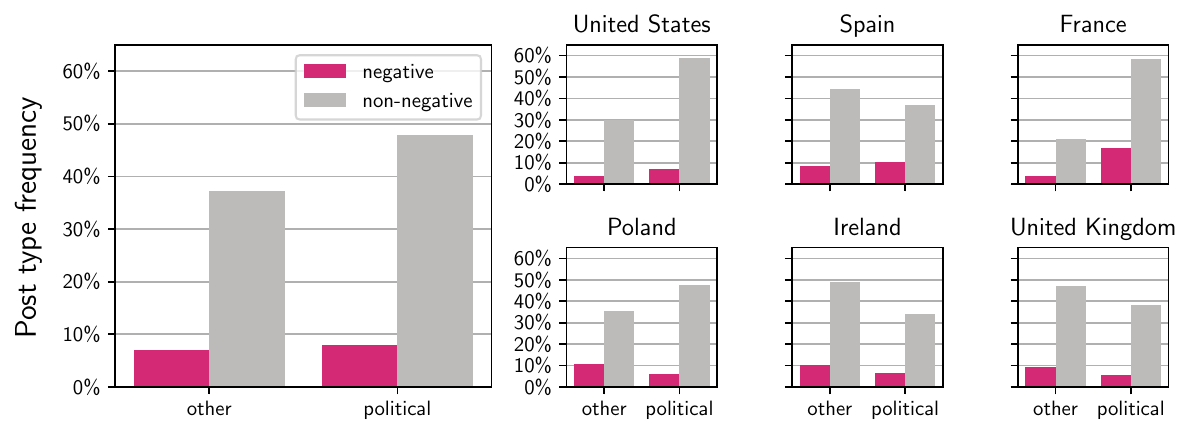}
    \tiny
    \sffamily
    \begin{tabularx}{\columnwidth}{X|rrr|rrr}
        \toprule
         & \multicolumn{3}{c}{} & \multicolumn{3}{c}{\% of} \\
         & outlets & posts & (\%) & political & negative & negative $\mid$ political \\
        \midrule
        United States & 37 & 2,274,210 & 37.4 & 66.5 & 10.7 & 10.5 \\
        Spain & 12 & 1,197,982 & 19.7 & 47.6 & 18.6 & 21.8 \\
        France & 8 & 355,697 & 5.8 & 75.7 & 19.7 & 21.4 \\
        Poland & 17 & 842,321 & 13.9 & 53.9 & 17.4 & 12.0 \\
        Ireland & 11 & 451,783 & 7.4 & 41.1 & 17.0 & 16.3 \\
        United Kingdom & 12 & 959,141 & 15.8 & 43.7 & 14.5 & 12.0 \\
        \midrule
        \textbf{Overall} & 97 & 6,081,134 & 100.0 & 56.1 & 14.8 & 13.9 \\
        \bottomrule
    \end{tabularx}
\end{subfigure}

\begin{subfigure}[t]{\columnwidth}
    \centering
    \caption{}
    \vspace{-.4em}
    \includegraphics[width=\textwidth]{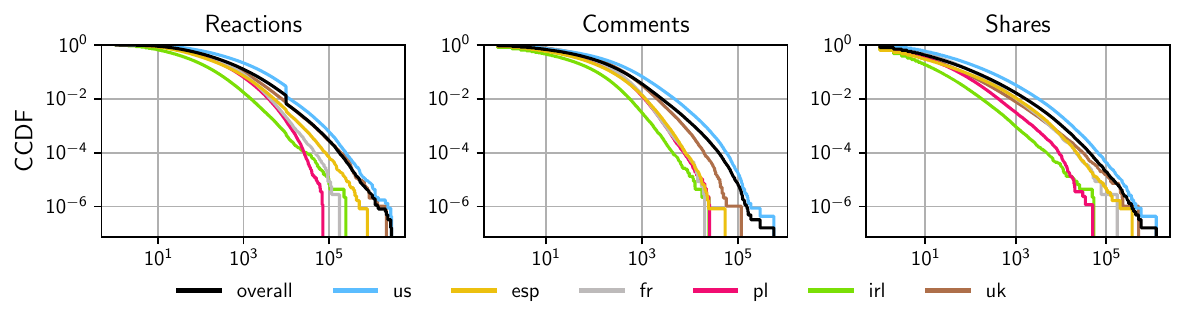}
    \tiny
    \sffamily
    \begin{tabularx}{\columnwidth}{l|XXX|XXX|XXX}
    \toprule
         & \multicolumn{3}{c}{Reactions} & \multicolumn{3}{c}{Comments} & \multicolumn{3}{c}{Shares} \\
         & mean & median & IQR & mean & median & IQR & mean & median & IQR \\
        \midrule
        United States & 1291.7 & 175.0 & 718 & 340.4 & 43.0 & 200.0 & 197.2 & 14.0 & 59 \\
        Spain & 368.8 & 48.0 & 175 & 94.2 & 15.0 & 69.0 & 62.1 & 2.0 & 13 \\
        France & 372.9 & 93.0 & 259 & 117.0 & 32.0 & 104.0 & 72.4 & 8.0 & 30 \\
        Poland & 295.8 & 62.0 & 197 & 104.0 & 25.0 & 90.0 & 30.2 & 4.0 & 14 \\
        Ireland & 116.4 & 19.0 & 66 & 45.1 & 7.0 & 33.0 & 12.8 & 1.0 & 6 \\
        United Kingdom & 590.5 & 87.0 & 276 & 179.4 & 38.0 & 136.0 & 57.8 & 3.0 & 16 \\
        \midrule
        \textbf{Overall} & 720.3 & 85.0 & 326 & 198.7 & 27.0 & 119.0 & 104.5 & 5.0 & 26 \\
        \bottomrule
    \end{tabularx}
\end{subfigure}

\begin{subfigure}[t]{\columnwidth}
    \centering
    \caption{}
    \vspace{-.4em}
    \includegraphics[width=\textwidth]{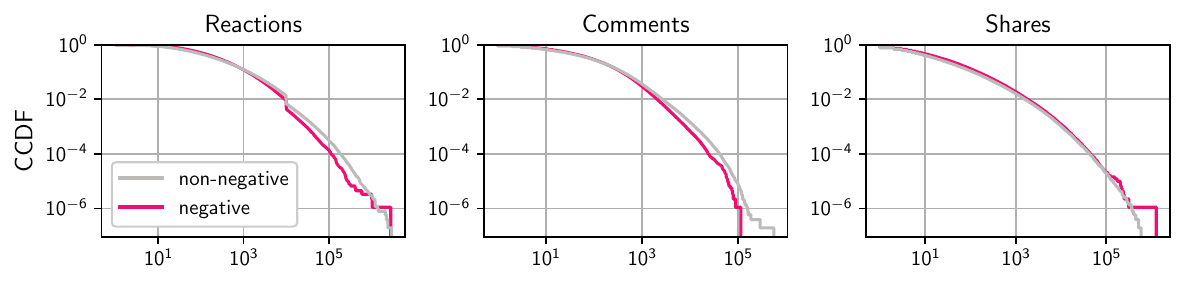}
    \tiny
    \sffamily
    \begin{tabularx}{\columnwidth}{ll|XXX|XXX|XXX}
    \toprule
     &  & \multicolumn{3}{c}{Reactions} & \multicolumn{3}{c}{Comments} & \multicolumn{3}{c}{Shares} \\
     &  & mean & median & IQR & mean & median & IQR & mean & median & IQR \\
    \midrule
    \multirow[t]{2}{*}{non-political} & negative & 613.7 & 107.0 & 355 & 134.6 & 24.0 & 91 & 95.4 & 7.0 & 28 \\
     & non-negative & 679.3 & 59.0 & 250 & 129.4 & 15.0 & 67 & 80.6 & 4.0 & 16 \\
    \multirow[t]{2}{*}{political} & negative & 624.7 & 106.0 & 346 & 203.2 & 41.0 & 145 & 143.2 & 10.0 & 40 \\
     & non-negative & 784.6 & 104.0 & 379 & 262.4 & 44.0 & 162 & 119.9 & 8.0 & 32 \\
    \bottomrule
    \end{tabularx}
\end{subfigure}

\caption{
    Basic descriptive statistics for the sample.
    \textbf{a}~Numbers of outlets and posts as well frequencies of political 
    and/or negative news posts, including frequencies of negative news posts among political news posts
    (negative $\mid$ political), overall and by country.
    \textbf{b}~Engagement metrics, overall and by country. Plots present empirical complementary
    cumulative distribution functions (CCDF).
    \textbf{c}~Overall engagement metrics for (non-)political and (non-)negative news posts with plots presenting CCDFs.
}
\label{fig:descriptives}
\end{figure}

Fig.~\ref{fig:descriptives} presents basic descriptive statistics based on raw data, that is,
using a single post as a unit of observation. These descriptives reveal several important patterns.
\footnote{
we emphasize that we do not analyze these raw data with simple methods like $\chi^2$ and proportion tests or standard (generalized) linear models because:
(1)~the sample is not random, and it is hard to conceive of a well-defined and practically usable notion of a population of news 
posts---see detailed discussion in Sec.~\ref{sec:methods:analysis};
(2)~the data are clustered along multiple dimensions 
(i.e.~countries, outlets and time)
meaning that the fundamental assumption of independence, required by all standard statistical test,
is severely broken, rendering such tests meaningless.
} 
First, negative posts are much less frequent than non-negative
posts, and this is relatively stable across both political and non-political news posts
as well as different countries. The few differences between negativity in political
and non-political news posts (also in different countries) are small in magnitude 
(Fig.~\ref{fig:descriptives}a). Second, as evident in Fig.~\ref{fig:descriptives}b-c,
engagement metrics have markedly right-skewed and heavy-tailed distributions with pronounced
differences in dispersion between countries (e.g.~the right tail of the distribution
for the U.S. is longer and thicker than for Ireland or Poland).
This insight informed our modeling strategy (see Sec.~\ref{sec:methods:analysis}).
Detailed descriptive statistics for raw data, including outlet-level information,
can be found in~\ref{app:sec:data}. 

\subsection{Prevalence of negative political and non-political news posts across countries}\label{sec:results:negativity}

As detailed in Sec.~\ref{sec:methods:analysis}, we use Generalized Linear Mixed Models (GLMMs) to estimate \enquote{population}-level quantities based on partial pooling~\cite[see ch.~12 in][]{gelmanDataAnalysisUsing2021}. In this approach, fixed effects capture the characteristics of
a typical post published by a typical news outlet on a typical day. This method avoids naive aggregation across posts, addressing potential under- or over-representation of particular outlets or time periods. Since observed posting frequencies may be different from future or unobserved frequencies, imbalances in the number of posts between news outlets or across time should have limited influence on the estimates. Instead, our estimates primarily reflect cross-outlet and temporal variation.
\footnote{ 
Apart from allowing for a more general out-of-sample interpretation, estimates 
obtained this way are both shielded from the overfitting that is typical for models with too many fixed effects as well as are relatively free from potential sample biases, which could potentially 
arise from outlet selection, incomplete posts, or transient time-dependent deviations 
from long-term averages.
}

\begin{figure*}[htb!]
\centering
\begin{minipage}[t]{.495\textwidth}

\begin{subfigure}[t]{\textwidth}
    \setlength{\abovedisplayskip}{0pt}
    \setlength{\belowdisplayskip}{0pt}
    \setlength{\abovedisplayshortskip}{0pt}
    \setlength{\belowdisplayshortskip}{0pt}
    \definecolor{avg}{HTML}{bf8a03}
    \definecolor{int}{HTML}{848383}
     
    \tiny\sffamily
    \centering
    \caption{}\vspace{1em}
    \textbf{Models for estimating marginal expectations}
    \rule{.925\textwidth}{1pt}
    \vspace{.5em}
    \begin{alignat*}{3}
        \setlength{\tabcolsep}{0pt}
        &\eta(\mathbb{P}(\text{\textbf{negative}})) &~\sim\quad
        &\text{\textcolor{avg}{political}} \times \text{\textcolor{avg}{country}}
        &~\mid
        \text{\textcolor{int}{outlet}} + \text{\textcolor{int}{time}}
        \\
        &\eta(\mathbb{E}(\text{\textbf{engagement}})) &\sim\quad
        &\text{\textcolor{avg}{political}} 
        \times \text{\textcolor{avg}{negativity}}
        \times \text{\textcolor{avg}{country}}
        &\mid
        \text{\textcolor{int}{outlet}} + \text{\textcolor{int}{time}}
    \end{alignat*}\smallskip
    \rule{.925\textwidth}{.1pt}
    \vspace{-1em}
    \begin{minipage}[t]{.9\textwidth}
    \tiny
        $\eta(\cdot)$ --- link function \hfill
        \textcolor{avg}{Fixed effects} \hfill 
        \textcolor{int}{Random effects} \hfill
    \end{minipage}
    \vspace{4.5em}
\end{subfigure}

\begin{subfigure}[t]{\textwidth}
    \centering
    \caption{}\vspace{0em}
    \includegraphics[width=.99\textwidth]{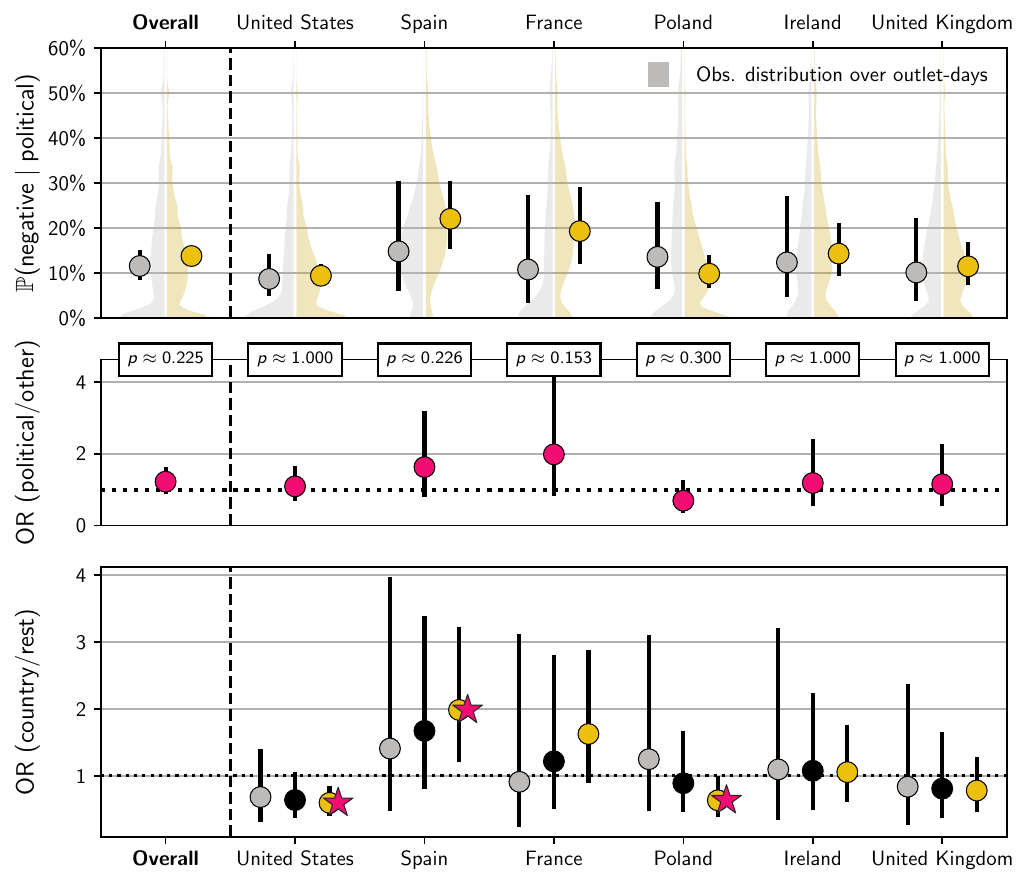}
\end{subfigure}

\end{minipage}
\hfill
\begin{minipage}[t]{.495\textwidth}

\begin{subfigure}[t]{\textwidth}
    \centering
    \caption{}\vspace{-.6em}
    \includegraphics[width=.93\textwidth]{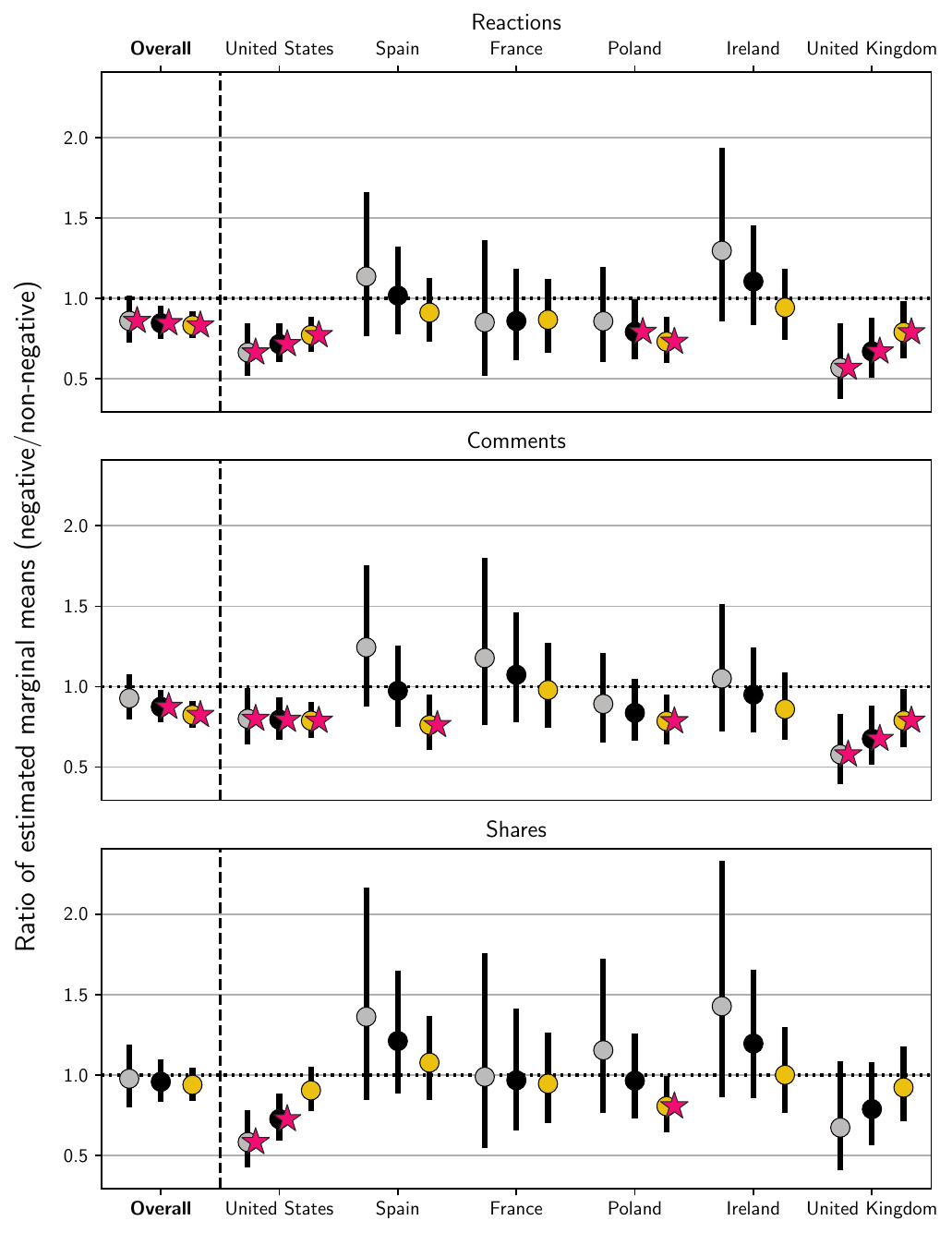}
\end{subfigure}
\end{minipage}

\centering
\includegraphics[width=.8\textwidth]{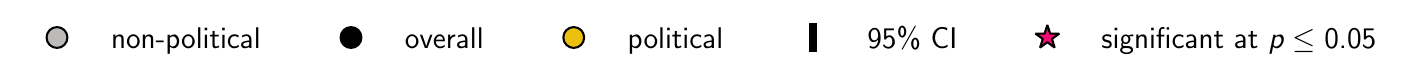}

\caption{
    Population-level estimates and contrasts based on Generalized Linear Mixed Models (GLMMs)
    (i.e.~with random intercept and slope effects for news outlets and time averaged out).
    The estimates and standard errors are derived from estimated marginal means for linear 
    predictors, with the overall estimate being the uniform average over countries.
    Logistic regression with mixed effects was used for predicting negativity,
    and negative binomial regression with mixed effects and non-constant 
    dispersion was used to model engagement counts.
    \textbf{a}~Simplified summary of the specification of GLMMs for predicting negativity
    and engagement counts (reactions, comments, shares). See Sec.~\ref{sec:methods:analysis}
    for more details.
    \textbf{b}~Predicted prevalence of negativity
    across political and non-political news posts 
    and across countries as well as within-country 
    (political vs. non-political) and between-country
    contrasts (country vs the average over the other countries)
    expressed in terms of odds ratios (OR).
    Density plots represent observed distributions of fractions of negative news posts 
    across outlet-days. All tests and confidence intervals related to a given contrast type
    (i.e.~political vs non-political or country/rest) are treated as a single family for the purpose
    of the multiple test adjustment.
    \textbf{c}~Comparisons in groups (political vs non-political and overall news posts) and by 
    countries in terms of relative difference between the expected engagement with negative 
    and non-negative news posts. All tests and confidence intervals for a given engagement metric 
    (i.e.~reactions, comments, or shares) are treated as a single test family and adjusted accordingly.
}
\label{fig:glmm}
\end{figure*}

Fig.~\ref{fig:glmm} plots the estimates of the overall and country-specific prevalence of negative political and non-political news posts as well as the relative impact of negativity on user 
engagement metrics in different countries.
Detailed numerical values of all estimates can be found in \ref{app:sec:results}.

As seen in Fig.~\ref{fig:glmm}b, there is little variation in terms of the prevalence of negativity in
among political and non-political Facebook posts by the analyzed news organizations, both overall and \textit{within} the specific countries. 
Indeed, the statistically insignificant tests of contrasts (Fig.~\ref{fig:glmm}b)
suggest there are no systematic differences in negativity in political vs. non-political 
posts, both on average and at the level of specific countries. In short, answering RQ1, political news posts are not more or less negative than news posts about non-political topics.

We test for specificity of individual countries by comparing their negativity levels with
averages taken across other countries (Fig.~\ref{fig:glmm}, country/rest contrasts).
The only significant differences are for political news,
with the U.S., which could be expected to have more negative news
than countries with less polarized multi-party systems, 
having less negative political news 
($\OR = 0.595~[0.432, 0.820]; p < 0.001$), 
similarly Poland ($\OR = 0.633~[0.418, 0.959], p \approx 0.016$),
and Spain having higher prevalence of negative political news
($\OR = 1.986~[1.234, 3.194], p < 0.001$).

Although news posts in the U.S. could be expected to be more negative than news in less polarized multi-party systems, the U.S. is the only country with significantly \textit{lower}
negativity in general ($\OR = 0.611~[0.376, 0.994]; p = 0.040$) and in political news posts in 
particular ($\OR = 0.583~[0.415, 0.818]; p < 0.001$). The only other country deviating from
the average is Spain, which has significantly higher levels of negativity in political news posts
($\OR = 2.03~[1.23, 3.36]; p < 0.001$) than the cross-country average. Addressing our RQ2, 
the prevalence of negativity in U.S. news posts is lower than in the other countries, higher in Spain, and largely similar in  the remaining countries tested for political and non-political posts.

\subsection{News negativity and engagement}\label{sec:results:engagement}

With regard to user engagement, Fig.~\ref{fig:glmm}c shows overall and country-specific ratios of estimated marginal means, or mean ratios (MR), for numbers of reactions, 
comments, and shares, between negative and non-negative news posts for political and non-political posts. 
The results clearly show that negative news posts generate significantly
\textit{less} engagement than non-negative news posts. This effect is particularly pronounced
for reactions and comments, for which overall (country-averaged) effects are statistically significant 
for all posts
(reactions: $\MR = 0.85~[0.77, 0.93]; p < 0.001$; 
comments: $\MR = 0.87~[0.79, 0.96]; p < 0.001$)
and for political posts specifically
(reactions: $\MR = 0.83~[0.77, 0.91]; p < 0.001$; 
comments: $\MR = 0.82~[0.76, 0.89]; p < 0.001$)
while the same effect is not significant for shares. Country-specific effects follow a similar negative pattern, with more statistically significant results for reactions and comments and fewer for 
shares (see Fig.~\ref{fig:glmm}c). 
Importantly, sharing in general is not affected by negativity,
apart from a few negative country-specific effects for the U.S. 
and Poland, while the link between negativity and reactions and 
commenting is much stronger---negativity decreases reactions for all
kinds of news and commenting for political news.
In short, answering RQ3, negativity in general is either unrelated
or linked to lower engagement. This depressing effect of negativity
applies primarily to reactions and comments, with effects for reactions
being universal while for comments more linked to political news.
At the same time, there is also a noticeable cross-country variation
in the magnitude of this effects---many country-specific effects are
null---but not in direction. Thus, our results suggests that negativity
effects on engagement are generally either null or depressing.
\footnote{
We again emphasize that the reported effects concern ratios of means, that is, the relationship between engagement counts for negative and other posts averaged over large numbers of posts by the analyzed news media organizations. 
This differs from estimating the average engagement ratio between two random posts, 
which would be much more influenced by the heavy-tailed distribution of engagement counts, 
leading to less stable results. 
Thus, our results should not be directly translated to comparisons between individual posts, 
which will always have substantially higher variance as well as mean values.
This is a direct consequence of the Central Limit Theorem and 
Jensen's inequality~\citep[][p.~46]{lehmannTheoryPointEstimation1998}.
}

\subsection{Relative volume of engagement with negative news posts}\label{sec:results:volume}

\begin{figure*}[htb!]
\centering
\begin{subfigure}[t]{.495\textwidth}
    \caption{}
    \centering
    \includegraphics[width=\textwidth]{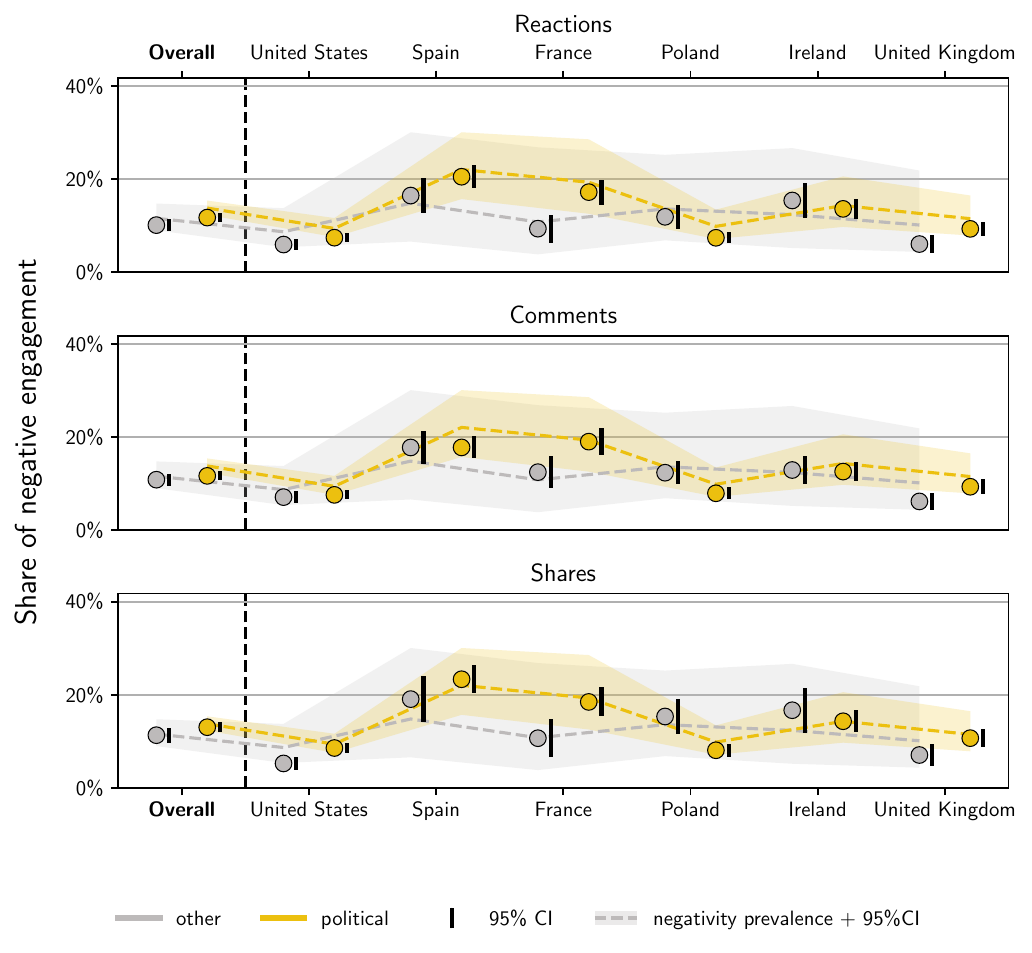}
\end{subfigure}
\hfill
\begin{subfigure}[t]{.495\textwidth}
    \caption{}\vspace{1em}
    \centering    
    \includegraphics[width=\textwidth]{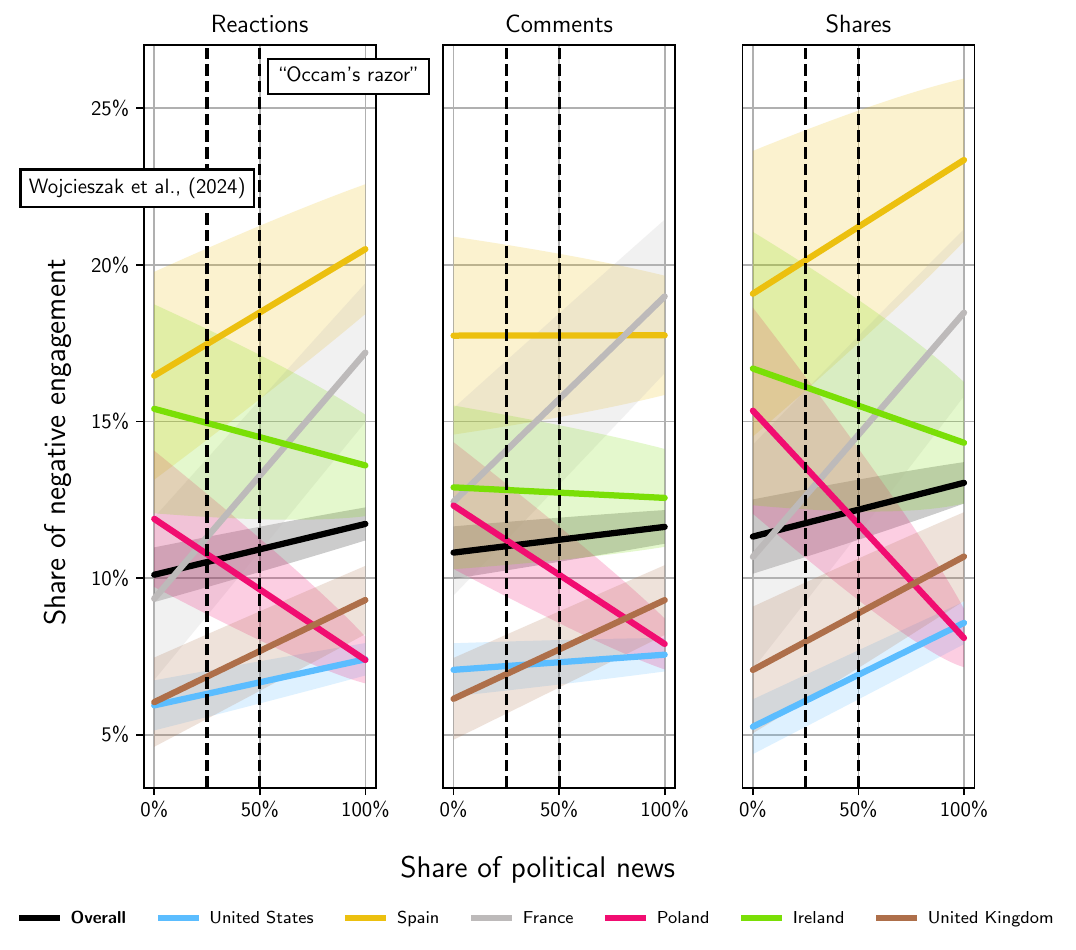}
\end{subfigure}
\caption{%
        Share of engagement with negative news posts.
        \textbf{a}~Shares across political and non-political news posts. Lines and bounds denote the 
        corresponding negativity prevalence and associated confidence intervals.
        \textbf{b}~Overall negativity shares assuming different shares of political news posts, $p$.
        Bounds denote confidence intervals and vertical lines correspond to two cases of special
        interest: $p = 0.5$ (\enquote{Occam's razor}, 
        and $p = 0.25$ \citep{wojcieszakNonNewsWebsitesExpose2024}.
    }
\label{fig:share}
\end{figure*}

We now test what proportion, or relative volume, of user 
engagement is associated with negative news posts accounting for the prevalence of negative and non-negative news posting in the six countries. For a given conditional prevalence of negative posts, 
$q = \Prob(\text{negative} \mid \ldots)$, where $\ldots$ represents a (possibly empty) sequence
of conditioning variables, we define the share of negative engagement as the weighted ratio of 
the expected engagement count for negative news posts and the total expected engagement count across both 
negative and non-negative news posts:
\begin{equation}\label{eq:share}
    \mathcal{S}(X \mid q, \ldots) = \frac{
        q\E[X \mid N{=}1, \ldots]
    }{
        q\E[X \mid N{=}1, \ldots] + (1{-}q)\E[X \mid N{=}0, \ldots]
    }
\end{equation}
where $X$ represents the engagement count (e.g.~reactions), $\E[X \mid \ldots]$ 
denotes the conditional expectation operator, and $N = 1$ or $N = 0$ 
indicates whether a post is negative or not. 
Finally, we obtain a general proportion of negative engagement, $\mathcal{S}(X \mid \ldots)$, 
by marginalizing out $q$ using the fact that Maximum Likelihood estimators are asymptotically normal. See Sec.~\ref{sec:methods:share} for more details, including the derivation of 
confidence intervals.

As shown in Fig.~\ref{fig:share}a, across all engagement forms---reactions, comments, and shares---the proportion of engagement with negative news posts---both political and non-political---is relatively
modest. That is, engagement with negative news posts closely follows their overall prevalence, ranging from  8\% to 22\%, and is relatively stable across countries.

As importantly, we combine the estimates for political and non-political news posts to assess the overall relative volume of engagement with negative news posts under different assumptions about the 
prevalence of news negativity in the population. Such a multi-scenario analysis is more robust and 
meaningful, as estimates of the prevalence of political news posts may vary greatly depending on sample construction or operational definition of what counts as \enquote{news} 
(see Sec.~\ref{sec:methods:analysis} for a related discussion).
Hence, we simulate the entire ranges of plausible outcomes
(see Sec.~\ref{sec:methods:share} for details) instead of focusing on specific cases.

Fig.~\ref{fig:share}b plots the results of the simulation for 
reactions, comments, and shares.
Vertical lines mark two cases of particular interest: (1)~\enquote{Occam's razor}-like assumption
using $p = \Prob(\text{political} \mid \ldots) = 1/2$ 
(this case is also similar to the overall prevalence in our raw data, $p \approx 0.56$); 
and (2)~empirical estimate from \cite{wojcieszakNonNewsWebsitesExpose2024} rounded to $p = 0.25$.
In general, the overall proportions are naturally similar to the proportions specific for 
political and non-political news posts, and the magnitude of cross-country differences is also similar.
What is more important, these results accommodate both the correlations between the prevalence
of political and negative news posts, as well as the link thereof with engagement, and show the combined
effects vis-à-vis the prevalence of political content. To illustrate, as Fig.~\ref{fig:glmm} shows,
in France the depressing impact of negativity on commenting is stronger for political than
non-political news posts, but because political news posts are negative slightly more often, this alone does not fully describe the ultimate \enquote{effect} with respect to varying prevalence of political news posts.
However, Fig.~\ref{fig:share}b clearly shows that the compound effect is positive, as indicated by 
the slope of the line, suggesting that in this case the overall consumption of negative news posts
is positively associated with the overall prevalence of political news posts.

More generally, this analysis shows that the overall relative volume of engagement with negative
content posted by news organizations, i.e.~across political and non-political news together, is low, ranging between $10.2\%$
(lower bound for reactions) and $13.1\%$ (upper bound for shares), and only weakly dependent on the 
population prevalence of political news posts.
We also find that the estimates for the U.K. and U.S. are consistently lower
than the average. Poland is close to the average for the likely range $p \in [0.25, 0.50]$,
while the estimates for the rest of the countries (France, Spain, and Ireland) are more diverse,
also depending on $p$, with Spain being the only consistent outlier positioned markedly above the
average. In sum, these results provide robust evidence of arguably low levels of overall engagement
with negative news posts.

\section*{Discussion}
Many observers worry that the social media environment incentivizes news organizations to publish negative news and that users disproportionally engage with such news on platforms \citep{robertsonNegativityDrivesOnline2023, watsonNegativeOnlineNews2024}. This, in turn, increases the visibility of negative stories and offers additional incentives for news organizations to continue producing and 
publishing them~\citep{mukerjee2023metrics}.
Because exposure to negative news may polarize policy opinions, deepen political divides \citep{levenduskyDoesMediaCoverage2016}, generate cynicism and political apathy \citep{mutzNewVideomalaiseEffects2005, cappellaNewsFramesPolitical1996}, and produce stress, anxiety, and other adverse psychological effects \citep{boukesNewsConsumptionIts2017}, these worries and evidence are concerning.

We provide systematic descriptive evidence from 97 news media organizations from six distinct democracies of the extent to which these organizations post negative versus non-negative news on Facebook and whether this differs depending on whether news is related to politics or not. In addition to showing the prevalence of political and non-political negativity, we also systematically test whether users indeed react to, comment on, and share negative news posts more and whether this differs for news posts about politics versus for news about celebrities, sports, or the weather. 

We offer five key findings. 
First, we demonstrate that the general prevalence of negativity
is relatively low, varying roughly between 5\% and 20\% 
(Fig.~\ref{fig:glmm}b). That is, a solid majority of news published on Facebook by numerous international news outlets is neutral or positive, reporting on neutral or positive events and/or offering neutral or positive perspective on these developments. Second, news posts about elections, politicians, healthcare, climate change, or social groups, among other political topics are \textit{not} inherently more negative on average than news posts about other issues; and this result is rather robust across countries. Although conflict, animosity, and clashes between politicians, parties, or interest groups characterize politics and even though news coverage of political polarization has increased over the years in the U.S.~\citep{levenduskyDoesMediaCoverage2016}, news posts about non-political events and developments are as negative as explicitly political posts by news organizations. 

These findings expand past work, which has not systematically differentiated between these different classes of content, suggesting that overall negativity levels are stable across news posts. Because we relied on an extensively validated and theoretically grounded classifiers, with our negativity classifier accounting for negativity coming from the event itself (e.g., an attack) and/or from the journalistic interpretation of an event (e.g., using a negative tone to describe an otherwise neutral event), we see these estimates as accurately representing the overall levels of negative news on Facebook across countries. 

These findings, we note, cannot speak to whether the same outlets have similar or different publishing patterns offline and on their websites. Some international work finds that editors tend to select emotionally charged news from websites for posting on platforms~\citep{lamotWhatMetricsSay2022} 
and that emotions, conflict, and negative consequences weigh more heavily in editors' decisions about what to post on Facebook than on the main news websites~\citep{lischkaLogicsSocialMedia2021}. 
If news outlets amplify negative content more on social media than on their websites, we may be overestimating the prevalence of negativity (which---we note---is quite low in our data to begin with). Future research needs to systematically analyze the prevalence of different types of coverage offline, on online websites, and on social media pages of news organizations internationally. Because the majority of the public, and especially the younger demographics, encounters news on social 
media~\citep{pewresearchcenterHowAmericansGet2024} and Facebook outpaces all other platforms as a regular news source over the past years~\citep{pewresearchcenterHowAmericansGet2024}, 
studying the different posts contributed by news outlets to Facebook and user engagement with these posts is important. 

Third, negativity in general and especially in political news is less prevalent in the U.S. posts than in the European countries on average, despite the fact that the U.S. could be expected to have higher negativity in its news posting due to the polarized politics and the commercial and polarized nature of its news media system. In turn, Spain's news is characterized by significantly higher negativity; otherwise, we find limited cross-country variation in negativity. Given the relatively small sample of countries, we cannot analytically account for different media or party systems, country-level polarization levels, GDP or inequality indicators, among other country-level factors that may matter to these findings. Although it is surprising that the U.S. lags behind the other nations analyzed in the level of negativity given its predominantly market-based system with minimal public service broadcasting, news media in Europe are also---and increasingly so---characterized by negativity bias in the selection and tonality of news---especially about the EU and local governments and by general contestation and 
critique~\citep{copelandRethinkingBritainEuropean2017,seddoneEuropeanDomesticPolitics2019}. 
It is thus possible that general news negativity, typically associated with American media, has been increasingly spreading to other countries.

Our fourth key finding regards the association of news negativity with user engagement. Because our dataset contains all the main forms of engagement available on Facebook, our results speak to the differential patterns of reacting to, commenting on, and sharing of negative news posts, both political and non-political. We show that negative news posts---especially about politics---attract \textit{less} engagement than non-negative posts. 
This is particularly true
for reactions and comments, with the average effects for shares also being negative but statistically insignificant  (Fig.~\ref{fig:glmm}c), and especially in the U.S., U.K., and---to a lesser extent--- Poland. In short, across countries and the three engagement forms, negativity does not pay off as much as previously
shown~\citep{watsonNegativeOnlineNews2024,robertsonNegativityDrivesOnline2023,trusslerConsumerDemandCynical2014}.

Despite the overall negativity bias in human cognition and judgement, this finding does align with some evidence that it is neutral or positive social media content that generates greater engagement~\citep{muddimanNegativityBiasBacklash2020, trillingNewsworthinessShareworthinessHow2017}, and is consistent with users' self-reported preferences for positive content—--such as accurate, nuanced, and educational~\citep{rathjePeopleThinkThat2023} and the general dissatisfaction with negative news \cite{skovsgaardConceptualizingNewsAvoidance2020} and partisan conflict~\citep{krupnikovOtherDivide2022}. 
Also, the fact that negativity has no effect on sharing and depresses reactions and commenting may also attenuate the worries about algorithmic prioritization of negative news. If platform algorithms prioritize negative, toxic, or uncivil content, the levels of liking, sharing, and potentially commenting on 
(politically) negative news posts (relative to other news posts) should be similar and, most importantly, at least not lower than those for non-negative news posts.

Fifth, by estimating the full range of possible scenarios and under different assumptions about 
the prevalence of political news posts on Facebook, we show that, across countries and different engagement metrics, 
the proportion of user engagement with negativity---across political and non-political news posts together--- is relatively low, ranging between 
6\% and 25\%. In short, our robust simulations find that engagement with negativity
constitutes a minority of the total user engagement with news on Facebook. In particular, averaging over the countries, it is bounded between 10.2\% and 13.1\% and only weakly depends on the prevalence of political news posts. 

We note that our statistical approach carefully incorporates uncertainty arising from 
cross-country, temporal, and cross-outlet variation, in an attempt at getting estimates that remain
meaningful and informative also beyond our sample. This is crucial because, 
as we argue in Sec.~\ref{sec:methods:analysis}, the traditional notion of a statistical population 
is ill-defined in the context of news posts on social media, and therefore it is hard to have a 
notion of \enquote{typical} news post. Hence, we propose that one should model 
\enquote{a typical post of a typical outlet on a typical day}, 
that is, be explicit about the fact 
that the observed news posts are products of the underlying populations of news outlets and transient 
time-based effects. This nuance is very important because most of the current evidence on negativity in news comes from studies which---while internally consistent and 
methodologically rigorous---focus on narrow contexts such as content of a single publisher
\citep[e.g.~\textit{Upworthy} in][]{robertsonNegativityDrivesOnline2023} or a single country and without accounting for temporal variation~\citep{heidenreichExploringEngagementEU2022}. 
Estimates from such studies---if interpreted in a context distinct from the very specific population (and time window) they were based
on---may be markedly biased and suffer from underestimated standard
errors. This is likely the reason why our study finds null result 
where some other studies reported significant effects.

More specifically, as we show for engagement metrics in
\ref{app:sec:heterogeneity},
there is a pronounced heterogeneity at the level of outlets in both
magnitudes and directions of negativity effects across all studied 
countries as well as political and non-political news posts.
Thanks to a relatively large and diverse selection of outlets across
six countries, 
our analytical approach accounts for this heterogeneity---which likely
reflects a diversity of content and audience profiles of different
outlets---and this is the reason why we observe more null results
than some prior studies. In other words, the evidence for 
increased engagement with negative news disappears once a sufficiently
broad spectrum of outlets and audiences is considered.
Importantly, this heterogeneity also means that a narrow selection
may easily lead to over- or underestimation of the
negativity effects depending on the kinds of selected outlets.
Last but not least, this observation also strongly suggests that
more in-depth studies focused on the interaction between negativity and different types of news outlets and audiences are needed to better
understand the nuances related to the production and consumption of 
negative news.

We also emphasize our methodological contribution, which is to offer two public, extensively validated, multilingual classifiers to at-scale classify large amounts of content as political or not and negative or not. We applied them to Facebook posts of numerous news media organizations in four languages (English, Spanish, French, and Polish) and obtained highly accurate 
classification (see performance metrics in Table~\ref{tab:performance}). Furthermore,
in ~\ref{app:sec:cls} we provide evidence that our models are likely to transfer their
classification ability to other languages and content types not present in the training data
with reasonable accuracy. 
We invite researchers to use these classifiers---and expand them to multimodal content prevalent on 
platforms---to start offering comprehensive and integrated evidence on (political) (news) negativity across countries, platforms, and contexts. Given the rather fragmented and methodologically limited nature of previous negativity detection methods, this, in our view, is an important advancement.

We note several areas for future research. First, we cannot ascertain the many ways in which the tested content may have affected social media users, their beliefs, attitudes, or emotions. Our project captures the prevalence of and user engagement with (political) negativity in news posts and we leave the study of the potential effects for future work. 
Second, although our data represent the most extensive dataset to date of negativity in (political) news, both in terms of the time frame and the number of countries, replications in other countries with different political systems and information environments are needed to determine how these results generalize beyond the six countries studied. After all, apart from the U.S., the other countries studied are all European, and thus represent a narrow slither of the world population and one that is the exception not the norm. Replication of our work in non-Western countries is needed. 

Relatedly, we focus on a single platform and cannot determine 
the extent to which similar patterns of posting and engagement occur on other platforms with 
different affordances. 
Comparing behavioral data from across different platforms is needed to better 
understand online negativity. Similarly, we encourage future work to systematically compare the levels of negativity in posts contributed to platforms by news outlets to negativity in social media posts of politicians, political parties, political influencers, or social media users themselves. Evidence on this relative prevalence of political / news negativity across different actors is needed. Scholars should also examine not only all posts from news organizations themselves, as we did, but also all public posts containing links to these news stories to capture full engagement with news stories---whether shared by news outlets themselves or by other social media users.  
 
We also acknowledge that our classification of posts is binary (either political or non-political) and it is possible that posting by news organizations as well as user attention and engagement with negative information may be conditional on specific topics (e.g., COVID-19), their personal relevance (e.g., price increases), geographical proximity (e.g., local vs. national vs. international reports), and/or their “threatening” (crime) vs. “non-threatening” (party platforms) nature. Accounting for these diverse political sub-topics could show a difference in levels of engagement with negativity. Similarly, we look at overall reactions that combine \enquote{like}, \enquote{love},
\enquote{care}, \enquote{haha}, \enquote{sad}, and \enquote{angry}
because these were aggregated during data collection and the subsequent checks for individual reactions proved them to be unstable. Inasmuch as negative content could increase certain reactions over others (in particular anger and sadness), we may be overlooking some differential engagements within the reactions themselves. Given the robust negative associations between negativity and the various categories of engagement, we suspect this not to be the case.

In sum, the above results challenge the prevalent concerns that news organizations---at least in the United States and the five European
countries studied---primarily post negative news, even despite the negative events that continually take place globally 
(e.g.~the COVID-19 pandemic or the war in Ukraine, which started during the analyzed years). 
The patterns we uncover---that political news posts are not more negative on average than non-political 
news posts and that social media users engage with negative news posts less than with neutral or positive 
content posted by news organizations---pushes back against the dominant narrative and some empirical evidence suggesting that 
negativity, polarization, and political conflict are covered disproportionately frequently in 
media coverage~\citep{levenduskyDoesMediaCoverage2016}. 
These results also leave some room for optimism. We trust that greater attention of scholars, journalists, media organizations, and social media users will be directed to positive events, developments, and online contents. Given the many challenges facing the world and the United States at the moment, putting negativity in perspective and attending to more positive news may be useful and needed.
\section{Materials and Methods}\label{sec:methods}

\subsection{Dataset}\label{sec:methods:data}

We used data on Facebook posts  ($N = \num{6081134}$) published by 97 major outlets from the U.S. 
and 5 European countries (United Kingdom, Poland, France, Spain and Ireland) between January 1, 
2020, and April 1, 2024. Details of further outlet selection criteria are discussed in 
SM:~\ref{app:sec:data:outlets}. 

For each post published by a news outlet, we obtained the text, type (e.g., status, video, link), 
posting time, number of reactions, comments, and shares The data were gathered through Sotrender, 
a social media analytics platform with direct access to the Facebook Graph API. 
The data contain no personally identifiable information, thereby preserving user privacy while 
allowing for robust posting and engagement analysis at the aggregate level
(see SM:~\ref{app:sec:data:collection} for more details on data extraction).

The dataset included:
\num{5093214} (83.8\%) links (content with an external link), 
\num{641168} (10.5\%) videos (short description plus video content),
\num{293377} (4.8\%) photos (short description plus static image),
\num{53373} (0.9\%) statuses (pure text content).
Classification of political and negative posts was based only on available text content,
that is, neither image, video or audio content was used.

\subsection{Classifiers}\label{sec:methods:classifiers}

To identify political and negative news, we developed two neural binary classifiers.
Both of them were constructed by adding a binary classification head to a pre-trained
base language model known as
XLM-RoBERTa-Large~\citep{conneauUnsupervisedCrosslingualRepresentation2020}, 
which is a transformer model with 561 millions of parameters based on 2.5 terabytes of 
CommonCrawl~\citep{patelIntroductionCommonCrawl2020} data for 100 languages. 
Both the head and the base model were fine-tuned and validated using custom datasets
of hand-annotated examples using separate random subsamples of the
dataset for optimization of model weights (training set), model hyperparameters (testing set),
and computing final performance metrics (validation set). This tripartite data splitting strategy
allowed for interpreting the reported performance metrics as proper measures of the out-of-sample
classification performance.

The performance of classifier was evaluated using a comprehensive set of three variants of
$F$-score~\citep{derczynskiComplementarityFscoreNLP2016}:
\begin{align}
    F_1^+ &= \frac{2\TP}{2\TP + \FP + \FN}     \label{eq:f1p} \\
    F_1^- &= \frac{2\TN}{2\TN + \FN + \FP}     \label{eq:f1m} \\
    F_1   &= \frac{2F_1^+F_1^-}{F_1^+ + F_1^-} \label{eq:f1}
\end{align}
where $\TP$, $\TN$, $\FP$ and $\FN$ are numbers of true positives, true negatives, false positives and false negatives respectively.
Eqs. \eqref{eq:f1p} and \eqref{eq:f1m} are the $F$-scores for predicting target (political or negative) 
and complement (other) labels, while Eq.~\eqref{eq:f1} is just the harmonic average of the first two.
Table~\ref{tab:performance} provides overall and country-specific performance metrics for the classifiers.
More information on the training dataset and model hyperparameters, as well as annotator 
instructions, including detailed definitions of political and negative news can be found in 
SM:~\ref{app:sec:cls}.

\begin{table}[htb!]
\caption{Classification performance metrics}
\centering
\tiny\sffamily
\begin{tabularx}{\columnwidth}{l|XXX|XXX}
    \toprule
    & \multicolumn{3}{c}{Political} & \multicolumn{3}{c}{Negativity} \\
    & $F_1$ & $F_1^+$ & $F_1^-$ & $F_1$ & $F_1^+$ & $F_1^-$ \\
    \midrule
    US  & 0.94 & 0.96 & 0.92   & 0.95 & 0.94 & 0.96 \\
    UK  & 0.88 & 0.87 & 0.89  & 0.88 & 0.89 & 0.88 \\
    IRL & 0.91 & 0.86 & 0.95   & 0.92 & 0.93 & 0.92 \\
    PL  & 0.87 & 0.85 & 0.90   & 0.90 & 0.91 & 0.88 \\
    FR  & 0.85 & 0.90 & 0.81   & 0.91 & 0.92 & 0.89 \\
    ESP & 0.84 & 0.78 & 0.90   & 0.89 & 0.90 & 0.88 \\
    \midrule
    \textbf{Overall} & 0.90 & 0.89 & 0.91  & 0.91 & 0.90 & 0.91 \\
    \bottomrule
\end{tabularx}
\label{tab:performance}
\end{table}

\subsection{Analytical strategy}\label{sec:methods:analysis}

While the samples for outlets are nearly complete and include most of their public posts, 
the dataset itself cannot be assumed to be representative of news posts
with respect to any salient variable that may be related to our outcomes of interest---namely, 
the prevalence of negative news and engagement counts (reactions, comments and shares).
As a result, estimates of central tendencies based on simple aggregation of raw data---treating individual posts as the unit of observation---may be biased and may not reflect long-term 
population averages. In particular, estimates of the prevalence of political news are likely to 
vary greatly depending on the operational definitions of what counts as \enquote{general news} and 
what as \enquote{political news}. For this reason we do not provide direct estimates of political news prevalence. Instead we either aggregate separate results for political and non-political news 
under a \enquote{maximum ignorance} (aka \enquote{Occam's razor}) assumption of 50/50 split, 
or simulate plausible results for all possible values of the prevalence of political news, 
$q \in [0, 1]$.

Crucially, the issue of representativity cannot be alleviated by using sampling weights, 
as no well-defined sampling frame exists. Firstly, to the best of our knowledge, exhaustive lists 
of news outlets operating on Facebook within specific countries (and time periods) are either hard 
to obtain or simply non-existent. Secondly, even if such lists were available, meaningful sampling 
would remain highly challenging. Should sampling weights be uniform or should they be based on 
publication volume, views, user engagement, or another criterion? Moreover, in many cases 
information needed for designing such sampling weights may either be inaccessible or require 
collecting full posting data for all listed outlets in the first place. 
Lastly, the boundary between what qualifies as a general news outlet and what should be included in 
the sampling frame is inherently blurry, and have to rely on arbitrary decisions in any practical 
application.

For these reasons, we opted for a different analytical strategy based on a careful analysis
and interpretation of Estimated Marginal Means (EMMs) derived from Generalized Linear Mixed 
Models~\citep[GLMMs;][]{batesFittingLinearMixedEffects2015,wangComputationApplicationGeneralized2022,brooksGlmmTMBBalancesSpeed2017} in the spirit of \texttt{lsmeans}/\texttt{emmeans} packages for
\texttt{R}~Language~\citep{lenthLeastSquaresMeansPackage2016}.
Namely, we use random effects to capture, and ultimately integrate out, location shifts specific
for time (country-specific daily effects) and individual outlets. 
This strategy aims at estimating \enquote{population-level} effects independently from 
idiosyncrasies of selected news outlets and time period. Most importantly, estimates obtained
this way can be seen as conceptually similar to a model without random effects based on a sample 
using weights such that the weight for an $i$'th posts of an $o$'th outlet in a day $t$ is:
\begin{equation}\label{eq:glmm:weight-interpretation}
    w_{i,o,t} \propto 1 / (N_oN_tn_{o,t})
\end{equation}
where $N_o, N_t$ are, respectively, the number of outlets and days in an ideal population and
$n_{o,t}$ is the number of posts produced by an $o$'th outlet in a day $t$. (However, note that
this correspondence is exact only in the case of a fully crossed design with independent factors
and sufficiently large groups.)

Thus, all estimated models were based on the same generic construction given by the following formula:
\begin{equation}\label{eq:glmm}
    \eta(\mu)
    = \eta\left(
        \mathbb{E}[\mathcal{Y} \mid \mathcal{B}_t = \mathbf{b}_t, \mathcal{B}_o = \mathbf{b}_o]
    \right) 
    = \mathbf{X}\mathbf{b} + \mathbf{Z}_t\mathbf{b}_t + \mathbf{Z}_o\mathbf{b}_o
\end{equation}
where $\eta(\cdot)$ is a link function, $\mathbb{E}[\cdot]$ expectation operator,
$\mathcal{Y}$ a random outcome vector, 
$\mathcal{B}_k \sim \mathcal{N}(0, \boldsymbol{\Sigma}_k)$ for $k = t, o$ random effects 
for, respectively, time and outlets following centered multivariate normal distribution, 
$\mathbf{X}$, $\mathbf{Z}_t$, $\mathbf{Z}_o$ model matrices for, respectively, fixed effects and 
random effects for time and outlets,  $\mathbf{b}$ a vector of estimated regression coefficients 
for fixed effects,  and $\mathbf{b}_t$, $\mathbf{b}_o$ estimated random effects (latent variables).
Crucially, the fixed effects correspond to countries in full interaction with whether a post is
political (in all cases) as well as negative (models for predicting engagement). 
This allows for deriving aggregate effects 
(e.g.~overall prevalence of negativity across all countries as well as political and non-political news) by averaging uniformly over the most granular groups representable using model fixed effects. 

For instance, when estimating the overall prevalence of negative news, $\mathbf{Xb}$ estimates
probabilities of negativity for specific countries and political/non-political news separately,
and only later those finely-grained estimates 
(in this case for $k$ countries there are $k \times 2$ of them) 
are uniformly averaged to obtain the final overall estimate.

The negativity model was formulated as a logistic regression with fixed and random effects
with the following specification~\citep[%
    using the syntax of the popular \texttt{lme4} package for~\texttt{R};
][]{batesFittingLinearMixedEffects2015}:
\begin{equation}\label{eq:glmm:negativity}
\small
\begin{split}
    \eta(\mu) \sim\ 
    &\texttt{country*political +} \\
    &\texttt{(1+political | outlet) +} \\
    &\texttt{(1+political | country:day)}
\end{split}
\end{equation}

The engagement models (reactions, comments and shares) were formulated as negative binomial 
regression models with fixed and random effects and non-constant dispersion, 
and hence each of them were based on two submodels---conditional mean ($\eta(\mu)$) and
dispersion ($\eta(\phi)$)---using the following specifications:
{\small
\begin{align}
    \eta(\mu) \sim\ 
    &\texttt{country*political*negativity +} \label{eq:glmm:engagement} \\
    &\texttt{(1+political*negativity | outlet) +} \nonumber\\
    &\texttt{(1+political*negativity | country:day)} \nonumber\\
    \eta(\phi) \sim\ 
    &\texttt{country + (1 | country:day)} \label{eq:glmm:engagement:disp}
\end{align}
}
In both cases the models were estimated using \texttt{glmmTMB} package for \texttt{R} 
language~\citep{brooksGlmmTMBBalancesSpeed2017}, which provides methods for fitting a broad
range of generalized linear mixed models using Maximum Likelihood Estimation (MLE).

Last but not least, when considering overall estimates we always average uniformly over
estimates specific for political/non-political news and/or countries, that is, we ignore any 
potential differences in group sizes in our raw data. This reflects our belief that there is 
no well-defined and stable notion of a general population of news posts, so it is more meaningful 
to look  at finely-grained subsets (e.g.~negative political news in the U.S.) and treat them 
equally when  averaging,  or study a wider range of aggregate values generated under different 
assumptions (see Sec.~\ref{sec:methods:share}). Most importantly, in our particular case such an 
aggregation strategy is valid as we implicitly assume that the potential causal flow is of the form:
(1)~country + political $\to$ negativity; 
(2)~country + political + negativity $\to$ engagement.
Further discussion of the models, including their validation, as well as the issue of
(un)biasedness and interpretation of our approach can be found in SM:~\ref{app:sec:glmm}.

\subsection{Estimation of the proportion of negative engagement}
\label{sec:methods:share}

In order to provide a robust description of the engagement with the negative news, we considered
relative volumes of engagement (reactions, comments and shares) with negative news defined 
in the following way.

For a given $q = \Prob(\text{negative} \mid \ldots)$, where $\ldots$ stands for a 
(potentially empty) sequence of conditioning variables, we define the share of engagement
using Eq.~\eqref{eq:share}. We derive its variance using 
multivariate delta method~\citep[][p.~61]{lehmannTheoryPointEstimation1998}
applied to
linear predictors (conditional expectations in the log scale) using a family of functions
indexed by $q$:
\begin{equation}
    h_q(x, y) = \frac{qe^x}{qe^x + (1-q)e^y}  
\end{equation}
where $x$ and $y$ are linear predictors for engagement counts linked to 
negative and non-negative posts.

Then, we use the fact our sample size is large and estimates
of the prevalence of negative news are obtained using Maximum Likelihood theory---they are
consistent, asymptotically normal and in general coincide with a Bayesian posterior for a flat
prior~\citep{lehmannTheoryPointEstimation1998}---so it can be assumed that 
$q \sim \mathcal{N}(\bar{q}, \sigma_{\bar{q}})$, where $\bar{q}$ is an estimate of the prevalence 
of negative news produced by our model from Sec.~\ref{sec:results:negativity} 
and $\sigma_{\bar{q}}$  its standard error. This allows us to get a general proportion of 
engagement with negative content for political/non-political news by integrating $q$ out:
\begin{equation}\label{eq:share-marginalization}
    \mathcal{S}(X \mid P) 
    = \int_0^1 \phi_{\bar{q}, \sigma_{\bar{q}}}(q)\mathcal{S}(X \mid q, P)dq
\end{equation}
where $\phi_{\bar{q}, \sigma_{\bar{q}}}$ is the appropriate normal density function
and $P$ denotes post type (political/non-political).

Finally, the overall share for a specific assumption about the prevalence of political news
can be obtained simply as:
\begin{equation}
    \mathcal{S}(X) = p\mathcal{S}(X \mid P = 1) + (1-p)\mathcal{S}(X \mid P = 0)
\end{equation}
where $P = 0, 1$ is the indicator of political/non-political news and 
$p = \Prob(P = 1) \in [0, 1]$ 
the assumed prevalence of political news.

\bibliographystyle{elsarticle-harv} 
\bibliography{references.bib,Political-Negativity.bib}

{
\small
\section*{Acknowledgments}\label{sec:acknowledgments}

We thank Maria Babińska for the help with annotating training examples.

\subsection*{Funding}

The authors gratefully acknowledge the support of the European Research Council 
(ERC Consolidator 101126218, NEWSUSE: Incentivizing Citizen Exposure to Quality News Online: Framework and Tools, PI Magdalena Wojcieszak) and of the Center for Excellence in Social Sciences at the University of Warsaw, which funded the project. Any opinions, findings, conclusions, or recommendations expressed in this material are those of the authors and do not necessarily reflect the 
views of the European Research Council.

Sotrender's tools for the data collection process were developed within the projects co-funded by the
National Centre for Research and Development (NCBR), Poland, project no. POIR.01.01.01-00-0952/17 and
MAZOWSZE/0061/19.

\subsection*{Author contributions}

S.T., D.B. and M.W. designed the research. D.B. and M.W. oversaw the data collection.
S.T. and M.W. annotated the training examples for the classifiers.
S.T. trained the classifiers and analyzed the data.
S.T., D.B. and M.W. wrote the paper. All authors discussed the results and commented on the
manuscript.

\subsection*{Competing interests}

The authors declare that they have no competing interests.

\subsection*{Data and materials availability}

Code and data necessary for reproducing the reported results
are available at the OSF under a permanent URL 
(\url{https://osf.io/79a46/?view_only=f7cad71ef9b54a5a866e2e579eb92e62})
and are identified with a DOI: \texttt{10.17605/OSF.IO/79A46}.
Code is also available from a dedicated Github repository
(\url{https://github.com/erc-newsuse/newsuse-study-polneg-repro}).
Classifiers in the format of the \texttt{transformers} 
library~\citep{wolfHuggingFacesTransformersStateoftheart2020}
can be found at Huggingface Hub:
political (\url{https://huggingface.co/sztal/erc-newsuse-political}),
negativity (\url{https://huggingface.co/sztal/erc-newsuse-negativity}).
Other details regarding reproducibility are discussed in the README file
available from both Github and OSF repositories.

\clearpage

\appendix
\setcounter{table}{0}
\onecolumn

\clearpage

\section{Dataset}\label{app:sec:data}

We used \num{6081134} Facebook posts published between January 1, 2020, and April 1, 2024, by 97 news media organizations in six countries. In particular, we use data on posts published by 37 major U.S.-based news outlets (\num{2274210} posts), 
12 outlets from the United Kingdom (\num{959141} posts), 11 from Ireland (\num{451783}), 17 from Poland (\num{842321} posts), 
8 from France (\num{355697} posts), and 12 news outlets from Spain (\num{1197982} posts).
See Table~\ref{app:tab:outlets} for the full list and some outlet-level descriptive statistics. 
For each post published by each news outlet, we obtained the text content, posting time, 
and the main engagement metrics --- numbers of reactions, comments, and shares
(see Table~\ref{app:tab:engagement} for descriptive statistics). 
The outlets were selected to provide a good coverage of each country's news media landscape with regard to quality and tabloid outlets and the political leaning. 

\begin{table*}[hbt!]
\caption{
Descriptive statistics for engagement metrics, overall and by country
}
\centering
\tiny\sffamily
\begin{tabularx}{\textwidth}{XXX|rrrr|rrrr|rrrr}
\toprule
 &  &  & \multicolumn{4}{c}{Reactions} & \multicolumn{4}{c}{Comments} & \multicolumn{4}{c}{Shares} \\
 &  &  & mean & std & median & IQR & mean & std & median & IQR & mean & std & median & IQR \\
\midrule
\multirow[t]{5}{*}{United States} & overall & overall & 355.8 & 2740.0 & 0.0 & 80.0 & 340.4 & 1667.4 & 43.0 & 200 & 197.2 & 2018.0 & 14.0 & 59 \\
 & \multirow[t]{2}{*}{political} & negative & 246.0 & 1556.4 & 0.0 & 53.0 & 368.7 & 1423.6 & 73.0 & 260 & 276.6 & 3713.3 & 24.0 & 90 \\
 &  & other & 334.5 & 2139.8 & 0.0 & 75.0 & 423.8 & 1953.1 & 65.0 & 262 & 202.6 & 1662.1 & 15.0 & 64 \\
 & \multirow[t]{2}{*}{other} & negative & 353.0 & 1982.6 & 0.0 & 101.0 & 275.2 & 1004.8 & 44.0 & 186 & 234.5 & 2475.9 & 18.0 & 72 \\
 &  & other & 424.1 & 3863.1 & 9.0 & 96.0 & 177.9 & 1052.7 & 17.0 & 86 & 163.1 & 2021.3 & 9.0 & 41 \\

 \midrule
\multirow[t]{5}{*}{Spain} & overall & overall & 270.2 & 2175.1 & 22.0 & 104.0 & 94.2 & 298.9 & 15.0 & 69 & 62.1 & 711.0 & 2.0 & 13 \\
 & \multirow[t]{2}{*}{political} & negative & 256.4 & 1156.7 & 29.0 & 130.0 & 107.2 & 264.8 & 24.0 & 91 & 103.4 & 952.8 & 4.0 & 25 \\
 &  & other & 227.6 & 1148.9 & 24.0 & 114.0 & 121.7 & 322.9 & 30.0 & 101 & 69.3 & 560.6 & 3.0 & 18 \\
 & \multirow[t]{2}{*}{other} & negative & 331.7 & 2658.3 & 37.0 & 173.0 & 86.3 & 276.0 & 15.0 & 66 & 74.9 & 592.8 & 4.0 & 20 \\
 &  & other & 297.5 & 2816.1 & 17.0 & 78.0 & 69.6 & 287.3 & 7.0 & 36 & 44.2 & 773.7 & 1.0 & 7 \\

 \midrule
\multirow[t]{5}{*}{France} & overall & overall & 352.4 & 1317.3 & 81.0 & 238.0 & 117.0 & 287.6 & 32.0 & 104 & 72.4 & 618.0 & 8.0 & 30 \\
 & \multirow[t]{2}{*}{political} & negative & 320.5 & 961.6 & 86.0 & 246.0 & 131.7 & 277.4 & 45.0 & 127 & 84.4 & 887.8 & 10.0 & 39 \\
 &  & other & 314.8 & 1025.2 & 80.0 & 226.0 & 122.0 & 278.7 & 37.0 & 112 & 73.3 & 568.7 & 9.0 & 30 \\
 & \multirow[t]{2}{*}{other} & negative & 424.2 & 1103.2 & 104.0 & 316.0 & 111.7 & 254.7 & 30.0 & 97 & 73.9 & 366.2 & 11.0 & 40 \\
 &  & other & 469.3 & 2085.3 & 76.0 & 256.5 & 92.3 & 321.7 & 14.0 & 58 & 60.1 & 513.7 & 6.0 & 21 \\

 \midrule
\multirow[t]{5}{*}{Poland} & overall & overall & 294.6 & 949.1 & 62.0 & 197.0 & 104.0 & 276.4 & 25.0 & 90 & 30.2 & 217.0 & 4.0 & 14 \\
 & \multirow[t]{2}{*}{political} & negative & 254.7 & 831.6 & 65.0 & 180.0 & 105.5 & 232.0 & 30.0 & 100 & 29.5 & 172.5 & 5.0 & 15 \\
 &  & other & 259.3 & 799.1 & 62.0 & 188.0 & 108.4 & 269.7 & 30.0 & 100 & 25.2 & 143.2 & 4.0 & 13 \\
 & \multirow[t]{2}{*}{other} & negative & 478.2 & 1374.4 & 101.0 & 324.0 & 117.0 & 303.8 & 25.0 & 94 & 65.9 & 466.0 & 7.0 & 28 \\
 &  & other & 292.9 & 987.8 & 54.0 & 179.0 & 93.9 & 283.4 & 17.0 & 71 & 26.3 & 183.3 & 3.0 & 11 \\

 \midrule
\multirow[t]{5}{*}{Ireland} & overall & overall & 115.4 & 869.0 & 19.0 & 65.0 & 45.1 & 154.3 & 7.0 & 33 & 12.8 & 162.3 & 1.0 & 6 \\
 & \multirow[t]{2}{*}{political} & negative & 94.5 & 269.1 & 31.0 & 82.0 & 57.1 & 114.3 & 15.0 & 57 & 15.2 & 77.2 & 3.0 & 8 \\
 &  & other & 95.1 & 366.3 & 24.0 & 69.0 & 58.2 & 130.7 & 15.0 & 55 & 14.1 & 85.0 & 2.0 & 8 \\
 & \multirow[t]{2}{*}{other} & negative & 142.8 & 492.5 & 33.0 & 108.0 & 34.3 & 131.7 & 6.0 & 26 & 16.4 & 176.0 & 2.0 & 8 \\
 &  & other & 126.8 & 1179.4 & 12.0 & 48.0 & 36.6 & 176.4 & 3.0 & 18 & 10.8 & 203.8 & 1.0 & 3 \\

 \midrule
\multirow[t]{5}{*}{United Kingdom} & overall & overall & 378.3 & 3779.8 & 46.0 & 181.0 & 179.4 & 619.7 & 38.0 & 136 & 57.8 & 1029.6 & 3.0 & 16 \\
 & \multirow[t]{2}{*}{political} & negative & 347.0 & 1745.7 & 66.0 & 235.0 & 189.8 & 455.2 & 54.0 & 184 & 81.0 & 1160.5 & 6.0 & 27 \\
 &  & other & 300.8 & 4142.2 & 51.0 & 183.0 & 172.5 & 501.7 & 50.0 & 156 & 45.7 & 1051.8 & 4.0 & 16 \\
 & \multirow[t]{2}{*}{other} & negative & 288.1 & 1419.9 & 51.0 & 193.0 & 128.1 & 426.9 & 33.0 & 98 & 54.9 & 604.0 & 4.0 & 19 \\
 &  & other & 463.0 & 3963.3 & 39.0 & 171.0 & 194.0 & 742.8 & 31.0 & 119 & 65.5 & 1062.0 & 3.0 & 14 \\

\midrule
\multirow[t]{5}{*}{Overall} & overall & overall & 315.9 & 2506.0 & 25.0 & 131.0 & 198.7 & 1071.6 & 27.0 & 119 & 104.5 & 1351.3 & 5.0 & 26 \\
 & \multirow[t]{2}{*}{political} & negative & 260.4 & 1300.4 & 29.0 & 140.0 & 203.8 & 867.4 & 41.0 & 147 & 143.2 & 2263.9 & 9.0 & 40 \\
 &  & other & 289.6 & 2154.3 & 23.0 & 129.0 & 262.1 & 1358.4 & 43.0 & 163 & 119.1 & 1221.8 & 7.0 & 33 \\
 & \multirow[t]{2}{*}{other} & negative & 340.2 & 1828.5 & 44.0 & 193.0 & 135.3 & 539.7 & 23.0 & 92 & 94.9 & 1211.4 & 6.0 & 28 \\
 &  & other & 357.1 & 3150.0 & 23.0 & 118.0 & 127.9 & 696.2 & 14.0 & 67 & 79.2 & 1272.9 & 3.0 & 16 \\
\bottomrule
\end{tabularx}
\label{app:tab:engagement}
\end{table*}

\begin{table*}[p]
\caption{Basic descriptive statistics for individual outlets}
\centering
\tiny\sffamily
\begin{tabularx}{\textwidth}{XX|rrr|rrr|rrr|rrr}
\toprule
 &  & \multicolumn{3}{c}{Posts} & \multicolumn{3}{c}{Reactions} & \multicolumn{3}{c}{Comments} & \multicolumn{3}{c}{Shares} \\
 &  & n & political & negative & mean & median & IQR & mean & median & IQR & mean & median & IQR \\
\midrule

\multirow[t]{37}{*}{United States} & ABC News & 91831 & 63.8 & 22.8 & 282.5 & 0.0 & 0.0 & 510.4 & 150.0 & 390.0 & 390.4 & 48.0 & 134.0 \\
 & AP & 36071 & 70.5 & 21.2 & 13.3 & 0.0 & 0.0 & 80.2 & 21.0 & 67.0 & 39.4 & 8.0 & 20.0 \\
 & Breitbart & 19553 & 71.6 & 8.8 & 1222.1 & 0.0 & 754.0 & 2040.4 & 288.0 & 1486.0 & 1077.6 & 84.0 & 454.0 \\
 & Business Insider & 225136 & 48.2 & 5.9 & 575.1 & 35.0 & 190.0 & 75.9 & 7.0 & 30.0 & 43.9 & 3.0 & 10.0 \\
 & BuzzFeed News & 18681 & 48.9 & 13.8 & 335.4 & 0.0 & 55.0 & 152.5 & 16.0 & 94.0 & 294.2 & 7.0 & 31.0 \\
 & CNBC & 71373 & 49.1 & 3.6 & 191.2 & 0.0 & 0.0 & 107.7 & 22.0 & 59.0 & 77.9 & 8.0 & 33.0 \\
 & CNN & 75132 & 59.6 & 17.3 & 662.3 & 0.0 & 358.2 & 863.8 & 356.0 & 739.2 & 562.4 & 70.0 & 241.0 \\
 & Daily Kos & 17222 & 90.2 & 6.0 & 177.5 & 27.0 & 133.0 & 80.9 & 32.0 & 82.0 & 95.3 & 19.0 & 56.0 \\
 & Democracy Now! & 20365 & 95.7 & 29.5 & 54.6 & 0.0 & 45.0 & 113.1 & 38.0 & 99.0 & 172.0 & 25.0 & 79.0 \\
 & Financial Times & 38488 & 74.0 & 5.6 & 191.4 & 65.0 & 130.0 & 41.1 & 13.0 & 33.0 & 28.8 & 8.0 & 16.0 \\
 & Forbes & 62174 & 46.6 & 3.8 & 94.8 & 0.0 & 18.0 & 37.5 & 5.0 & 11.0 & 16.1 & 3.0 & 5.0 \\
 & Fox News & 79222 & 71.1 & 15.4 & 110.0 & 0.0 & 0.0 & 2804.0 & 974.0 & 2051.0 & 1046.3 & 194.0 & 605.8 \\
 & HuffPost & 76766 & 42.2 & 8.0 & 1574.4 & 283.0 & 1225.0 & 357.9 & 98.0 & 342.0 & 177.6 & 17.0 & 81.0 \\
 & Jacobin magazine & 14066 & 96.0 & 4.8 & 333.8 & 148.0 & 340.0 & 35.3 & 13.0 & 35.0 & 67.4 & 24.0 & 57.0 \\
 & MSNBC & 67510 & 94.8 & 8.5 & 111.7 & 0.0 & 0.0 & 461.7 & 285.0 & 394.0 & 220.6 & 40.0 & 126.0 \\
 & Mother Jones & 10613 & 90.6 & 9.5 & 79.0 & 0.0 & 53.0 & 228.3 & 115.0 & 221.0 & 241.9 & 69.0 & 183.0 \\
 & NEWSMAX & 78018 & 88.9 & 5.9 & 635.9 & 0.0 & 356.0 & 545.9 & 206.0 & 410.0 & 273.9 & 43.0 & 136.0 \\
 & NPR & 53528 & 65.8 & 13.3 & 0.0 & 0.0 & 0.0 & 619.4 & 221.0 & 564.0 & 542.0 & 83.0 & 310.0 \\
 & Newsweek & 128534 & 56.0 & 13.8 & 429.1 & 34.0 & 171.0 & 43.3 & 5.0 & 22.0 & 37.2 & 2.0 & 6.0 \\
 & One America News Network & 21732 & 87.2 & 9.5 & 33.0 & 0.0 & 0.0 & 171.7 & 72.0 & 152.0 & 2.6 & 0.0 & 0.0 \\
 & PBS & 11331 & 40.0 & 5.6 & 1028.9 & 296.0 & 665.5 & 215.0 & 44.0 & 144.0 & 257.8 & 46.0 & 121.0 \\
 & Reuters & 175545 & 74.6 & 12.3 & 45.4 & 0.0 & 0.0 & 71.9 & 11.0 & 40.0 & 49.8 & 10.0 & 23.0 \\
 & THE WEEK & 31993 & 73.0 & 7.8 & 50.0 & 5.0 & 25.0 & 27.2 & 4.0 & 18.0 & 26.1 & 2.0 & 8.0 \\
 & The Daily Caller & 81690 & 71.7 & 11.1 & 367.2 & 23.0 & 136.0 & 465.3 & 91.0 & 325.0 & 336.6 & 28.0 & 130.0 \\
 & The Economist & 68578 & 72.2 & 6.6 & 260.2 & 63.0 & 117.0 & 60.6 & 9.0 & 27.0 & 42.8 & 7.0 & 17.0 \\
 & The Epoch Times & 60368 & 59.4 & 12.8 & 261.6 & 13.0 & 39.0 & 58.1 & 3.0 & 20.0 & 49.6 & 1.0 & 6.0 \\
 & The Hill & 95438 & 91.9 & 10.5 & 13.2 & 0.0 & 0.0 & 372.1 & 155.0 & 291.0 & 184.7 & 14.0 & 74.0 \\
 & The New Republic & 14695 & 81.4 & 3.7 & 29.2 & 9.0 & 17.0 & 6.2 & 1.0 & 4.0 & 7.9 & 3.0 & 5.0 \\
 & The New York Times & 94699 & 54.7 & 13.9 & 160.3 & 0.0 & 0.0 & 317.4 & 100.0 & 251.0 & 156.2 & 27.0 & 68.0 \\
 & The New Yorker & 53795 & 34.2 & 3.5 & 643.3 & 76.0 & 246.0 & 53.5 & 7.0 & 28.0 & 154.3 & 11.0 & 35.0 \\
 & The Wall Street Journal & 48386 & 61.0 & 7.2 & 154.6 & 0.0 & 0.0 & 141.9 & 24.0 & 78.0 & 56.4 & 7.0 & 14.0 \\
 & The Young Turks & 117777 & 80.7 & 11.1 & 148.1 & 13.0 & 79.0 & 176.6 & 31.0 & 90.0 & 199.8 & 16.0 & 49.0 \\
 & Truthdig & 3470 & 91.8 & 8.9 & 23.2 & 9.0 & 11.0 & 14.5 & 2.0 & 6.0 & 3.9 & 1.0 & 4.0 \\
 & USA TODAY & 50501 & 51.1 & 14.8 & 95.3 & 0.0 & 0.0 & 248.7 & 89.0 & 150.0 & 179.2 & 58.0 & 80.0 \\
 & Vox & 29552 & 70.2 & 6.7 & 12.6 & 0.0 & 0.0 & 99.4 & 25.0 & 86.0 & 165.5 & 16.0 & 49.0 \\
 & Washington Post & 79168 & 71.4 & 14.3 & 1603.9 & 296.0 & 1113.0 & 366.5 & 133.0 & 325.0 & 203.0 & 20.0 & 81.0 \\
 & Yahoo News & 51209 & 81.9 & 16.9 & 11.7 & 0.0 & 0.0 & 257.9 & 73.0 & 219.0 & 126.4 & 14.0 & 54.0 \\
 \midrule
\multirow[t]{12}{*}{Spain} & 20minutos.es & 92502 & 38.4 & 22.1 & 129.8 & 22.0 & 60.0 & 44.0 & 7.0 & 30.0 & 20.0 & 2.0 & 7.0 \\
 & ABC.es & 138236 & 52.5 & 21.9 & 263.7 & 42.0 & 147.0 & 83.3 & 16.0 & 64.0 & 90.4 & 6.0 & 24.0 \\
 & Antena 3 & 82481 & 12.3 & 8.8 & 127.7 & 13.0 & 41.0 & 21.4 & 1.0 & 7.0 & 19.3 & 1.0 & 4.0 \\
 & El Mundo & 71101 & 55.6 & 24.7 & 271.6 & 20.0 & 107.0 & 153.8 & 35.0 & 121.0 & 56.3 & 1.0 & 12.0 \\
 & El País & 114666 & 56.5 & 19.9 & 843.5 & 101.0 & 406.0 & 164.3 & 34.0 & 134.0 & 176.9 & 13.0 & 59.0 \\
 & La Vanguardia & 125816 & 34.0 & 20.8 & 406.6 & 12.0 & 120.0 & 156.8 & 30.0 & 122.0 & 67.9 & 0.0 & 11.0 \\
 & Público & 97315 & 80.3 & 18.9 & 152.4 & 0.0 & 62.0 & 123.6 & 37.0 & 110.0 & 50.0 & 0.0 & 10.0 \\
 & RTVE & 99718 & 52.3 & 18.1 & 105.6 & 16.0 & 38.0 & 20.4 & 2.0 & 12.0 & 13.3 & 1.0 & 4.0 \\
 & Telecinco & 85475 & 19.2 & 20.0 & 419.1 & 65.0 & 251.0 & 142.8 & 38.0 & 128.0 & 100.2 & 4.0 & 23.0 \\
 & elDiario.es & 66136 & 80.5 & 15.3 & 178.4 & 10.0 & 103.0 & 110.4 & 37.0 & 104.0 & 49.8 & 0.0 & 15.0 \\
 & laSexta & 106756 & 49.1 & 18.3 & 55.1 & 1.0 & 17.0 & 32.2 & 5.0 & 22.0 & 15.5 & 0.0 & 2.0 \\
 & okdiario.com & 117780 & 42.2 & 13.8 & 158.7 & 18.0 & 82.0 & 75.8 & 14.0 & 55.0 & 50.6 & 2.0 & 16.0 \\
 \midrule
\multirow[t]{8}{*}{France} & Causeur.fr & 14221 & 78.5 & 8.5 & 139.9 & 43.0 & 116.0 & 43.5 & 11.0 & 42.0 & 26.6 & 6.0 & 19.0 \\
 & Le Figaro & 65560 & 61.7 & 22.7 & 767.0 & 227.0 & 593.0 & 229.6 & 87.0 & 219.0 & 114.8 & 19.0 & 56.0 \\
 & Le Monde & 84869 & 73.7 & 20.8 & 355.8 & 111.0 & 254.0 & 132.6 & 42.0 & 126.0 & 52.8 & 10.0 & 26.0 \\
 & Libération & 75203 & 73.6 & 20.3 & 136.8 & 27.0 & 68.0 & 51.2 & 14.0 & 42.0 & 19.2 & 2.0 & 8.0 \\
 & Marianne & 29786 & 82.2 & 16.1 & 196.6 & 54.0 & 148.0 & 66.7 & 20.0 & 63.0 & 76.8 & 9.0 & 32.0 \\
 & Mediapart & 28094 & 90.5 & 18.8 & 442.5 & 99.0 & 308.0 & 122.9 & 22.0 & 104.0 & 217.0 & 19.0 & 77.0 \\
 & Reporterre, le média de l'écologie & 12075 & 77.6 & 16.6 & 198.8 & 79.0 & 176.0 & 29.0 & 9.0 & 26.0 & 80.1 & 18.0 & 50.0 \\
 & Valeurs actuelles & 45889 & 84.0 & 26.2 & 259.5 & 62.0 & 234.0 & 110.0 & 44.0 & 115.0 & 56.0 & 6.0 & 30.0 \\
 \midrule
\multirow[t]{17}{*}{Poland} & DoRzeczy & 37498 & 87.7 & 8.7 & 28.5 & 12.0 & 21.0 & 13.7 & 6.0 & 13.0 & 2.5 & 1.0 & 2.0 \\
 & Dziennik Gazeta Prawna & 10330 & 80.3 & 2.1 & 2.9 & 1.0 & 3.0 & 0.6 & 0.0 & 0.0 & 0.7 & 0.0 & 1.0 \\
 & FAKT24.pl & 67062 & 25.7 & 35.2 & 601.3 & 165.0 & 441.0 & 208.9 & 66.0 & 201.0 & 78.5 & 11.0 & 37.0 \\
 & Gazeta Wyborcza & 66024 & 61.6 & 11.0 & 478.0 & 147.0 & 445.0 & 132.6 & 45.0 & 131.0 & 53.9 & 10.0 & 37.0 \\
 & Gazeta.pl & 72655 & 46.8 & 17.1 & 282.1 & 76.0 & 219.0 & 103.6 & 30.0 & 94.0 & 28.6 & 5.0 & 16.0 \\
 & Gość Niedzielny & 24832 & 60.7 & 11.0 & 118.2 & 37.0 & 80.0 & 17.6 & 3.0 & 11.0 & 14.4 & 4.0 & 10.0 \\
 & Najwyższy CZAS! & 32713 & 78.4 & 12.6 & 38.0 & 14.0 & 27.0 & 7.3 & 3.0 & 6.0 & 7.2 & 1.0 & 4.0 \\
 & Newsweek Polska & 36594 & 56.9 & 7.2 & 170.1 & 40.0 & 140.0 & 70.6 & 18.0 & 66.0 & 21.4 & 4.0 & 13.0 \\
 & Onet & 90820 & 33.0 & 20.1 & 247.0 & 55.0 & 163.0 & 94.8 & 21.0 & 77.0 & 24.9 & 3.0 & 12.0 \\
 & Polityka & 26822 & 67.1 & 4.9 & 115.8 & 33.0 & 85.0 & 37.7 & 14.0 & 36.0 & 17.3 & 3.0 & 8.0 \\
 & Rzeczpospolita & 42965 & 79.3 & 10.7 & 104.6 & 35.0 & 87.0 & 40.8 & 14.0 & 39.0 & 10.6 & 3.0 & 7.0 \\
 & Super Express & 49133 & 25.5 & 27.6 & 383.8 & 60.0 & 194.0 & 127.8 & 24.0 & 98.0 & 42.9 & 4.0 & 13.0 \\
 & TVN24 & 94177 & 56.5 & 23.5 & 592.4 & 171.0 & 458.0 & 172.3 & 71.0 & 165.0 & 41.0 & 8.0 & 25.0 \\
 & Tygodnik Sieci & 37644 & 81.7 & 12.0 & 97.3 & 41.0 & 96.0 & 44.9 & 20.0 & 48.2 & 7.4 & 2.0 & 6.0 \\
 & Warszawska Gazeta & 6205 & 74.4 & 13.5 & 48.8 & 24.0 & 49.0 & 14.5 & 6.0 & 15.0 & 7.0 & 1.0 & 5.0 \\
 & Wirtualna Polska & 86972 & 37.5 & 13.3 & 179.8 & 36.0 & 108.0 & 81.5 & 16.0 & 61.0 & 16.9 & 2.0 & 7.0 \\
 & tvp.info & 59875 & 68.8 & 16.1 & 333.1 & 140.0 & 287.0 & 179.7 & 90.0 & 181.0 & 36.2 & 8.0 & 22.0 \\
 \midrule
\multirow[t]{11}{*}{Ireland} & Gript & 17516 & 85.3 & 9.1 & 119.4 & 49.0 & 97.0 & 38.8 & 14.0 & 38.0 & 36.4 & 9.0 & 24.0 \\
 & Independent.ie & 48886 & 38.2 & 16.2 & 214.5 & 48.0 & 161.0 & 85.5 & 19.0 & 83.0 & 16.8 & 3.0 & 9.0 \\
 & Irish Examiner & 54265 & 42.5 & 11.3 & 65.4 & 8.0 & 27.0 & 19.0 & 2.0 & 12.0 & 5.6 & 1.0 & 3.0 \\
 & Irish Independent & 8051 & 27.4 & 17.1 & 184.5 & 28.0 & 102.0 & 55.6 & 11.0 & 42.0 & 10.6 & 1.0 & 5.0 \\
 & JOE.ie & 33321 & 20.6 & 9.9 & 316.8 & 38.0 & 143.0 & 115.4 & 21.0 & 85.0 & 29.1 & 2.0 & 8.0 \\
 & RTÉ & 7103 & 14.6 & 0.9 & 192.3 & 19.0 & 61.0 & 22.4 & 2.0 & 12.0 & 21.5 & 2.0 & 7.0 \\
 & Sunday World & 47683 & 29.7 & 34.2 & 65.9 & 15.0 & 37.0 & 26.4 & 5.0 & 19.0 & 7.6 & 1.0 & 4.0 \\
 & The Irish Sun & 71675 & 14.0 & 19.4 & 65.2 & 8.0 & 38.0 & 19.7 & 2.0 & 11.0 & 9.1 & 0.0 & 3.0 \\
 & The Irish Times & 81533 & 54.7 & 13.3 & 107.0 & 23.0 & 69.0 & 47.2 & 10.0 & 43.0 & 10.9 & 1.0 & 6.0 \\
 & TheJournal.ie & 25261 & 69.8 & 20.7 & 125.6 & 41.0 & 102.0 & 81.0 & 28.0 & 86.0 & 22.1 & 3.0 & 11.0 \\
 & breakingnews.ie & 56489 & 55.3 & 18.3 & 51.5 & 13.0 & 38.0 & 26.2 & 4.0 & 21.0 & 6.2 & 1.0 & 3.0 \\
 \midrule
\multirow[t]{12}{*}{United Kingdom} & Daily Mail & 82743 & 8.1 & 20.3 & 449.5 & 0.0 & 94.0 & 350.3 & 75.0 & 219.0 & 24.9 & 0.0 & 4.0 \\
 & Financial Times & 38488 & 74.0 & 5.6 & 191.4 & 64.5 & 130.0 & 41.1 & 13.0 & 33.0 & 28.8 & 8.0 & 16.0 \\
 & GB News & 71298 & 73.5 & 8.2 & 194.9 & 19.0 & 126.0 & 203.4 & 90.0 & 201.0 & 26.7 & 0.0 & 8.0 \\
 & Manchester Evening News & 76544 & 20.3 & 28.2 & 179.1 & 0.0 & 83.0 & 204.2 & 52.0 & 157.0 & 36.4 & 0.0 & 11.0 \\
 & Sky News & 82121 & 66.8 & 22.3 & 608.3 & 113.0 & 356.0 & 281.3 & 104.0 & 259.0 & 101.6 & 10.0 & 35.0 \\
 & The Guardian & 59650 & 51.9 & 10.8 & 713.1 & 229.0 & 567.0 & 244.6 & 111.0 & 238.0 & 90.7 & 18.0 & 51.0 \\
 & The Independent & 134647 & 37.0 & 14.7 & 780.8 & 116.0 & 393.0 & 227.9 & 54.0 & 191.0 & 130.0 & 7.0 & 32.0 \\
 & The Sun & 159607 & 8.4 & 14.7 & 309.1 & 46.0 & 157.0 & 174.8 & 27.0 & 103.0 & 78.0 & 4.0 & 21.0 \\
 & The Telegraph & 127078 & 64.3 & 10.6 & 279.4 & 51.0 & 145.0 & 91.1 & 24.0 & 74.0 & 24.4 & 3.0 & 8.0 \\
 & The Times and The Sunday Times & 57813 & 68.6 & 13.5 & 76.4 & 18.0 & 40.0 & 38.3 & 9.0 & 32.0 & 7.4 & 2.0 & 4.0 \\
 & The i Paper & 60799 & 62.7 & 11.1 & 59.9 & 7.0 & 27.0 & 20.3 & 2.0 & 10.0 & 5.3 & 0.0 & 2.0 \\
 & UnHerd & 8353 & 86.1 & 2.6 & 15.9 & 7.0 & 14.0 & 4.6 & 1.0 & 4.0 & 2.8 & 1.0 & 2.0 \\

\bottomrule
\end{tabularx}
\label{app:tab:outlets}
\end{table*}

\subsection{Data extraction via Facebook API}\label{app:sec:data:collection}

Posts data, including text content and interaction metrics, was obtained through 
Sotrender--- a social media analytics platform with direct access to the Facebook Graph API.
We employed a two-tiered data collection approach to ensure comprehensive and temporally
relevant data: 
\begin{enumerate}
    \item\textbf{Ongoing Data Collection.} 
    The primary method for collecting data was a real-time integration
    with the Facebook API, whereby Sotrender retrieved posts and updated interaction metrics
    on an hourly basis. This method allowed for close tracking of user engagement in the
    initial period following a post’s publication. Post engagement data were
    refreshed over a two-week period, which represents the typical window of peak user interaction
    on Facebook posts.
    \item\textbf{Historical Data Retrieval.} 
    For posts not captured through ongoing data collection, Sotrender utilized the Facebook API’s
    historical data retrieval option. This method allows for the acquisition of engagement metrics
    for posts beyond the two-week tracking window.
\end{enumerate}

\subsection{Outlet selection}\label{app:sec:data:outlets}

The selection of 97 news media organizations was conducted by experts from each country to ensure representation of the largest news outlets with a strong and active presence on Facebook, characterized by frequent daily postings. The selection criteria included reach and influence, ensuring the inclusion of the most prominent media in each country; Facebook activity, prioritizing outlets that publish at least 4 posts per day on average; ideological balance, incorporating both mainstream and politically right- and left-leaning outlets; and diversity in format, covering both serious journalism and tabloid-style reporting. Media with low or highly irregular Facebook activity were not included to maintain data consistency and analytical robustness. A detailed list of selected media outlets and their characteristics is provided in Table~\ref{app:tab:outlets}.

\section{Classifiers}\label{app:sec:cls}

To identify political and negative news, we developed two neural binary classifiers.
Both of them were constructed by adding a binary classification head to a pre-trained
base language model known as
XLM-RoBERTa-Large~\citep{conneauUnsupervisedCrosslingualRepresentation2020}, 
which is a transformer model with 561 millions of parameters optimized based on
2.5 terabytes of CommonCrawl data for 100 languages. 
Both the head and the base model were fine-tuned and validated using custom datasets
of hand-annotated examples using separate random subsamples of the
dataset for optimization of model weights (training set), model hyperparameters (testing set),
and computing final performance metrics (validation set). This tripartite data splitting strategy
allowed for interpreting the reported performance metrics as proper measures of the out-of-sample
classification performance.

Table~\ref{app:tab:setsizes} presents the datasets used for training, testing and validation
(classification performance is discussed in the Main Text).
The following two subsections discuss details specific for individual classifiers.
Both classifiers were trained for up to 10 epochs with early stopping after 2 epochs without
improvement on the testing set. Training and inference were done using Huggingface
\texttt{transformers} library for Python~\citep{wolfHuggingFacesTransformersStateoftheart2020}.
Table~\ref{app:tab:models} presents main specifications and hyperparameters used for training
both models. All parameters and configuration options not mentioned in the table used default
values defined in the \textit{Trainer API} of the \texttt{transformers} library.
Training hyperparameters were optimized (with respect to performance on the testing set)
based on efficient search over 100 trials using \texttt{optuna}
package~\citep{akibaOptunaNextgenerationHyperparameter2019}. This hyperparameter search was
conducted only for the political classifier and the identified optimal parameter values were used
also for training the negativity classifier---which was justified by the fact that both used
the same backbone model.

\begin{table}[ht]
\centering
\tiny\sffamily
\begin{minipage}[t]{.46\textwidth}
\centering
\caption{Sizes of training, testing and validation sets}
\begin{tabularx}{\columnwidth}{c|X|rrr|rrr|rrr}
    \toprule
    && \multicolumn{3}{c}{Training} & \multicolumn{3}{c}{Testing} & \multicolumn{3}{c}{Validation}\\
    & Country & $N$ & $N(+)$ & $N(-)$ & $N$ & $N(+)$ & $N(-)$ & $N$ & $N(+)$ & $N(-)$ \\
    \midrule
    \multirow{6}{*}{\rotatebox[origin=c]{90}{Political}}
    & US  & 1324 & 928 & 396   & 171 & 122 & 49   & 140 & 98 & 42  \\
    & UK  & 749  & 344 & 405   & 106 & 50  & 56   & 83  & 36 & 47  \\
    & IRL & 1098 & 292 & 806   & 135 & 43  & 92   & 140 & 37 & 103 \\
    & PL  & 1576 & 698 & 878   & 187 & 84  & 103  & 231 & 93 & 138 \\
    & FR  & 759  & 248 & 511   & 93  & 58  & 35   & 87  & 58 & 29  \\
    & ESP & 660  & 427 & 233   & 79  & 21  & 58   & 90  & 28 & 62  \\
    \midrule
    \multirow{6}{*}{\rotatebox[origin=c]{90}{Negativity}}
    & US  & 751 & 257 & 494   & 93  & 38 & 55   & 105 & 41 & 64 \\
    & UK  & 691 & 325 & 366   & 96  & 53 & 43   & 85  & 42 & 43 \\
    & IRL & 795 & 385 & 410   & 101 & 49 & 52   & 92  & 48 & 44 \\
    & PL  & 686 & 341 & 345   & 71  & 36 & 35   & 80  & 45 & 35 \\
    & FR  & 669 & 325 & 344   & 76  & 38 & 38   & 88  & 53 & 35 \\
    & ESP & 660 & 342 & 318   & 95  & 45 & 50   & 82  & 44 & 38 \\
    \bottomrule
    \multicolumn{11}{l}{$N$ - all cases} \\
    \multicolumn{11}{l}{$N(+)$ - target label cases} \\
    \multicolumn{11}{l}{$N(-)$ - complement label cases}
\end{tabularx}
\label{app:tab:setsizes}
\end{minipage}
\hfill
\begin{minipage}[t]{.45\textwidth}
\caption{Models' specifications and hyperparameters}
\begin{tabularx}{\columnwidth}{l|X|X}
    \toprule
    & Political & Negativity \\
    \midrule
    Backbone & XLM-Roberta-Large & XLM-Roberta-Large \\
    Labels & POLITICAL/OTHER & NEGATIVE/OTHER \\
    Batch size & 16 & 16 \\
    learning rate & \num{1e-6} & \num{1e-6} \\
    learning rate scheduler & cosine & cosine \\
    warmup ratio & 0.25 & 0.25 \\
    weight decay & 0.004 & 0.004 \\
    random seed & 1884749421 & 1884749421 \\
    gradient checkpointing & yes & yes \\
    \bottomrule
\end{tabularx}
\label{app:tab:models}
\end{minipage}
\end{table}

Training, testing and validation data were sampled from the analyzed dataset of Facebook posts.
Moreover, US and PL sets for the political classifier used labeled headlines of news articles
studies by~\cite{wojcieszakNonNewsWebsitesExpose2024}.
In all cases domain names, URLs, emails and HTML tags have been stripped from text data used for
training. This was to prevent potential leakage of information associated with proper names, domain URL etc., and unrelated to the semantics of input texts.
Examples with less than 3 words, using whitespace tokenization, were excluded from the training.
Only Facebook posts with both post and snippet text (full textual context presented to users) 
were used for training, i.e.~posts for which only part of textual content was harvested were 
excluded from the training sample.

\subsection{Political classifier}\label{app:sec:cls:pol}
The annotators were instructed to consider \enquote{politics} rather broadly as: 
\begin{quote}
    references to political figures, policies, or elections \emph{as well as} issues such as 
    climate change, immigration, healthcare, gun control, sexual assault, racial, gender, sexual, 
    ethnic, and religious minorities, the regulation of large tech companies, 
    as well as crimes involving guns (e.g., school shootings) or with political implications (e.g., large criminal organizations, such as cartels or mafias).
\end{quote}

Moreover, the annotators were provided with some auxiliary definitions (Table~\ref{app:tab:poldefs})
targeting specific edge cases that were challenging to consistently classify.
The average pairwise inter-annotator accuracy based on 708 comparisons was 89\%.

\begin{table}[ht]
\caption{Auxiliary definitions of political news}
\centering
\tiny\sffamily
\begin{tabularx}{\columnwidth}{l|X} 
    \toprule
    Context & Instructions \\
    \midrule
    Coronavirus & 
    \textbf{Political.} Content related to political implications of the pandemics, public health
    policies or general political consequences, lockdowns etc. 
    as well as reports of deaths, ICU cases or increasing cases. \newline
    \textbf{Non-political.} Anything else, in particular neutral reports of new cases without
    any assessment of the state of affairs related to the pandemic; pandemics lifestyle advice etc.
    \\
    \midrule
    Crimes & 
    \textbf{Political.} Content containing \textit{assessments} of police actions 
    (e.g., police brutality);
    Stating explicitly that crime was committed using a firearm;
    The victims are women, minorities, children or members of other protected groups;
    The perpetrators are immigrants;
    Crime with political implications, e.g.~activity of cartels, mafias, crimes done
    by politicians.
    \newline
    \textbf{Non-political.} Otherwise.
    \\
    \midrule
    Justice system & 
    \textbf{Political.} Post containing assessments of state actions;
    About crime cases described in the "Crimes" section above.
    \newline
    \textbf{Non-political.} Otherwise.
    \\
    \midrule
    Monarchies & 
    \textbf{Political.} Posts concerned with the reigning queen or king;
    Reports of \textit{political} activities of any members of royal families;
    Assessments, both explicit and implicit, of the functioning or institutions of the monarchy.
    \newline
    \textbf{Non-political} Otherwise, in particular gossips and lifestyle pieces concerned
    with members of royal families as long as they do not contain a critique or significant praise
    of the institution of monarchy.
    \\
    \bottomrule
\end{tabularx}
\label{app:tab:poldefs}
\end{table}

Note that while the base model is multilingual and supports 100 languages, our fine-tuning process
used data for only 6 countries and 4 languages given that our core goal was to train a model that
is highly performant in the four languages and in the six countries considered in this study. 
However, to ensure that our model performs across countries and languages and to estimate the
efficacy of our model in applications to contexts not seen during training, we used a dataset of
labeled headlines from articles published by Dutch media 
outlets~\cite{wojcieszakNonNewsWebsitesExpose2024}. This provided us with additional validation
data for a language not present in the main training dataset and a relatively rare content type
(most training examples are social media posts, not article headlines). 
The obtained $F_1$-scores for predicting target label (POLITICAL), complement label (OTHER) and 
the harmonic average of the two were correspondingly: 0.805, 0.797 and 0.800, indicating a 
significant ability of the model to transfer its \enquote{knowledge} to new unseen contexts. 
While the cited metrics are noticeably lower than the validation metrics reported in 
Table~\ref{tab:performance} in the Main Text, they are arguably high enough to be useful in 
practice; that is, the classifier could be used to produce acceptably accurate estimates of 
prevalence of political news within a specific corpus of news texts.

\subsection{Negativity Classifier} \label{app:sec:negatclass}

To define negativity, we followed the conceptual work
by~\cite{lengauerNegativityPoliticalNews2011}, who proposed definitions and measurements of
negativity in news. Our operationalization and classifier define negativity by
combining \enquote{the mere dissemination of negative news} 
(exogenous negativity coming into the news from outside, that is, from the topic itself) 
and \enquote{endogenous negativity imposed on news by journalists through their usage of language}~\citep[p.~181]{lengauerNegativityPoliticalNews2011}. 
Although we recognize that there are theoretical and practical reasons to distinguish between 
media-initiated negativity (e.g.,~harshly criticizing the opposition) and other-initiated
negativity (e.g.,~covering tragedies or reporting on one politician attacking another),
we do not differentiate between these two for methodological and analytical reasons. 
For one, these two are unlikely to be reliably captured and distinguished in manual and automated
labeling without subjective interpretations~\citep[see~][]{lengauerNegativityPoliticalNews2011}.
In addition, our large scale analyses are concerned with the level of overarching negativity
in political and non-political news in a comparative perspective, where researchers may not always
be familiar with the complexities of the relationships between journalists, news outlets, and the
political context. 

More concretely, the annotators were asked to focus on the following aspects of negativity
in news with the following instructions:

\begin{itemize}
\item \textbf{Negative sentiment.} 
    We want to capture “traditional” negative sentiment, that is, language with 
    clearly negative overall tone or dominated by negative sentiment expressed towards particular 
    objects (individuals, institutions etc.). Ambiguous sentiment, including 
    mixed sentiment towards different objects (i.e.~some positive and some negative) should \textit{not} be coded as negative.
    
    {\footnotesize
    \begin{itemize}
        \item \textbf{Examples.}
        \begin{itemize}
            \item This is so stupid!
            \item John is a pathetic speaker.
        \end{itemize}
        \item \textbf{Non-examples.}
        \begin{itemize}
            \item Disgusting dessert after an otherwise very pleasant meal.
            \item Unlike Robert, who is great, John is a pathetic speaker.
        \end{itemize}
    \end{itemize}
    }

\item \textbf{Crimes.} In general all non-petty, and in particular violent, crimes should be considered negative. However, there are some edge cases, explained below.

    {\footnotesize
    \begin{itemize}
        \item \textbf{Negative.} Reports on new crimes should be considered negative.
        \item \textbf{Other.} A frequent category of news articles are follow-up articles related
        to a past crime, which focus on later developments only tangentially related to the crime, 
        such as actions of law enforcement or justice. Such news, unless they focus on the details 
        of the crime and/or have also negative emotional/linguistic tone, should not be considered 
        negative. Moreover, speculations detached from factual reporting on any actual 
        crime also should not be negative.

        \item \textbf{Examples.}
        \begin{itemize}
            \item \textbf{Petty crime.}
            A man attacked and bit a garda who drove him home after he was found lying in his own vomit on the side of a road in north Co Dublin Garda bitten after driving home man
            found lying in vomit beside road.
            \item \textbf{Sensational but tangential to crime.} Last weekend, Zimbabwean media 
            outlets reported that a man wanted for the murder of three people was allegedly living 
            in Ireland as an asylum seeker. Gardai have said they are investigating, but they have 
            not found the fugitive in Ireland. Garda investigate whether Zimbabwean wanted for 3 
            murders is in Ireland as asylum seeker - Gript.
            \item \textbf{Follow-ups to old crimes.} Co Wexford woman who stabbed her boyfriend
            spared jail time.

        \end{itemize}
    \end{itemize}
    }

\item \textbf{Accidents.} 
    Similar to crimes, reporting on seroius accidents is in general negative, but follow-ups not focused on the accident itself are 
    not negative. However, mentions of burial/commemoration events with explicit mentions of 
    victims should be negative.

\item \textbf{Wars and clashes.} 
     The labeling depends on the details of the reporting:
    
    {\footnotesize
    \begin{itemize}
        \item \textbf{Negative.} When reporting mentions civilian casualties and/or attacks on
        civilian targets (e.g.~hospitals) or adverse consequences of military actions on civilian 
        populations. Moreover, reporting on military clashes mentioning excess violence/cruelty 
        (e.g.~tortures on prisoners of war) is also considered negative.
        \item \textbf{Other.} News on military actions and clashes between armed forces without mentioning adverse consequences for civilians.
        \item \textbf{Justification.} Military actions detached from civilian life are not 
        unambiguously negative, and their perception may depend on political sympathies and other 
        attitudes of a reader, and the classifier should be as unbiased as possible in this 
        respect.
    \end{itemize}
    }

\item \textbf{Negative events.} 
     By this we mean any news on events that are typically perceived as negative, in particular
     serious and/or violent crimes, disasters and accidents, especially ones with human casualties 
     and/or huge material losses, social, political and economic events leading to mass disruptions of daily life, e.g. high inflation resulting in higher costs of living. 
     As there are ambiguous edge cases, here is a list of most typical categories 
     and corresponding justifications:
    
    {\footnotesize
    \begin{itemize}
        \item \textbf{Examples.}
        \begin{itemize}
            \item The president added, however, that “terrorists are still using weapons and are damaging people’s property”, and that “counter-terrorist actions” should be continued. 

            Kazakhstan leader says constitutional order restored amid deadly protests.
        \end{itemize}
        \item \textbf{Justification.} Some news report on both negative and positive events. For instance, in the example above, the positive event is the restoration of constitutional order, and the negative is that the clashes are still ongoing, so ultimately the post
        should not be considered negative.
        \end{itemize}
    }

\item \textbf{Negative events in a distant past.}
    Examples include true crimes stories about mysterious unsolved crimes from many years ago, 
    or articles on drastic historical events.
    
    {\footnotesize
    \begin{itemize}
        \item \textbf{Negative.} Only if the emotional/linguistic tone is very negative and/or 
        direct adverse consequences of contemporary relevance are mentioned.
        \item \textbf{Other.} In all other cases.
    \end{itemize}
    }

\item \textbf{COVID.} Depending on details:
{\footnotesize
\begin{itemize}
    \item \textbf{Negative.} Reports mentioning deaths (or ICU patients) 
    and/or new cases in an alarming fashion.
    \item \textbf{Other.} Neutral reports on new cases without mentioning deaths and/or
    alarming or negative language. In particular, reports on decreasing incidence rates are not 
    negative.
\end{itemize}
}

\item \textbf{Advice.} In particular there are health advice articles which may mention death, 
but rather in abstract terms, and therefore should not be considered negative.
    
\end{itemize}

\subsection{Consistency of post and full content classification}
\label{app:sec:cls-consistency}

We tested the consistency between labels produced by our classifiers
based on post content and full article content using a random
sample of 212 articles constructed to be balanced with respect to
countries and post-based political and negativity labels.
The labels were identical in 82.7\% [77.7\%--87.8\%] 
of cases for the political classifier 
and 77.9\% [72.3\%--83.5\%].
of cases for the negativity classifier.
Note that, since both classifiers are roughly 90\% accurate,
the obtained results are close to what should be expected when
outcomes of two binary random variables with $p = 0.90$ probability
of \enquote{success}:
\begin{equation*}
    p^2 + (1-p)^2 = 0.82 
\end{equation*}
Thus, the reported results are likely to be similar to results 
that could obtained for full article content.

\subsection{Examples}\label{app:sec:examples}

Tables~\ref{app:tab:examples} presents a random selection of post classifications for the
U.S. and U.K. and split by political/non-political as well as negative/non-negative posts
with 5 examples for each combination.

\begin{table}[htb!]
\caption{Examples of post classification (US and UK)}
\centering\tiny\sffamily
\begin{tabularx}{\textwidth}{lll|X}
\toprule
\multirow[t]{20}{*}{United States} & \multirow[t]{10}{*}{non-political} & negative & “could have gotten it from a kid.”

Children potentially exposed to monkeypox at Illinois day care \\
 &  & negative & The bottom of a 13-story condo tower outside Miami has emerged as a possible failure point, according to experts who examined footage of the collapse.

Possible Failure Point Emerges in Miami Building Collapse \\
 &  & negative & Houston rapper Big Pokey died Sunday after collapsing during a performance, a representative confirmed to USA TODAY Sunday.

Rapper Big Pokey dies at 48 after collapsing on stage \\
 &  & negative & The "disturbing" decision came after officials found more than 60 empty alcohol bottles and cans aboard the vessel.

Officials rule out alcohol as factor in Florida teen's boat-crash death \\
 &  & negative & Here's what the note said:

Dog rescuers find bundle on doorstep with heartbreaking note that ‘tore at our heartstrings’ \\
 &  & non-negative & Brainless Troll Gets SCHOOLED By Scholar \\
 &  & non-negative & Celebrate the holiday season, reflect on the year that was and look ahead with NPR Presents! Beginning Dec. 14, join us for virtual events, holiday features from NPR hosts and more. Sign up for 12 days of NPR Presents.

12 Days of NPR Presents \\
 &  & non-negative & A cartoon by Pia Guerra and Ian Boothby. \#NewYorkerCartoonsSee more from this week’s issue: http://nyer.cm/6lynrLf \\
 &  & non-negative & This is the first time you can watch TikTok on TV. \\
 &  & non-negative & 'HE IS EXHAUSTING': Woman’s hilarious adoption ad for foster dog goes viral.

Woman’s hilarious adoption ad for ‘hellion’ foster dog goes viral: ‘He is exhausting’ \\
 & \multirow[t]{10}{*}{political} & negative & An unidentified missile struck southern Kyiv. A witness said she ‘heard a terrible explosion of unbelievable power.’ More on the Ukraine crisis: https://reut.rs/3vulr2P

This content isn't available right now \\
 &  & negative & Wisconsin's governor declared a new public health emergency and extended a face mask mandate into November to fight a coronavirus flareup in his state https://reut.rs/300nvPk

As U.S. surpasses 200,000 COVID-19 deaths, Wisconsin sounds alarm over surges in cases \\
 &  & negative & Advocates plead for stronger federal action as nursing homes see a mounting death toll from the novel coronavirus.

Advocates demand stronger federal action as nursing homes engulfed by pandemic \\
 &  & negative & The White House coronavirus task force has not held a daily press briefing in over a month, even though about 800 to 1,000 U.S. residents are dying of COVID-19 each day.

White House Coronavirus Task Force Goes Quiet as Mike Pence Flouts Social Distancing \\
 &  & negative & The U.S. Supreme Court has rejected a last-minute appeal from an Arizona prisoner to delay his execution for his murder conviction in the 1984 killing of an 8-year-old girl, clearing the way for the state’s second execution in less than a month.

Arizona set to execute Frank Atwood, who killed girl in 1984 \\
 &  & non-negative & Trust the "science."

Pro-Maskers May Be Laying The Groundwork To Bring Back COVID Restrictions \\
 &  & non-negative & Brad D. Smith was six years old when a plane carrying the 1970 Marshall University football team crashed a mile from his home near campus in southern West Virginia, killing all 75 people on board. His cousins rushed to aid their dying neighbors as volunteer firefighters. “I watched the flames burn outside my window,” Smith remembers. “And then I watched this community rise from the ashes.”Half a century later, the recently retired Intuit CEO’s community is waging new battles, with an opioid epidemic raging and the coal economy that once made Governor Jim Justice a billionaire on the verge of extinction. So, after 36 years away, Smith decided to take the country roads back home to West Virginia, the place he belongs—and into the President’s House at Marshall, his alma mater, which he took over in January 2022. He brought back with him a sizable fortune, accumulated over nearly four decades in business. According to Forbes’ ranking of the richest person in each state, released Thursday for the first time since 2019, he’s West Virginia’s wealthiest resident, worth \$700 million.Forbes estimates that roughly half of his fortune is comprised of 943,000 Intuit shares and options he still holds. That’s after selling 2.4 million shares during his tenure as CEO from 2008 to 2018 (and as chairman until January 2022), netting him about \$300 million (after taxes and the cost of option exercises). He takes the mantle as the state’s richest from Jim Justice, whose wealth has been weighed down by debt. Smith is worth some \$250 million more than the governor, who dropped from the ranks of the world’s billionaires in 2021, when it was revealed that he’d personally guaranteed \$850 million of loans to his coal businesses by Credit Suisse via a now insolvent intermediary, Greensill Capital. (Justice also owns the iconic Greenbrier Resort in White Sulphur Springs, West Virginia, and other real estate assets in Appalachia; he disputes Forbes’ estimate of his fortune.) Smith declined to comment on Forbes’ estimate of his net worth.WATCH: https://trib.al/mElsZPC \\
 &  & non-negative & WATCH: Jacob Soboroff demands evidence from Ric Grenell, Trump adviser and former acting director of national intelligence, to back up his assertions about votes in Nevada.

MSNBC reporter confronts Trump campaign's Ric Grenell on allegations of Nevada voter fraud \\
 &  & non-negative & An NFT of Ukraine's flag has raised over \$6.7 million for the country's defenses as cryptocurrency donations continue to flood in following the Russian invasion.

NFT backed by Pussy Riot member raises \$6.7 million for Ukraine \\
 &  & non-negative & Cairo's 'City of the Dead' has been razed by the Egyptian government, which is undertaking a massive modern development project.

Egypt is trying to make Cairo look like Dubai. It's taken 10 years and cost \$58 billion. \\

\midrule

\multirow[t]{20}{*}{United Kingdom} & \multirow[t]{10}{*}{non-political} & negative & National Weather Service service issues warnings to boaters as ‘whirling’ columns of ‘air and water mist’ seen offshore

Terrifying waterspouts spotted off coast of Florida \\
 &  & negative & Storm Ciarán has made landfall, bringing strong winds and heavy rain across parts of the UK until Friday morning.Yellow and amber weather warnings are in place, including a danger to life warning due to very strong winds hitting the south coast of England and Wales on Thursday. Some areas could see “significant flooding”, the Environment Agency said

Storm Ciarán: how will flights and trains be affected? \\
 &  & negative & Paramedics attempted to revive Hawkins in his hotel room in Bogota, Colombia, but there was no response, authorities have said.

Foo Fighters’ Taylor Hawkins ‘one of last true rock star drummers of our time’ \\
 &  & negative & Such a tragedy. RIP

Haunting video emerges of 'talented' boy pilot, 16, who died alone \\
 &  & negative & This is so heartbreaking for him and the rest of the cast. Thoughts and prayers go out to them. \\
 &  & non-negative & Here's what the house is REALLY like.

'I was on X Factor - Simon Cowell tried to get me to wear an outrageous outfit' \\
 &  & non-negative & A mother was astounded to be charged 20p by a primary school to replace a pencil they say her 10-year-old son broke.“I couldn’t believe what the letter said – I thought it was ridiculous.”READ MORE HERE https://www.independent.co.uk/news/uk/home-news/school-charges-pupil-broken-pencil-b2222889.html \\
 &  & non-negative & The student cycled the long distance after feeling homesick \\
 &  & non-negative & Just published: front page of the Financial Times, UK edition, Thursday 10 March https://on.ft.com/3I1jTzE \\
 &  & non-negative & "I wasn’t defending the joke"

Victoria Coren Mitchell clarifies view on Jimmy Carr joke after supporting comedian \\
 & \multirow[t]{10}{*}{political} & negative & Care homes learnt a ‘painful lesson’ from SARS, and quickly sprung into action to make sure the same thing didn’t happen with Covid-19 \\
 &  & negative & Awful news.

Pregnant mum and baby die after Russian airstrike on hospital \\
 &  & negative & Children's drawings lay scattered on the ground strewn with shattered glass and debris in a severely damaged village school northwest of Kyiv on Wednesday

Watch: School destroyed as Russia bombards Kyiv despite pledge to withdraw \\
 &  & negative & "There's a real worry she's facing a custodial sentence." \\
 &  & negative & A London resident who was shot dead in Pakistan had reportedly rejected two men’s marriage proposals prior to her death

UK woman shot dead in Pakistan after ‘refusing two men’s marriage proposals’ \\
 &  & non-negative & How much should we talk about pay transparency and is it worth it? In the latest episode of the Working It podcast, host Isabel Berwick discusses different approaches to the topic https://on.ft.com/3FO7MWe \\
 &  & non-negative & Are you up to date with the latest booster policy from the PM? Boris Johnson is urging Britons to get the booster jab to avoid future restrictions.https://www.gbnews.uk/news/how-to-book-a-covid-booster-vaccine/179831

How to book a Covid booster vaccine? \\
 &  & non-negative & Lawyers for the British socialite have submitted a motion under seal seeking a new trial after a juror told the media he was a victim of sexual abuse.

Ghislaine Maxwell files for retrial in sex trafficking case weeks after conviction \\
 &  & non-negative & Poland has accused Moscow of ‘gas imperialism’ after Gazprom cut off supplies but insisted that the central European nation insisted it would be able to cope without Russian hydrocarbons. Bulgaria, however, only has one month of gas in reserve.

Poland will not need Russian gas from autumn, says prime minister \\
 &  & non-negative & Officers say 'any potential disorder will be dealt with swiftly'

GMP statement ahead of Bolton and Wigan clash with police patrols stepped up \\
\bottomrule
\end{tabularx}
\label{app:tab:examples}
\end{table}

\clearpage
\section{GLMM models}\label{app:sec:glmm}

This appendix complements the discussion of our analytical approach in \nameref{sec:methods}~(\ref{sec:methods:analysis}).

\subsection{Model interpretation and bias}\label{app:sec:glmm:model}

For the sake of clarity, let us begin with restating the assumed data generating process:
\begin{equation}
    \mathcal{Y} 
    = \mathbf{X}\mathbf{\beta} + \mathbf{Z}\mathcal{B} + \mathbf{\epsilon}
\end{equation}
where $\mathcal{Y}$ is the random vector of outcome values (in the linear predictor scale),
$\mathbf{X}$ and $\mathbf{\beta}$ the design matrix and parameter vector for the fixed effects,
$\mathbf{Z}$ and $\mathcal{B} \sim \mathcal{N}(0, \boldsymbol{\Sigma)}$ the design matrix and 
random parameter vector for the random effects, and $\mathbf{\epsilon}$ the random vector of 
residuals. Moreover, the $\mathbf{Z}\mathcal{B}$ term can be decomposed into separate time and 
outlet terms, $\mathbf{Z}_t\mathcal{B}_t + \mathbf{Z}_o\mathcal{B}_o$, but whenever possible 
without a loss of clarity we will use the more compact notation for the sake of simplicity.
Crucially, $\mathbf{Z}\mathcal{B} = \mathbf{Z}_t\mathcal{B}_t + \mathbf{Z}_o\mathcal{B}_o$ term
captures all information characteristic for specific outlets and days, allowing the fixed effects,
$\mathbf{X}\mathbf{\beta}$, to be a proper representation of \enquote{population}-level effects
unaffected by disparities in the number of raw observations (posts) corresponding to each group
(outlet or day). In other words, $\mathbf{Z}\mathcal{B}$ denotes effects specific for individual 
outlets and days, while $\mathbf{X}\mathbf{\beta}$ represents the remaining systematic effects
that are the same across groups. And, as far as the goal is to gain general knowledge extending 
beyond the observed sample, it is the latter that are of practical importance, while group-specific
effects are effectively nuisance variables, which must be controlled, but ultimately removed
from the analysis.

Since random effects are assumed to follow a centered multivariate normal distribution 
the \enquote{population-level} estimates should correspond to some general and long-term averages 
under the following assumptions: 
\begin{enumerate}[label={(A\arabic*)}]
    \item Within-group means and variances (on the linear predictor scale) 
    can be estimated effectively. \label{A:effective}
    \item Analyzed time period is long enough. \label{A:time}
    \item Selected outlets are distributed symmetrically around the population average 
    for a given outcome. \label{A:outlets}
\end{enumerate}

Due to the moderately large size of the dataset (${\sim}7$ million observations) \ref{A:effective}
is likely to hold. Moreover, by the design of mixed models, estimates for groups with limited 
information are regularized and shrinked towards the 
\enquote{population} mean~\citep{gelmanDataAnalysisUsing2021}.
\ref{A:time} may be also considered justified, at least to some significant extent, 
given that our data  spans more than 4 years. Crucially, if an analyzed time window is too short, 
any transient deviation from the long term average will be incorporated in fixed effects, 
since random effects are forced to have zero mean and on, short timescales, there may be just very 
little variation around the mean.
For instance, if our analysis was based solely on data from 2020, it would
be likely to be biased due to an exceptional, but transient, perturbation, that is, the COVID
pandemic. However, since our time window spans more than four years, any transient perturbations
have much more space for either regressing to the long-term average or canceling each other out
(by being distributed symmetrically around the average), so it is arguably likely that by including
random time effects our analysis gives a relatively good and unbiased estimates of long-term
trends. Note that the use of random effects is crucial, since it makes our analysis independent
from potential over- or under-representation (or over- or under-sampling) of any subinterval
within our studied time window.

At the same time assumption \ref{A:outlets} is probably satisfied to different degrees for 
different outcomes, but this cannot be verified empirically as, by design, estimates of random effects, $\mathbf{b}_t$, $\mathbf{b}_o$, are zero-centered in expectation, so any systematic 
deviation from the overall tendency in the sample gets incorporated in the fixed effects.
However, in general, as long as the outlets are sampled roughly symmetrically around the overall
\enquote{population} tendency (i.e.~an average over all outlets within a sufficiently long time 
window), the fixed effects obtained after removing outlet-specific information should be 
approximately consistent. This assumption cannot be directly tested, but our sample of outlets was 
collected with this \enquote{symmetry} in mind.

That said, it is possible that some country-specific estimates that we report are not consistent
and, therefore, biased (we note that all Maximum Likelihood estimates are biased, 
but in well-specified models they are consistent, so the bias is negligible in large samples). 
But, importantly, such a risk is much lower for the overall estimates
averaging over countries. It is so, because, by the virtue of averaging over country-level
estimates, it is possible that biases occurring for specific countries will cancel each other out.
In other words, if the country-level biases are roughly symmetric around zero
(i.e.~values for some countries are under- and for some overestimated),
then they will cancel each other during summation and averaging. Thus, we argue that the overall
estimates that we report may be considered particularly robust. The same applies---although
of course to a proportionally lesser extent---to overall estimates averaging over estimates
specific for political and non-political news.

For similar reasons we also refrain from estimating prevalence of political news, as we expect
that such estimates may be particularly sensitive to arbitrary sample characteristics
(e.g.~outlet selection, operational definition of what is considered \enquote{political} etc.).
Instead, we either construct \enquote{overall} estimates by averaging (uniformly) over estimates
specific for political and non-political news---a variant of the \enquote{Occam's razor}
principle---or study the entire range of scenarios under different assumptions about $p$,
that is, the prevalence of political news. Furthermore, when constructing overall estimates
with respect to countries, we also take simple averages over country-specific estimates.
This allows us to obtain a kind of average over national contexts, instead of producing estimates
biased towards the specificity of countries with the highest number of observations in the sample
(in our case that would be the U.S.).

\subsection{Model details}\label{app:sec:glmm:models}

The Generalized Linear Mixed Models (GLMMs) that we used for modeling 
Facebook reactions were defined as negative binomial regression with 
mixed effects and non-constant dispersion and were fitted
with Maximum Likelihood Estimation (MLE) using \texttt{glmmTMB} 
package for \texttt{R}~\citep{brooksGlmmTMBBalancesSpeed2017}. 
Their general specifications are discussed
in \nameref{sec:methods}~(\ref{sec:methods:analysis}) and here we 
provide a more detailed summary of the main model parameters
(Table~\ref{app:tab:models}).

\begin{table*}[htb!]
\caption{
    Estimated parameters of the generalized linear mixed models for predicting
    negativity (logistic regression) and counts of reactions, comments and shares
    (zero-inflated negative binomial regression)
}
\centering\tiny\sffamily

\begin{tabularx}{\textwidth}{ll|rrr|rrr|rrr|rrr}
\toprule
 &  & \multicolumn{3}{c}{Negativity} & \multicolumn{3}{c}{Reactions} & \multicolumn{3}{c}{Comments} & \multicolumn{3}{c}{Shares} \\
 &  & b & se & p & b & se & p & b & se & p & b & se & p \\
\midrule

\multirow[t]{24}{*}{fixed} & (Intercept = US) & -2.354 & 0.181 & 7.928e-39 & 6.567 & 0.173 & 0.000e+00 & 4.587 & 0.195 & 7.563e-123 & 4.454 & 0.174 & 2.893e-144 \\
 & ESP & 0.604 & 0.364 & 0.097 & -1.038 & 0.349 & 0.003 & -0.681 & 0.393 & 0.083 & -1.275 & 0.351 & 2.753e-04 \\
 & FR & 0.240 & 0.427 & 0.575 & -1.130 & 0.410 & 0.006 & -0.781 & 0.461 & 0.090 & -0.915 & 0.411 & 0.026 \\
 & PL & 0.504 & 0.322 & 0.117 & -1.743 & 0.308 & 1.535e-08 & -1.063 & 0.347 & 0.002 & -1.980 & 0.310 & 1.578e-10 \\
 & IRL & 0.395 & 0.377 & 0.295 & -1.919 & 0.361 & 1.062e-07 & -1.182 & 0.406 & 0.004 & -2.225 & 0.362 & 8.200e-10 \\
 & UK & 0.170 & 0.365 & 0.641 & -0.796 & 0.349 & 0.023 & -0.111 & 0.393 & 0.777 & -1.407 & 0.351 & 5.977e-05 \\
 & political & 0.087 & 0.139 & 0.534 & -0.278 & 0.073 & 1.296e-04 & 0.336 & 0.058 & 6.014e-09 & -0.416 & 0.094 & 9.044e-06 \\
 & negative &  &  &  & -0.410 & 0.072 & 1.385e-08 & -0.223 & 0.065 & 6.512e-04 & -0.539 & 0.089 & 1.151e-09 \\
 & ESP $\times$ political & 0.400 & 0.280 & 0.153 & 0.057 & 0.144 & 0.693 & 0.212 & 0.115 & 0.065 & 0.482 & 0.187 & 0.010 \\
 & FR $\times$ political & 0.597 & 0.329 & 0.070 & 0.308 & 0.169 & 0.069 & 0.175 & 0.134 & 0.193 & 0.665 & 0.219 & 0.002 \\
 & PL $\times$ political & -0.452 & 0.248 & 0.068 & 0.195 & 0.128 & 0.128 & -0.030 & 0.102 & 0.770 & 0.362 & 0.166 & 0.029 \\
 & IRL $\times$ political & 0.082 & 0.291 & 0.777 & 0.031 & 0.149 & 0.835 & 0.120 & 0.119 & 0.314 & 0.433 & 0.194 & 0.025 \\
 & UK $\times$ political & 0.055 & 0.281 & 0.846 & -0.140 & 0.145 & 0.333 & -0.163 & 0.115 & 0.155 & 0.337 & 0.187 & 0.072 \\
 & ESP $\times$ negative &  &  &  & 0.537 & 0.142 & 1.524e-04 & 0.441 & 0.128 & 5.476e-04 & 0.848 & 0.174 & 1.119e-06 \\
 & FR $\times$ negative &  &  &  & 0.249 & 0.167 & 0.137 & 0.386 & 0.151 & 0.011 & 0.528 & 0.206 & 0.010 \\
 & PL $\times$ negative &  &  &  & 0.256 & 0.127 & 0.044 & 0.109 & 0.115 & 0.343 & 0.683 & 0.156 & 1.185e-05 \\
 & IRL $\times$ negative &  &  &  & 0.669 & 0.148 & 6.044e-06 & 0.272 & 0.134 & 0.042 & 0.895 & 0.182 & 8.359e-07 \\
 & UK $\times$ negative &  &  &  & -0.155 & 0.144 & 0.280 & -0.324 & 0.129 & 0.012 & 0.144 & 0.176 & 0.415 \\
 & political $\times$ negative &  &  &  & 0.151 & 0.065 & 0.021 & -0.015 & 0.054 & 0.781 & 0.439 & 0.080 & 4.694e-08 \\
 & ESP $\times$ political $\times$ negative &  &  &  & -0.371 & 0.127 & 0.003 & -0.475 & 0.104 & 4.824e-06 & -0.674 & 0.156 & 1.646e-05 \\
 & FR $\times$ political $\times$ negative &  &  &  & -0.133 & 0.150 & 0.377 & -0.170 & 0.124 & 0.170 & -0.483 & 0.185 & 0.009 \\
 & PL $\times$ political $\times$ negative &  &  &  & -0.311 & 0.114 & 0.006 & -0.113 & 0.094 & 0.233 & -0.798 & 0.140 & 1.360e-08 \\
 & IRL $\times$ political $\times$ negative &  &  &  & -0.470 & 0.133 & 4.203e-04 & -0.185 & 0.110 & 0.093 & -0.794 & 0.164 & 1.289e-06 \\
 & UK $\times$ political $\times$ negative &  &  &  & 0.178 & 0.129 & 0.167 & 0.325 & 0.106 & 0.002 & -0.124 & 0.159 & 0.433 \\
\multirow[t]{10}{*}{country:outlet} & $\sigma(\text{(Intercept = US)})$ & 1.094 &  &  & 1.050 &  &  & 1.182 &  &  & 1.051 &  &  \\
 & $\sigma(\text{political})$ & 0.841 &  &  & 0.428 &  &  & 0.340 &  &  & 0.558 &  &  \\
 & $\sigma(\text{negative})$ &  &  &  & 0.418 &  &  & 0.376 &  &  & 0.513 &  &  \\
 & $\sigma(\text{political $\times$ negative})$ &  &  &  & 0.371 &  &  & 0.301 &  &  & 0.455 &  &  \\
 & $\rho(\text{(Intercept = US), political})$ & -0.897 &  &  & -0.224 &  &  & -0.144 &  &  & -0.323 &  &  \\
 & $\rho(\text{(Intercept = US), negative})$ &  &  &  & 0.012 &  &  & -0.046 &  &  & 0.037 &  &  \\
 & $\rho(\text{(Intercept = US), political $\times$ negative})$ &  &  &  & -0.051 &  &  & -0.051 &  &  & 0.057 &  &  \\
 & $\rho(\text{political, negative})$ &  &  &  & 0.577 &  &  & 0.552 &  &  & 0.629 &  &  \\
 & $\rho(\text{political, political $\times$ negative})$ &  &  &  & -0.750 &  &  & -0.652 &  &  & -0.763 &  &  \\
 & $\rho(\text{negative, political $\times$ negative})$ &  &  &  & -0.851 &  &  & -0.799 &  &  & -0.875 &  &  \\
\multirow[t]{10}{*}{country:day} & $\sigma(\text{(Intercept = US)})$ & 0.316 &  &  & 0.480 &  &  & 0.399 &  &  & 0.821 &  &  \\
 & $\sigma(\text{political})$ & 0.449 &  &  & 0.688 &  &  & 0.537 &  &  & 0.744 &  &  \\
 & $\sigma(\text{negative})$ &  &  &  & 0.673 &  &  & 0.585 &  &  & 0.832 &  &  \\
 & $\sigma(\text{political $\times$ negative})$ &  &  &  & 0.694 &  &  & 0.598 &  &  & 0.909 &  &  \\
 & $\rho(\text{(Intercept = US), political})$ & -0.671 &  &  & -0.417 &  &  & -0.454 &  &  & -0.261 &  &  \\
 & $\rho(\text{(Intercept = US), negative})$ &  &  &  & -0.388 &  &  & -0.321 &  &  & -0.144 &  &  \\
 & $\rho(\text{(Intercept = US), political $\times$ negative})$ &  &  &  & 0.381 &  &  & 0.326 &  &  & 0.205 &  &  \\
 & $\rho(\text{political, negative})$ &  &  &  & 0.649 &  &  & 0.586 &  &  & 0.524 &  &  \\
 & $\rho(\text{political, political $\times$ negative})$ &  &  &  & -0.613 &  &  & -0.549 &  &  & -0.561 &  &  \\
 & $\rho(\text{negative, political $\times$ negative})$ &  &  &  & -0.799 &  &  & -0.785 &  &  & -0.783 &  &  \\

\midrule
\multicolumn{14}{c}{Dispersion model} \\
\midrule
 
\multirow[t]{6}{*}{fixed} & (Intercept = US) &  &  &  & -0.693 & 0.004 & 0.000e+00 & -0.757 & 0.003 & 0.000e+00 & -0.943 & 0.008 & 0.000e+00 \\
 & ESP &  &  &  & -0.101 & 0.005 & 8.089e-82 & -0.250 & 0.005 & 0.000e+00 & -0.719 & 0.011 & 0.000e+00 \\
 & FR &  &  &  & 0.270 & 0.006 & 0.000e+00 & 0.144 & 0.005 & 1.510e-175 & 0.001 & 0.011 & 0.947 \\
 & PL &  &  &  & 0.108 & 0.005 & 3.833e-92 & 0.018 & 0.005 & 1.754e-04 & -0.059 & 0.011 & 3.511e-08 \\
 & IRL &  &  &  & -0.117 & 0.005 & 1.205e-101 & -0.344 & 0.005 & 0.000e+00 & -0.316 & 0.011 & 1.983e-184 \\
 & UK &  &  &  & -0.000 & 0.005 & 0.939 & 0.019 & 0.005 & 6.169e-05 & -0.435 & 0.011 & 0.000e+00 \\
country:day & (Intercept = US) &  &  &  & 0.141 &  &  & 0.126 &  &  & 0.296 &  &  \\

\bottomrule
\multicolumn{14}{l}{Number of observations: \num{6081134};} \\
\multicolumn{14}{l}{b - estimated coefficient; se - standard error; p - p-value} \\
\multicolumn{14}{l}{
    fixed - fixed effects' coefficients; 
    country:outlet - estimated parameters of random effects' distribution for outlets;
    country:day - estimated parameters of random effects' distribution for days
    (separate effects per country)
} \\
\multicolumn{14}{l}{%
    $\sigma(\cdot)$ - standard deviation of a random effects' distribution;
    $\rho(\cdot, \cdot)$ - correlation between random effects;
} \\
\end{tabularx}

\end{table*}

\subsection{Validation}\label{app:sec:glmm:validation}

The rest of this appendix discusses validation of the individual models. We focus on the assumptions
that are: (1)~empirically testable, (2)~crucial for the consistency of the models. Hence, we study
distributions of random effects---while remembering about the relative robustness of MLE-based
mixed models to moderate violations of the distributional 
assumptions~\citep{mccullochMisspecifyingShapeRandom2011}---and their correlations with
predictors, which should be null~\citep{mcneishUnnecessaryUbiquityHierarchical2017}. 
Last but not least, in each case we demonstrate consistency
of model-simulated response distributions with the raw data. All simulations were based on
drawing $R$ samples from the model-based response distribution for each observation (post).
Importantly, this is different from studying \enquote{population}-level estimates discussed
in the Main Text, as the goal here is to assess the extent to which the model reproduces
sample-based distributions when conditioned exactly on predictor and random effect values observed
in the sample.

\clearpage
\subsubsection{Negativity}\label{app:sec:glmm:validation:negativity}

In general, distributions of estimated random effects are roughly symmetric and close to normal
in all cases (Fig.~\ref{app:fig:validation:negativity}a-b). 
That said, time-specific effects are more regular, while outlet-specific effects
violate the normality assumption to a noticeably larger degree. However, none of the deviations
seem to be strong enough to be problematic given the relative robustness of GLMMs to 
misspecification of random effects' distributions~\citep{mccullochMisspecifyingShapeRandom2011}.
Similarly, correlations between predictors and time-effects are effectively zero, while the
same correlations for outlet-effects are somewhat larger 
(Fig.~\ref{app:fig:validation:negativity}c)---but still small enough to likely not
affect the consistency of the model in any consequential manner. Last but not least,
as demonstrated in Fig.~\ref{app:fig:validation:negativity}d, simulated response distributions
are in all cases consistent with the observed values.

\begin{figure}[tb!]
\centering
\begin{subfigure}[t]{.66\textwidth}
    \centering
    \begin{subfigure}[t]{.495\textwidth}
        \caption{}
        \vspace{-1.2em}
        \centering
        \includegraphics[width=.925\textwidth]{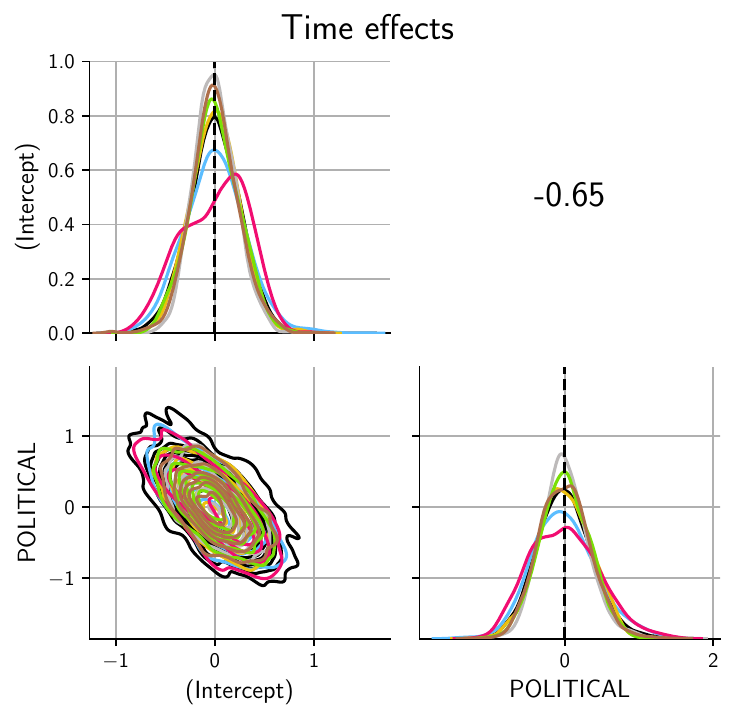}
    \end{subfigure}
    \hfill
    \begin{subfigure}[t]{.495\textwidth}
        \caption{}
        \vspace{-1.2em}
        \centering
        \includegraphics[width=.925\textwidth]{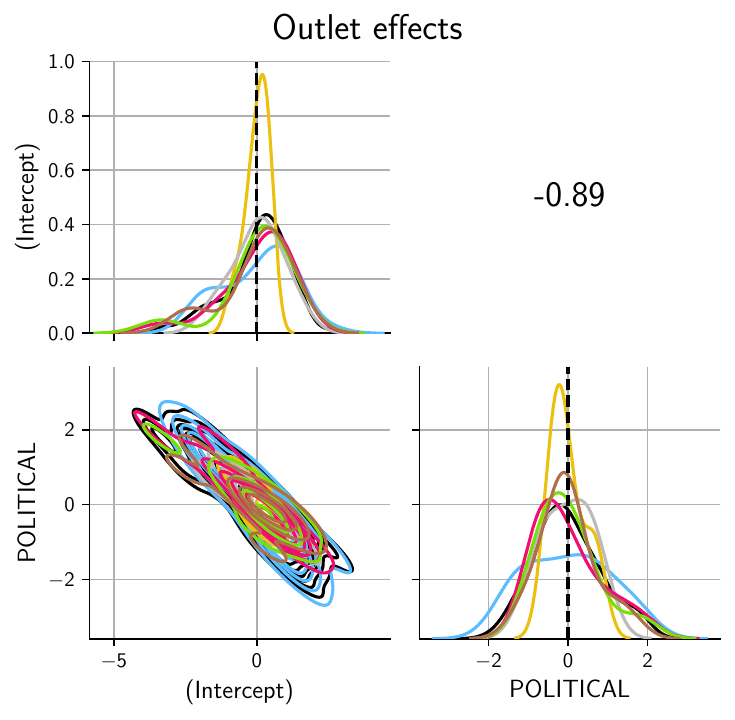}
    \end{subfigure}
\end{subfigure}
\begin{subfigure}[t]{.33\textwidth}
    \caption{}
    \vspace{-1.2em}
    \centering 
    \includegraphics[width=.925\textwidth]{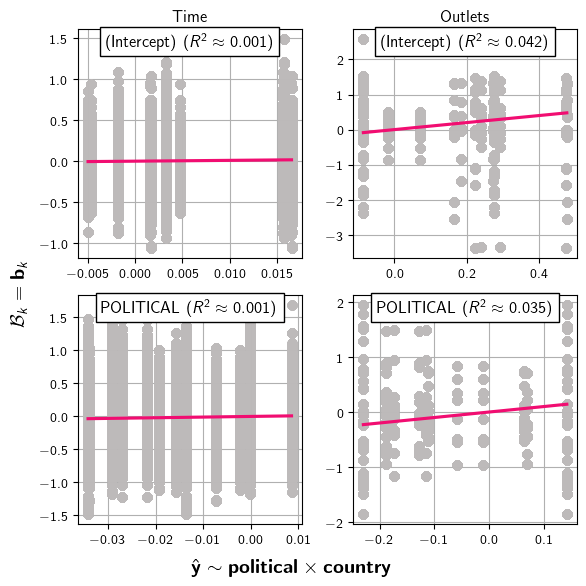}
\end{subfigure}
\includegraphics[width=.6\textwidth]{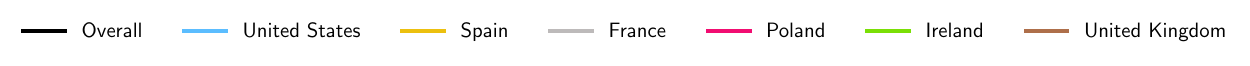}
\begin{subfigure}[t]{\textwidth}
    \caption{}
    \centering
    \includegraphics[width=\textwidth]{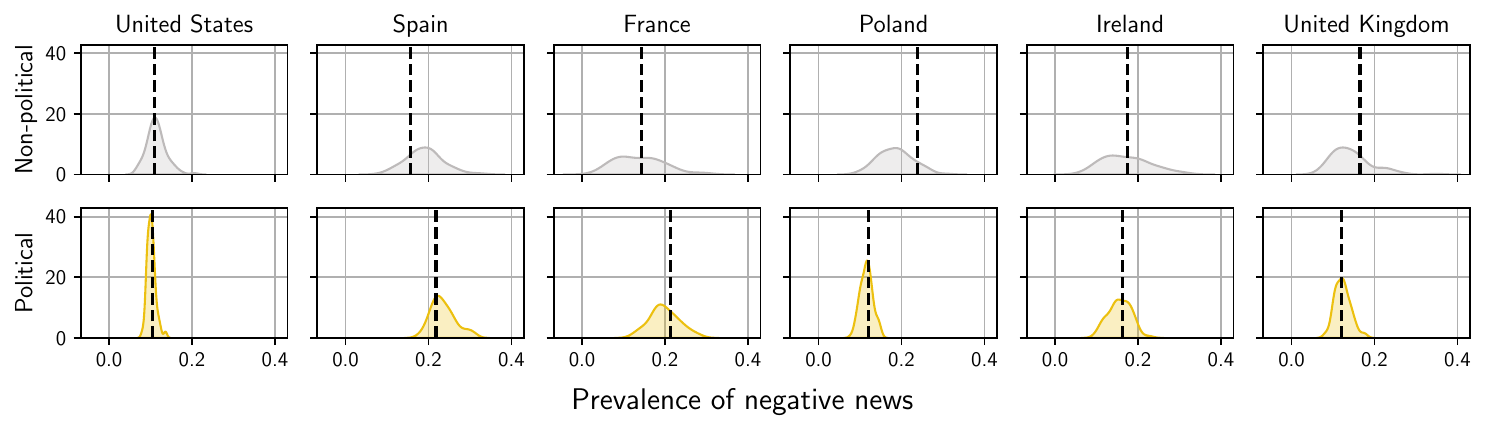} 
\end{subfigure}
\caption{%
    Validation of the GLMM for prevalence of negativity.
    \textbf{a-b}~Distributions of time-specific and outlet-specific random effects. Above diagonal
    panels show Pearson correlations between estimates for specific effects.
    \textbf{c}~Correlations between (fixed effect) predictors and estimated random effects.
    \textbf{d}~Distributions of simulated negativity prevalence ($R = 100$) with vertical lines 
    denoting the observed values based on raw data.
}
\label{app:fig:validation:negativity}
\end{figure}

\clearpage
\subsubsection{Engagement}\label{app:sec:glmm:validation:engagemen}

The distributions of random effects for reactions (Fig.~\ref{app:fig:validation:reactions}a-b) are all
roughly symmetric and normal-like. Time effects are particularly well-structured while outlet
effects are more irregular but neither seem to be severely non-Gaussian. Again, correlations
with predictors (Fig.~\ref{app:fig:validation:reactions}c) are close to null for time and positive,
but very low, for outlets, which points to good model specification. 
Fig.~\ref{app:fig:validation:reactions}c presents Complementary Cumulative Distribution Functions
(CCDFs) for observed and simulated distributions ($R = 10$ draws per observation). Clearly,
in most cases simulated distributions follow the observed ones very closely, with some,
but mostly minor, deviations in the tails. In particular, the observed dispersion seems to be 
somewhat higher among non-negative, non-political news. This suggest that a model with dispersion 
specific for news types could provide a slightly better fit, but it would require an even larger 
dataset to do it effectively, so we decided to use a simpler---but still evidently very 
effective---model. A second minor inconsistency are higher fractions of zeros in simulated data. 
This points to a  slight inadequacy of the negative binomial model, which cannot represent the 
fact that news content rarely gets exactly zero attention. However, this issue is of little 
consequence for any downstream analyses due to general high level of consistency between the 
distributions.

\begin{figure}[htb!]
\centering
\begin{subfigure}{.9\textwidth}
\begin{subfigure}[t]{.79\textwidth}
    \centering
    \begin{subfigure}[t]{.495\textwidth}
        \caption{}
        \centering
        \vspace{-1em}
        \includegraphics[width=.9\textwidth]{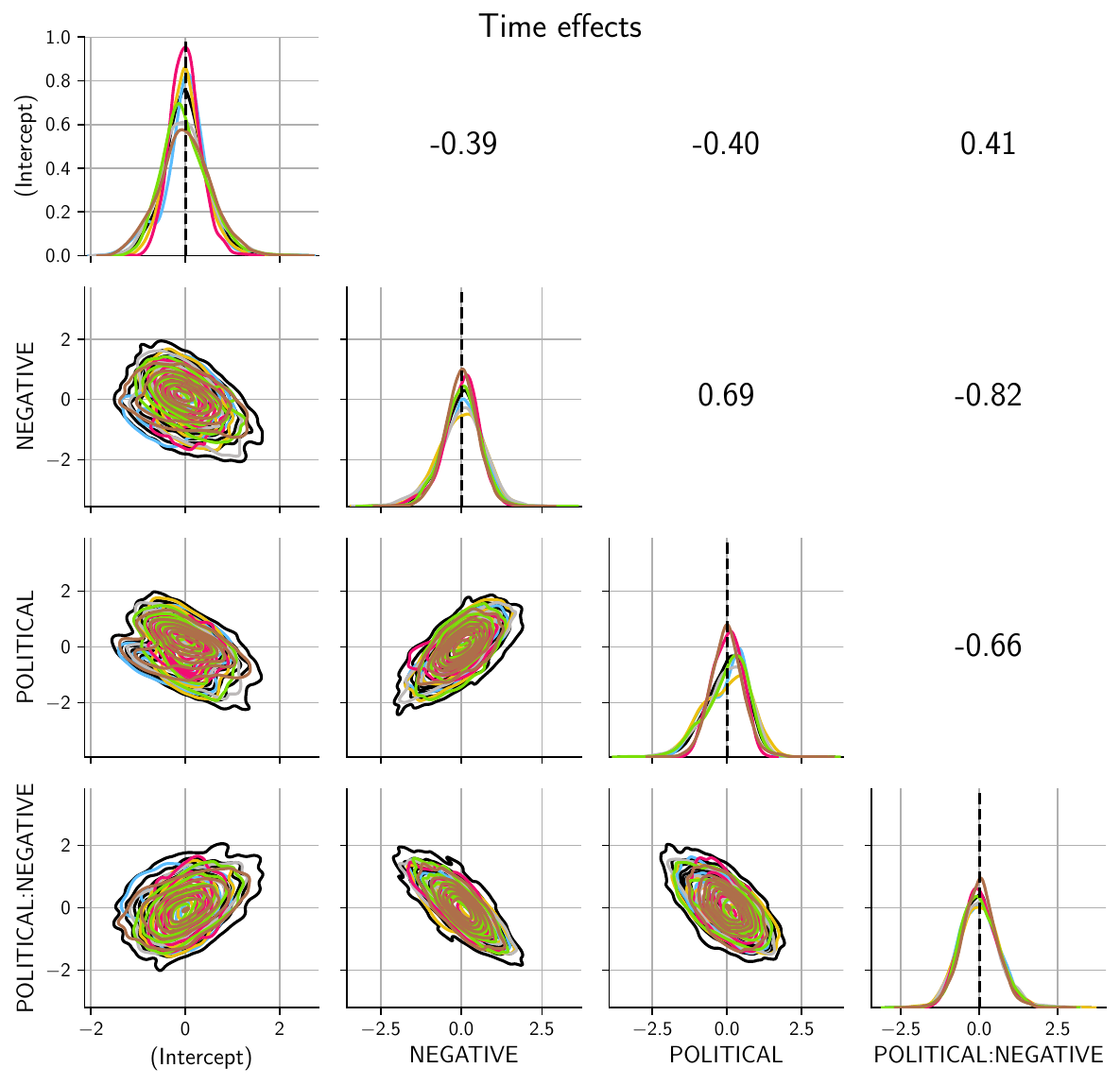}
    \end{subfigure}
    \hfill
    \begin{subfigure}[t]{.495\textwidth}
        \caption{}
        \centering
        \vspace{-1em}
        \includegraphics[width=.9\textwidth]{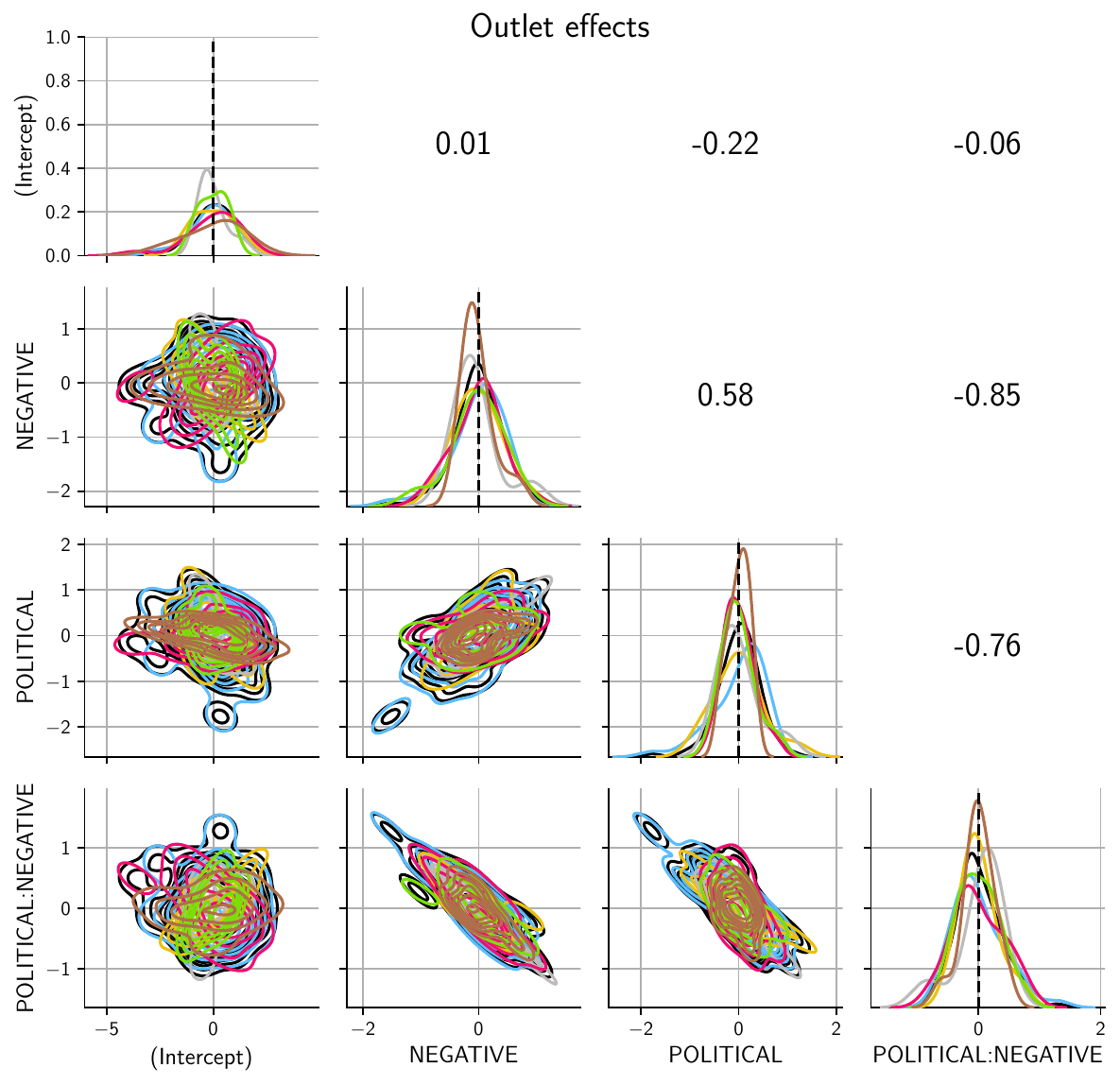}
    \end{subfigure}
\end{subfigure}
\hfill
\begin{subfigure}[t]{.195\textwidth}
    \caption{}
    \centering 
    \vspace{-1em}
    \includegraphics[width=.9\textwidth]{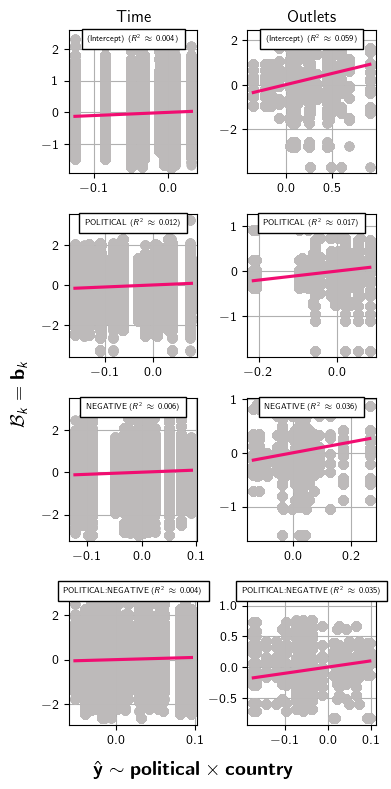}
\end{subfigure}
\end{subfigure}
\includegraphics[width=.6\textwidth]{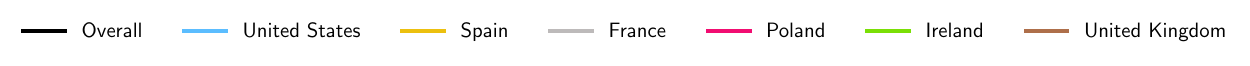}
\hfill
\begin{subfigure}[t]{\textwidth}
    \caption{}
    \centering 
    \vspace{-2em}
    \includegraphics[width=.9\textwidth]{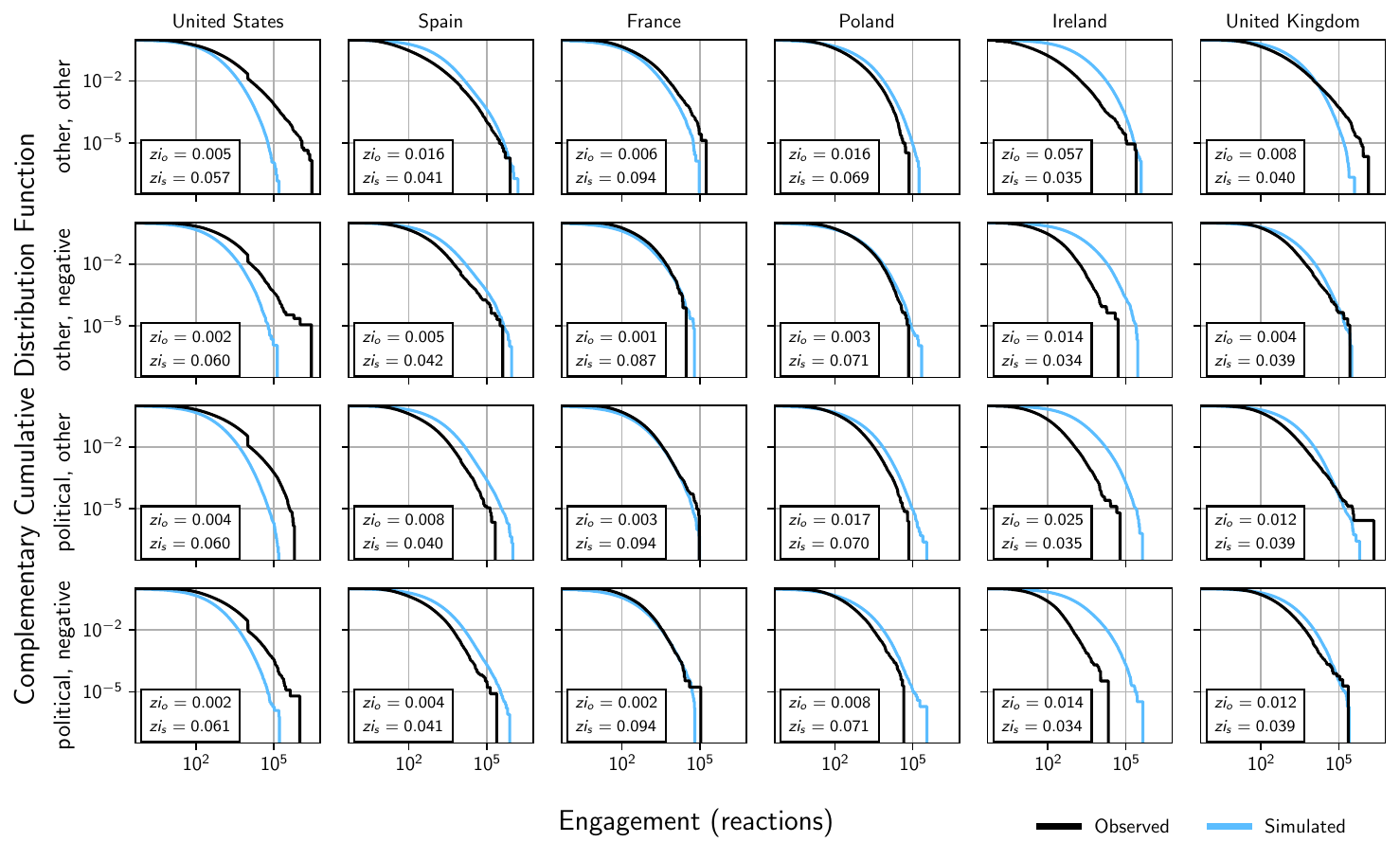}
\end{subfigure}
\caption{
    \textbf{a-b}~Distributions of estimated random effects for time and outlets.
    \textbf{c}~Correlation of estimated random effects with predictors.
    \textbf{d}~Observed and simulated CCDFs with $zi_o$ and $zi_s$ scores denoting the frequencies
    of zeros within observed and simulated data.
}
\label{app:fig:validation:reactions}
\end{figure}

The model for comments (Fig.~\ref{app:fig:validation:comments}) 
performs very similarly to the model for reactions and have largely
\enquote{well-behaved} distributions of random effects, with outlet effects being more erratic
but in no case pathological to the extent that would be problematic given the relative robustness
of GLMMs~\citep{mccullochMisspecifyingShapeRandom2011}. Correlations between predictors and
random effects are also either effectively null or low. Furthermore, simulated distributions
follow the observed ones quite closely with some relatively minor exceptions in the tails.
These deviations could be, in principle, be addressed by using a more complex dispersion model,
but that would require a larger dataset to be effective, and it is rather unlikely it would
lead to any qualitative differences in the results vis-à-vis our research questions. Similarly,
the model produces a bit too many zeros, but, again, this is non-consequential for our research
questions, since, in general, the observed distributions are reproduces are rather faithfully.

\begin{figure}[htb!]
\centering
\begin{subfigure}{.9\textwidth}
\begin{subfigure}[t]{.79\textwidth}
    \centering
    \begin{subfigure}[t]{.495\textwidth}
        \caption{}
        \centering
        \vspace{-1em}
        \includegraphics[width=.9\textwidth]{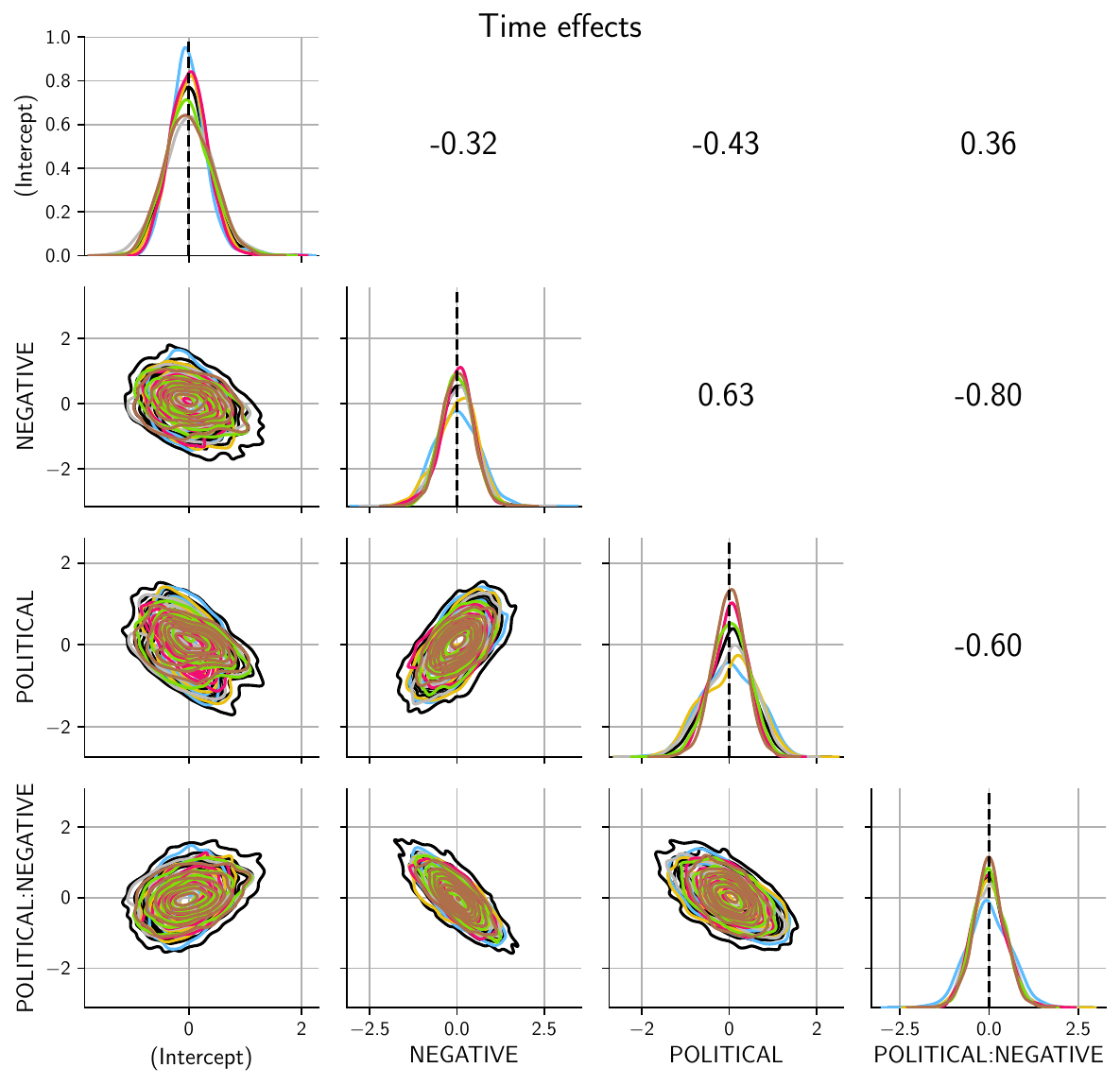}
    \end{subfigure}
    \hfill
    \begin{subfigure}[t]{.495\textwidth}
        \caption{}
        \centering
        \vspace{-1em}
        \includegraphics[width=.9\textwidth]{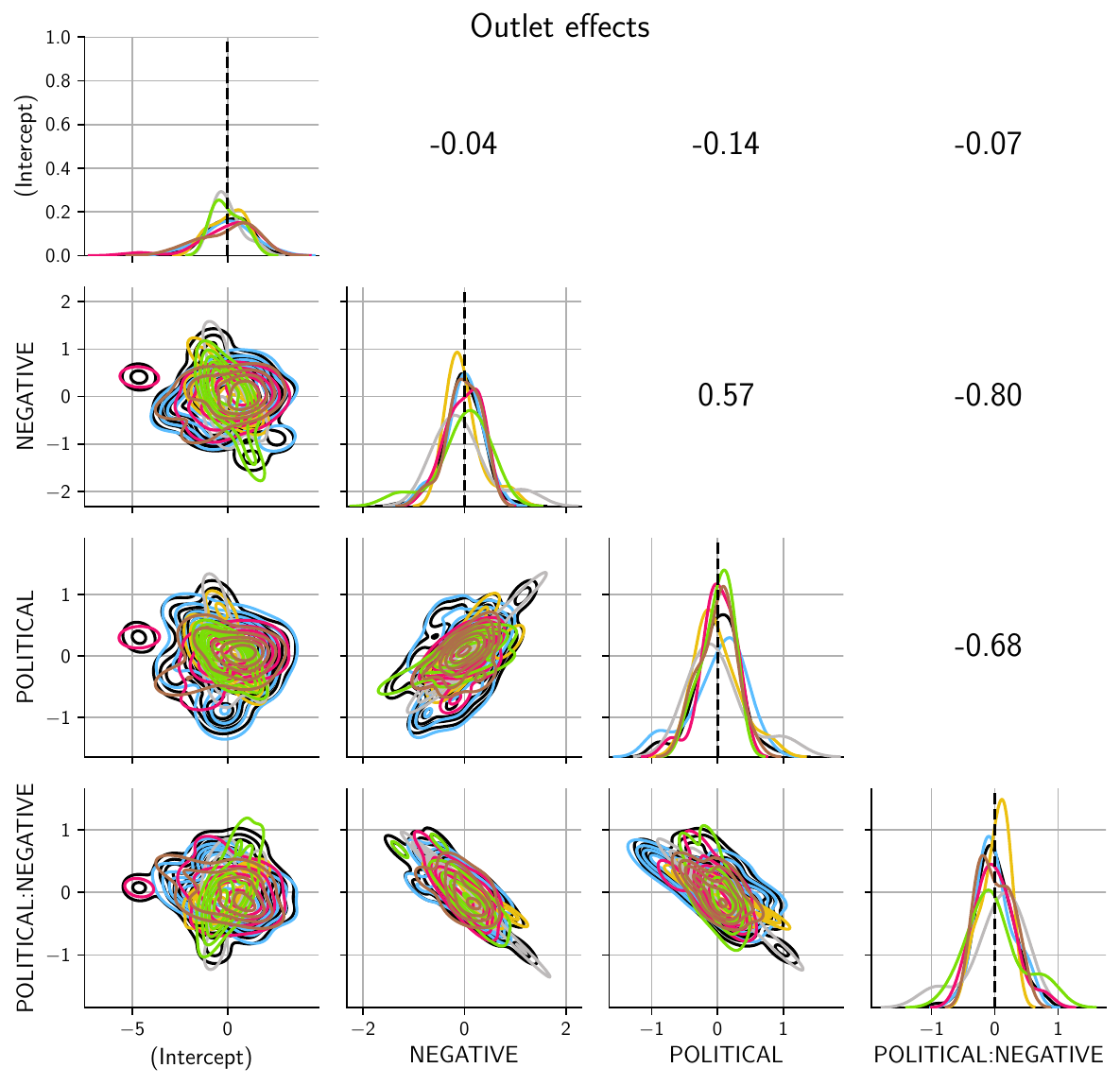}
    \end{subfigure}
\end{subfigure}
\hfill
\begin{subfigure}[t]{.195\textwidth}
    \caption{}
    \centering 
    \vspace{-1em}
    \includegraphics[width=.9\textwidth]{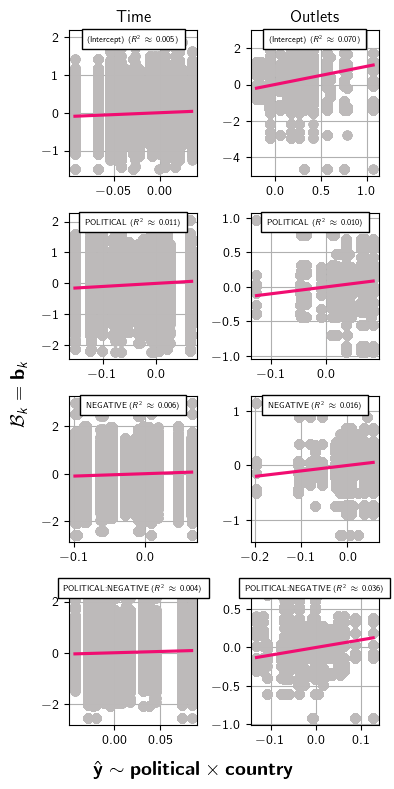}
\end{subfigure}
\end{subfigure}
\includegraphics[width=.7\textwidth]{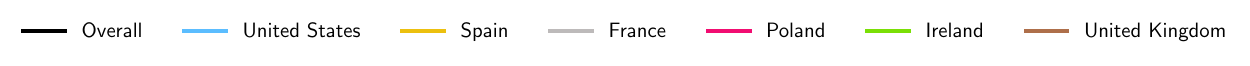}
\hfill
\begin{subfigure}[t]{\textwidth}
    \caption{}
    \centering 
    \vspace{-2em}
    \includegraphics[width=.9\textwidth]{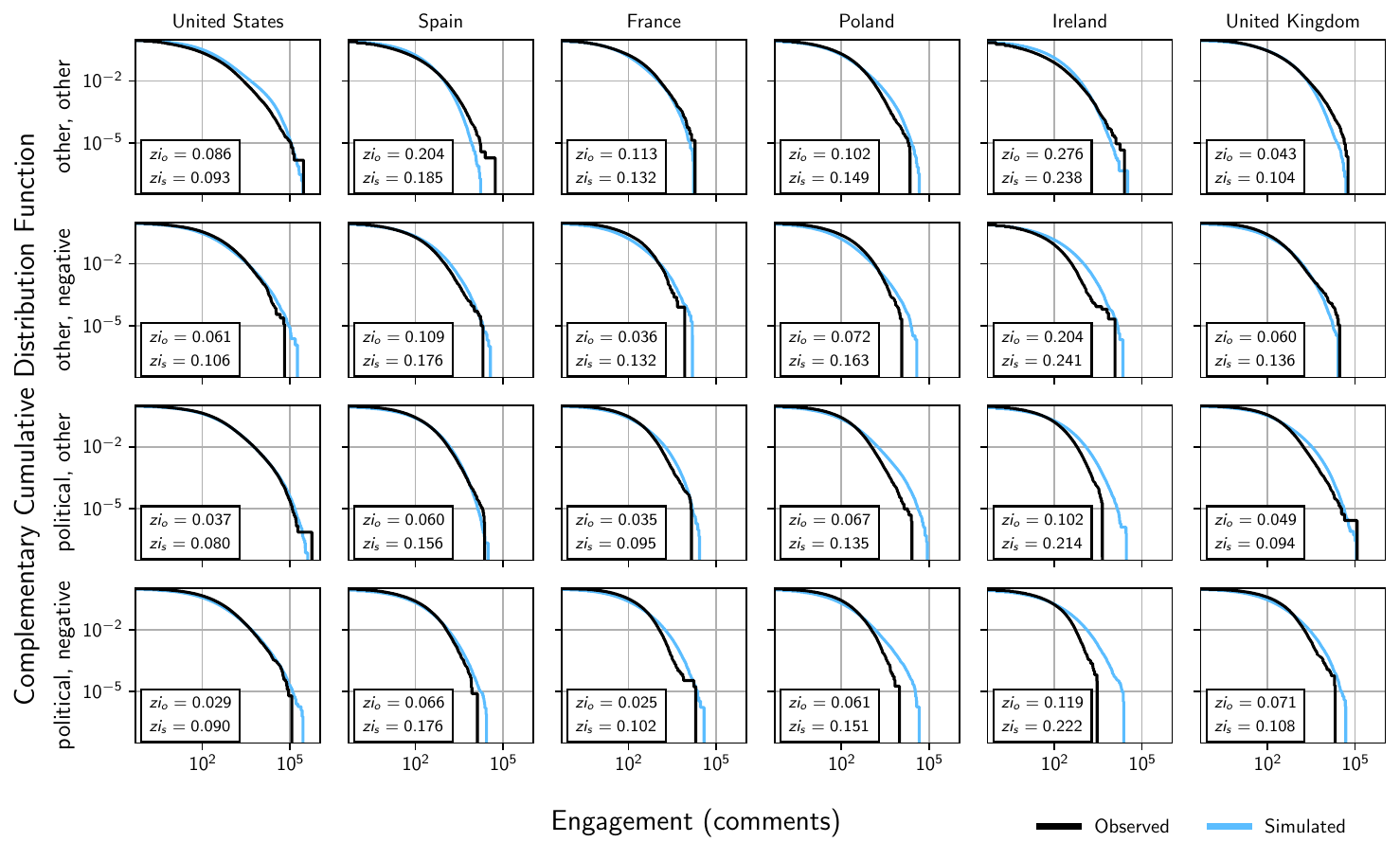}
\end{subfigure}
\caption{
    \textbf{a-b}~Distributions of estimated random effects for time and outlets.
    \textbf{c}~Correlation of estimated random effects with predictors.
    \textbf{d}~Observed and simulated CCDFs with $zi_o$ and $zi_s$ scores denoting the frequencies
    of zeros within observed and simulated data.
}
\label{app:fig:validation:comments}
\end{figure}

The situation in the case of the model for shares is analogous. This is not surprising,
of course, as all three models have the same structure, and were fitted on similar dependent
variables in the same dataset. Thus, we present the results (Fig.~\ref{app:fig:validation:shares})
without any additional commentary.

\begin{figure}[htb!]
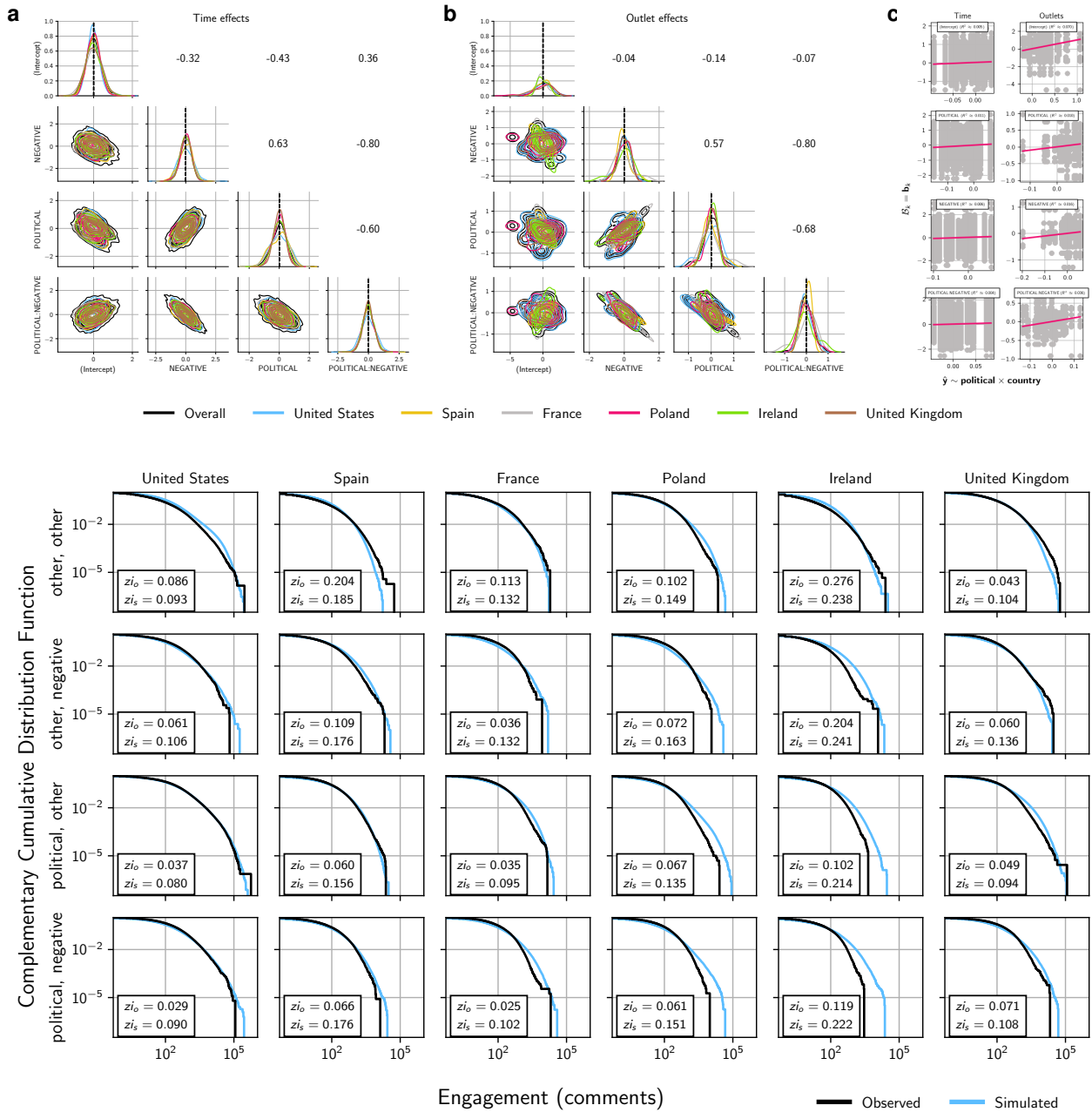

\centering
\begin{subfigure}{.9\textwidth}
\begin{subfigure}[t]{.79\textwidth}
    \centering
    \begin{subfigure}[t]{.495\textwidth}
        \caption{}
        \centering
        \vspace{-1em}
        \includegraphics[width=.9\textwidth]{validation/shares/random-effects-time}
    \end{subfigure}
    \hfill
    \begin{subfigure}[t]{.495\textwidth}
        \caption{}
        \centering
        \vspace{-1em}
        \includegraphics[width=.9\textwidth]{validation/shares/random-effects-outlets}
    \end{subfigure}
\end{subfigure}
\hfill
\begin{subfigure}[t]{.195\textwidth}
    \caption{}
    \centering 
    \vspace{-1em}
    \includegraphics[width=.9\textwidth]{validation/shares/zero-cov}
\end{subfigure}
\end{subfigure}
\includegraphics[width=.7\textwidth]{validation/shares/legend}
\hfill
\begin{subfigure}[t]{\textwidth}
    \caption{}
    \centering 
    \vspace{-2em}
    \includegraphics[width=.9\textwidth]{validation/shares/simulation-ccdf}
\end{subfigure}
\caption{
    \textbf{a-b}~Distributions of estimated random effects for time and outlets.
    \textbf{c}~Correlation of estimated random effects with predictors.
    \textbf{d}~Observed and simulated CCDFs with $zi_o$ and $zi_s$ scores denoting the frequencies
    of zeros within observed and simulated data.
}
\label{app:fig:validation:shares}
\end{figure}

Lastly, we note that the issue of too high frequencies of zeros in the simulated data could be,
probably, solved by using hurdle models (with truncated negative binomial distribution).
However, as already noted, this would rather not lead to any substantial differences in the
downstream analyses, while introducing a lot of extra parameters and, as a result,
unnecessarily dilute the available information.

\section{Outlet-level heterogeneity of negativity effects}
\label{app:sec:heterogeneity}

Here we show detailed evidence for high outlet-level heterogeneity
of negativity effects across the six studied countries. These results
are important as they explain why studies based on markedly narrower
outlet samples sometimes produce significant results where we report
null outcomes.

\begin{figure}[p]
\begin{subfigure}[t]{\textwidth}
    \caption{}
    \centering    
    \includegraphics[width=.6\textwidth]{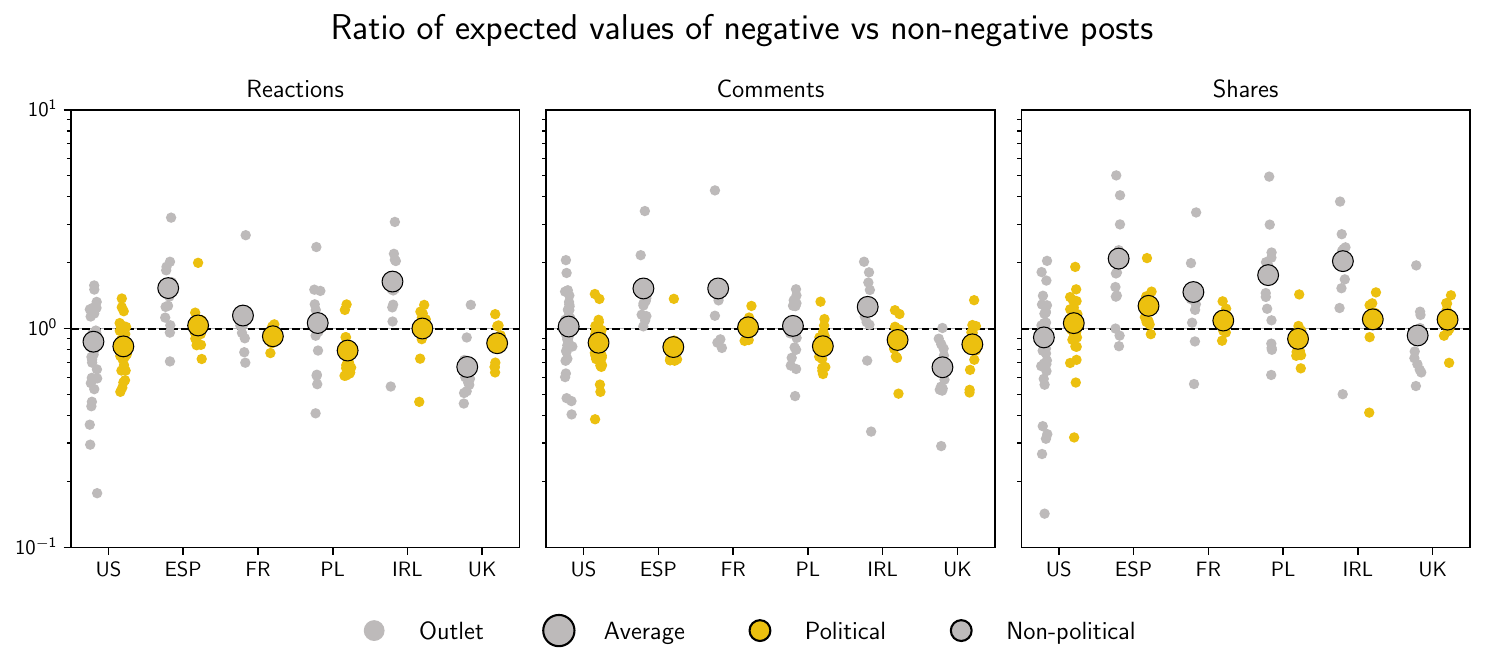}
\end{subfigure}
\begin{subfigure}[t]{\textwidth}
    \caption{}
    \centering    
    \includegraphics[width=.6\textwidth]{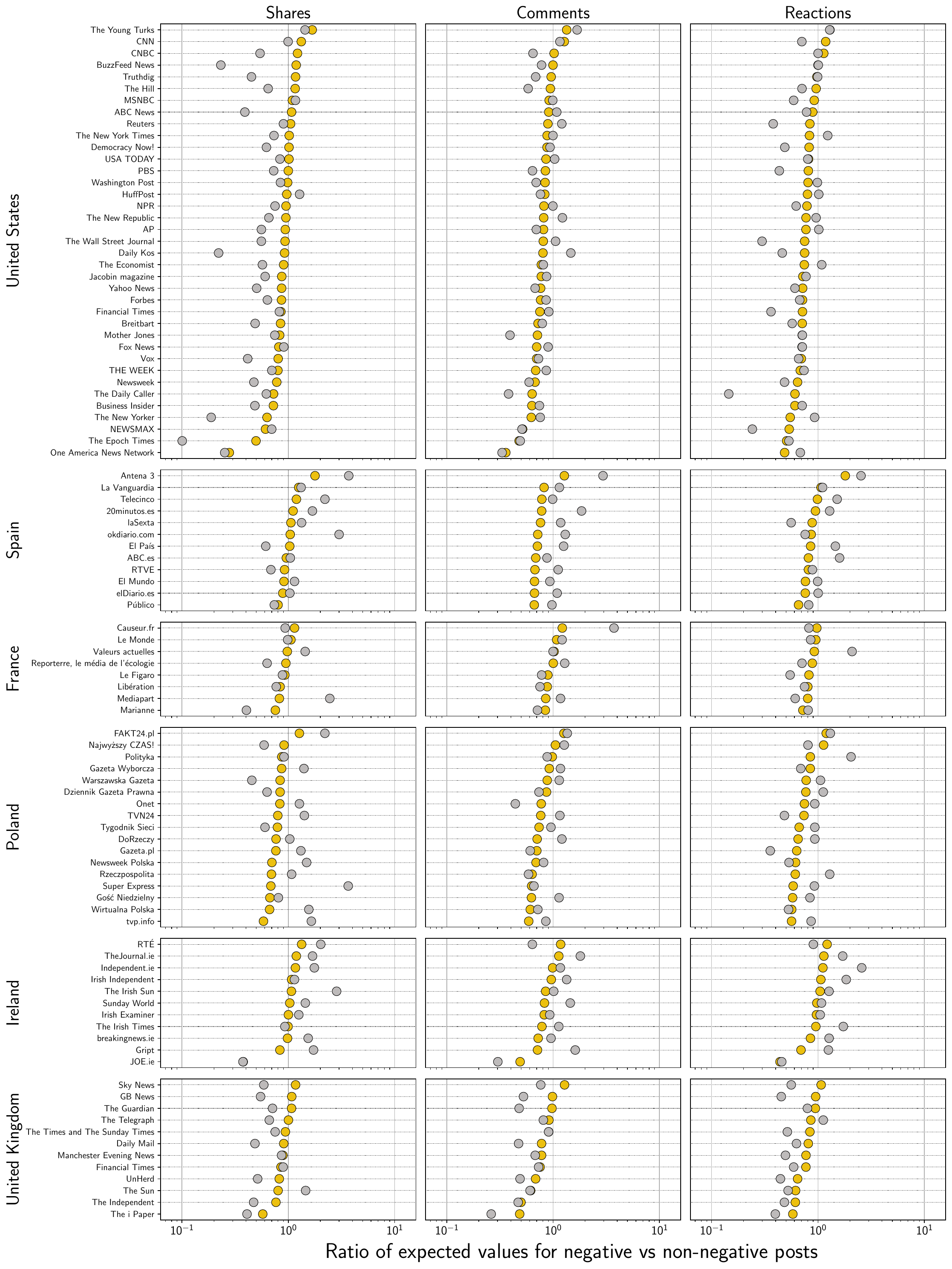}
\end{subfigure}
\caption{
    Heterogeneity of outlet-level negativity effects.
    \textbf{a}~Ratios of expected engagement counts for negative
    and non-negative posts for outlets and on average.
    \textbf{b}~Ratios of expected engagement counts for specific
    outlets sorted in descending order by ratios associated with
    political news.
}
\label{app:fig:heterogeneity}
\end{figure}

As evident on Fig.~\ref{app:fig:heterogeneity}, in all countries,
irrespective of their average locations, there are outlets---in most
cases multiple---that with either markedly negative or positive
effects of negativity on engagement, and the heterogeneity is
particularly pronounced in the case of non-political news.
This is the reason why our analysis yields some null results
that seem surprising in the light of some previous works such
as~\citep{robertsonNegativityDrivesOnline2023,heidenreichExploringEngagementEU2022}. Namely, our study considers a much broader spectrum
of news outlets (and therefore also content and audience profiles),
and once such a broad spectrum is included, it becomes apparent
that some of the effects that appear to be significant for more
idiosyncratic outlet samples, do not hold in a broader perspective.

\section{Detailed results}\label{app:sec:results}

Here we present exact numerical values for all point and interval estimates
reported and discussed in the Main Text.

\subsection{Population-level estimates}\label{app:sec:results:glmm}

Table~\ref{app:tab:emm:negativity} presents values of the estimates of prevalence of negativity
across political and/or other news as well as different countries. All point estimates are
accompanied by 95\% confidence intervals using simultaneous inference~\citep{hothornSimultaneousInferenceGeneral2008}
corresponding to $p$-value correction with Holm-Bonferroni
method~\citep{holmSimpleSequentiallyRejective1979}. Overall estimates were derived by taking
simple arithmetic averages over political/non-political or country estimates.

\begin{table}[hbt!]
\caption{Point and interval estimates of negativity probabilities and odds ratios (political / other)} 
\centering\tiny\sffamily

\begin{tabularx}{\columnwidth}{XX|XXX|XXX}
\toprule
 &  & \multicolumn{3}{c}{Probability} & \multicolumn{3}{c}{Contrast (political / non-political)} \\
 &  & $\mathbb{P}$ & 2.5\% & 97.5\% & OR & 2.5\% & 97.5\% \\
\midrule
\multirow[t]{2}{*}{United States} & non-political & 0.087 & 0.054 & 0.138 & 1.091 & 0.749 & 1.587 \\
 & political & 0.094 & 0.075 & 0.116 &  &  &  \\
\multirow[t]{2}{*}{Spain} & non-political & 0.148 & 0.066 & 0.301 & 1.628 & 0.846 & 3.130 \\
 & political & 0.221 & 0.157 & 0.301 &  &  &  \\
\multirow[t]{2}{*}{France} & non-political & 0.108 & 0.038 & 0.268 & 1.982 & 0.888 & 4.422 \\
 & political & 0.193 & 0.125 & 0.285 &  &  &  \\
\multirow[t]{2}{*}{Poland} & non-political & 0.136 & 0.068 & 0.252 & 0.694 & 0.400 & 1.205 \\
 & political & 0.098 & 0.071 & 0.134 &  &  &  \\
\multirow[t]{2}{*}{Ireland} & non-political & 0.124 & 0.052 & 0.267 & 1.184 & 0.597 & 2.351 \\
 & political & 0.143 & 0.097 & 0.206 &  &  &  \\
\multirow[t]{2}{*}{United Kingdom} & non-political & 0.101 & 0.043 & 0.218 & 1.152 & 0.597 & 2.224 \\
 & political & 0.115 & 0.079 & 0.165 &  &  &  \\
 \midrule
\multirow[t]{2}{*}{\textbf{Overall}} & non-political & 0.116 & 0.090 & 0.147 & 1.222 & 0.943 & 1.584 \\
 & political & 0.138 & 0.123 & 0.154 &  &  &  \\
\bottomrule
\end{tabularx}

\label{app:tab:emm:negativity}
\end{table}

\begin{table}[hbt!]
\caption{Country vs rest contrasts for the prevalence of negativity}
\centering\tiny\sffamily

\begin{tabularx}{\columnwidth}{XX|XXX}
\toprule
 &  & OR & 2.5\% & 97.5\% \\
\midrule
\multirow[t]{3}{*}{United States} & political & 0.595 & 0.432 & 0.820 \\
 & non-political & 0.682 & 0.341 & 1.366 \\
 & overall & 0.637 & 0.398 & 1.019 \\
\multirow[t]{3}{*}{Spain} & political & 1.986 & 1.234 & 3.194 \\
 & non-political & 1.408 & 0.503 & 3.940 \\
 & overall & 1.672 & 0.833 & 3.357 \\
\multirow[t]{3}{*}{France} & political & 1.625 & 0.925 & 2.855 \\
 & non-political & 0.910 & 0.268 & 3.085 \\
 & overall & 1.216 & 0.532 & 2.779 \\
\multirow[t]{3}{*}{Poland} & political & 0.633 & 0.418 & 0.959 \\
 & non-political & 1.249 & 0.508 & 3.071 \\
 & overall & 0.889 & 0.484 & 1.635 \\
\multirow[t]{3}{*}{Ireland} & political & 1.055 & 0.643 & 1.729 \\
 & non-political & 1.095 & 0.377 & 3.185 \\
 & overall & 1.075 & 0.522 & 2.215 \\
\multirow[t]{3}{*}{United Kingdom} & political & 0.780 & 0.484 & 1.254 \\
 & non-political & 0.837 & 0.298 & 2.348 \\
 & overall & 0.808 & 0.402 & 1.623 \\
\bottomrule
\end{tabularx}

\label{app:tab:emm:negativity-country}
\end{table}

Table~\ref{app:tab:emm:engagement} presents analogous values for estimates of ratios of the
mean engagement counts for negative and non-negative news. We restrain from analyzing engagement
in absolute terms (e.g.~in terms of differences between means), since for heavy-tailed distributions
these may not be very meaningful, that is, any exact estimated value may fluctuate quite a bit
between different sample realizations.

\begin{table}[hbt!]
\caption{Ratios of the means of engagement counts for negative and non-negative posts}
\centering\tiny\sffamily

\begin{tabularx}{\columnwidth}{ll|XXX|XXX|XXX}
\toprule
 &  & \multicolumn{3}{c}{Reactions} & \multicolumn{3}{c}{Comments} & \multicolumn{3}{c}{Shares} \\
 &  & MR & 2.5\% & 97.5\% & MR & 2.5\% & 97.5\% & MR & 2.5\% & 97.5\% \\
\midrule
\multirow[t]{3}{*}{United States} & political & 0.772 & 0.687 & 0.867 & 0.788 & 0.700 & 0.888 & 0.904 & 0.791 & 1.034 \\
 & non-political & 0.664 & 0.533 & 0.827 & 0.800 & 0.656 & 0.976 & 0.583 & 0.446 & 0.763 \\
 & overall & 0.716 & 0.619 & 0.827 & 0.794 & 0.690 & 0.915 & 0.726 & 0.610 & 0.864 \\
\multirow[t]{3}{*}{Spain} & political & 0.911 & 0.748 & 1.110 & 0.762 & 0.623 & 0.933 & 1.077 & 0.860 & 1.349 \\
 & non-political & 1.135 & 0.784 & 1.645 & 1.244 & 0.892 & 1.737 & 1.362 & 0.864 & 2.149 \\
 & overall & 1.017 & 0.795 & 1.302 & 0.974 & 0.766 & 1.239 & 1.211 & 0.901 & 1.629 \\
\multirow[t]{3}{*}{France} & political & 0.867 & 0.682 & 1.103 & 0.979 & 0.765 & 1.253 & 0.946 & 0.719 & 1.244 \\
 & non-political & 0.851 & 0.538 & 1.347 & 1.178 & 0.778 & 1.784 & 0.989 & 0.562 & 1.739 \\
 & overall & 0.859 & 0.634 & 1.163 & 1.074 & 0.798 & 1.445 & 0.967 & 0.672 & 1.393 \\
\multirow[t]{3}{*}{Poland} & political & 0.730 & 0.617 & 0.864 & 0.786 & 0.661 & 0.934 & 0.806 & 0.664 & 0.978 \\
 & non-political & 0.857 & 0.624 & 1.178 & 0.892 & 0.670 & 1.189 & 1.154 & 0.782 & 1.704 \\
 & overall & 0.791 & 0.641 & 0.977 & 0.837 & 0.682 & 1.028 & 0.964 & 0.749 & 1.242 \\
\multirow[t]{3}{*}{Ireland} & political & 0.942 & 0.762 & 1.166 & 0.860 & 0.690 & 1.071 & 1.001 & 0.784 & 1.278 \\
 & non-political & 1.296 & 0.876 & 1.917 & 1.051 & 0.737 & 1.497 & 1.427 & 0.882 & 2.312 \\
 & overall & 1.105 & 0.851 & 1.435 & 0.951 & 0.736 & 1.228 & 1.195 & 0.873 & 1.637 \\
\multirow[t]{3}{*}{United Kingdom} & political & 0.790 & 0.647 & 0.964 & 0.790 & 0.644 & 0.968 & 0.922 & 0.733 & 1.159 \\
 & non-political & 0.568 & 0.390 & 0.829 & 0.579 & 0.412 & 0.814 & 0.673 & 0.423 & 1.070 \\
 & overall & 0.670 & 0.522 & 0.860 & 0.676 & 0.530 & 0.863 & 0.788 & 0.583 & 1.064 \\
 \midrule
\multirow[t]{3}{*}{\textbf{Overall}} & political & 0.832 & 0.769 & 0.900 & 0.824 & 0.760 & 0.894 & 0.939 & 0.858 & 1.028 \\
 & non-political & 0.860 & 0.741 & 0.998 & 0.928 & 0.811 & 1.061 & 0.977 & 0.814 & 1.173 \\
 & overall & 0.846 & 0.767 & 0.933 & 0.874 & 0.794 & 0.963 & 0.958 & 0.851 & 1.078 \\
\bottomrule
\end{tabularx}

\label{app:tab:emm:engagement}
\end{table}

\subsection{Volume of engagement linked to negative news}\label{app:sec:results:volume}

Table~\ref{app:tab:volume} presents estimates of the relative volumes of engagement linked to
negative news for all three metrics---reactions, comments and shares. 
As explained in Sec.~\ref{sec:methods:share}, the confidence intervals were derived using
multivariate delta method.

\begin{table}[htb!]
\centering\tiny\sffamily
\caption{
Relative volume of engagement linked to negative content in political and non-political news
}

\begin{tabularx}{\textwidth}{XX|XXX|XXX|XXX}
\toprule
 &  & \multicolumn{3}{c}{Reactions} & \multicolumn{3}{c}{Comments} & \multicolumn{3}{c}{Shares} \\
 &  & estimate & 2.5\% & 97.5\% & estimate & 2.5\% & 97.5\% & estimate & 2.5\% & 97.5\% \\
\midrule
\multirow[t]{2}{*}{United States} & non-political & 0.059 & 0.051 & 0.067 & 0.071 & 0.062 & 0.079 & 0.053 & 0.044 & 0.061 \\
 & political & 0.074 & 0.069 & 0.079 & 0.076 & 0.070 & 0.081 & 0.086 & 0.079 & 0.093 \\
\multirow[t]{2}{*}{Spain} & non-political & 0.165 & 0.131 & 0.198 & 0.177 & 0.146 & 0.209 & 0.191 & 0.145 & 0.236 \\
 & political & 0.205 & 0.184 & 0.226 & 0.178 & 0.158 & 0.197 & 0.233 & 0.207 & 0.259 \\
\multirow[t]{2}{*}{France} & non-political & 0.093 & 0.067 & 0.119 & 0.124 & 0.094 & 0.154 & 0.107 & 0.071 & 0.143 \\
 & political & 0.172 & 0.150 & 0.194 & 0.190 & 0.165 & 0.214 & 0.185 & 0.158 & 0.211 \\
\multirow[t]{2}{*}{Poland} & non-political & 0.119 & 0.097 & 0.141 & 0.123 & 0.103 & 0.143 & 0.153 & 0.121 & 0.186 \\
 & political & 0.074 & 0.066 & 0.081 & 0.079 & 0.071 & 0.087 & 0.081 & 0.072 & 0.090 \\
\multirow[t]{2}{*}{Ireland} & non-political & 0.154 & 0.121 & 0.187 & 0.129 & 0.103 & 0.155 & 0.167 & 0.123 & 0.211 \\
 & political & 0.136 & 0.120 & 0.152 & 0.126 & 0.110 & 0.141 & 0.143 & 0.124 & 0.163 \\
\multirow[t]{2}{*}{United Kingdom} & non-political & 0.060 & 0.046 & 0.075 & 0.061 & 0.048 & 0.075 & 0.071 & 0.050 & 0.091 \\
 & political & 0.093 & 0.082 & 0.104 & 0.093 & 0.082 & 0.104 & 0.107 & 0.093 & 0.121 \\
 \midrule
\multirow[t]{2}{*}{\textbf{Overall}} & non-political & 0.101 & 0.092 & 0.110 & 0.108 & 0.100 & 0.117 & 0.113 & 0.101 & 0.125 \\
 & political & 0.117 & 0.112 & 0.123 & 0.116 & 0.111 & 0.122 & 0.130 & 0.124 & 0.137 \\
\bottomrule
\end{tabularx}

\label{app:tab:volume}
\end{table}

Table~\ref{app:tab:volume-overall} presents simulated relative volumes of engagement linked
to negative news under two different assumptions about the prevalence of political news.

\begin{table}[htb!]
\centering\tiny\sffamily
\caption{
    Relative volume of engagement linked to negative news for different prevalence of political
    news
}

\begin{tabularx}{\textwidth}{XXr|rrr|rrr|rrr}
\toprule
 &  &  & \multicolumn{3}{c}{Reactions} & \multicolumn{3}{c}{Comments} & \multicolumn{3}{c}{Shares} \\
 &  &  & estimate & 2.5\% & 97.5\% & estimate & 2.5\% & 97.5\% & estimate & 2.5\% & 97.5\% \\
 &  & \% political &  &  &  &  &  &  &  &  &  \\
\midrule
\multirow[t]{2}{*}{United States} & Wojcieszak et al. (2024) & 25 & 0.063 & 0.056 & 0.070 & 0.072 & 0.064 & 0.080 & 0.061 & 0.053 & 0.069 \\
 & Occam's razor & 50 & 0.067 & 0.060 & 0.073 & 0.073 & 0.066 & 0.080 & 0.069 & 0.061 & 0.077 \\
\multirow[t]{2}{*}{Spain} & Wojcieszak et al. (2024) & 25 & 0.175 & 0.144 & 0.205 & 0.177 & 0.148 & 0.206 & 0.201 & 0.160 & 0.243 \\
 & Occam's razor & 50 & 0.185 & 0.157 & 0.212 & 0.177 & 0.151 & 0.204 & 0.212 & 0.175 & 0.249 \\
\multirow[t]{2}{*}{France} & Wojcieszak et al. (2024) & 25 & 0.113 & 0.088 & 0.138 & 0.141 & 0.112 & 0.169 & 0.126 & 0.093 & 0.160 \\
 & Occam's razor & 50 & 0.133 & 0.109 & 0.157 & 0.157 & 0.130 & 0.184 & 0.146 & 0.114 & 0.177 \\
\multirow[t]{2}{*}{Poland} & Wojcieszak et al. (2024) & 25 & 0.108 & 0.088 & 0.127 & 0.112 & 0.094 & 0.130 & 0.135 & 0.106 & 0.164 \\
 & Occam's razor & 50 & 0.096 & 0.080 & 0.113 & 0.101 & 0.086 & 0.117 & 0.117 & 0.093 & 0.141 \\
\multirow[t]{2}{*}{Ireland} & Wojcieszak et al. (2024) & 25 & 0.150 & 0.120 & 0.180 & 0.128 & 0.104 & 0.152 & 0.161 & 0.122 & 0.200 \\
 & Occam's razor & 50 & 0.145 & 0.119 & 0.171 & 0.127 & 0.106 & 0.149 & 0.155 & 0.121 & 0.189 \\
\multirow[t]{2}{*}{United Kingdom} & Wojcieszak et al. (2024) & 25 & 0.069 & 0.055 & 0.082 & 0.069 & 0.057 & 0.082 & 0.080 & 0.061 & 0.099 \\
 & Occam's razor & 50 & 0.077 & 0.064 & 0.089 & 0.077 & 0.065 & 0.089 & 0.089 & 0.071 & 0.106 \\
\multirow[t]{2}{*}{\textbf{Overall}} & Wojcieszak et al. (2024) & 25 & 0.105 & 0.097 & 0.113 & 0.110 & 0.102 & 0.118 & 0.118 & 0.107 & 0.128 \\
 & Occam's razor & 50 & 0.109 & 0.102 & 0.116 & 0.112 & 0.105 & 0.119 & 0.122 & 0.112 & 0.131 \\
\bottomrule
\end{tabularx}

\label{app:tab:volume-overall}
\end{table}


\end{document}